\documentclass[structabstract]{aa}  
\usepackage{tabularx}
\usepackage{graphicx}
\usepackage{natbib}
\usepackage{txfonts}
\usepackage{longtable}
\usepackage{multirow}
\usepackage{lscape}
\usepackage{amssymb}
\usepackage{ulem}
\newcommand{\hersc}{{\it Herschel}}
\newcommand{\lab}{LABOCA}
\newcommand{\APEX}{{\it APEX}}
\newcommand{\JCMT}{{\it JCMT}}
\newcommand{\spitz}{{\it  Spitzer}}
\newcommand{\iso}{{\it ISO}}
\newcommand{\iras}{{\it IRAS}}

\newcommand{\lsun}{$L_\odot$}
\newcommand{\msun}{$M_\odot$}
\newcommand{\zsun}{$Z_\odot$}
\newcommand{\mic}{$\mu$m}

\renewcommand*{\figurename}{ Fig.}
\renewcommand*{\tablename}{Table}
\newlength{\pointwidth}
\DeclareGraphicsExtensions{.pdf,.png,.jpg, .eps}

\def\revised{}
\begin{document}

  \title{Revealing the cold dust in low-metallicity environments: \\
  	I - Photometry analysis of the Dwarf Galaxy Survey with \hersc}
  
  \author{A. R\'emy-Ruyer\inst{1} 
  	\and S.C. Madden\inst{1}
  	\and F. Galliano\inst{1}
  	\and S. Hony\inst{1}
  	\and M. Sauvage\inst{1}
  	\and G. J. Bendo\inst{2}  
  	\and H. Roussel\inst{3}
  	\and M. Pohlen\inst{4}
  	\and M. W. L. Smith\inst{4}   
  	\and M. Galametz\inst{5}   
  	\and D. Cormier\inst{1}
  	\and V. Lebouteiller\inst{1}  
  	\and R. Wu\inst{1} 
  	\and M. Baes\inst{6}  
  	\and M. J. Barlow\inst{7}  
  	\and M. Boquien\inst{8}  
  	\and A. Boselli\inst{8}  
  	\and L. Ciesla\inst{9}  
  	\and I. De Looze\inst{6}  
  	\and O. \L. Karczewski\inst{7}  
  	\and P. Panuzzo\inst{10}   
  	\and L. Spinoglio\inst{11}
	\and M. Vaccari\inst{12}  
  	\and C.D. Wilson\inst{13} 
  	\and the \hersc-SAG2 consortium.
	}

	      \institute{Laboratoire AIM, CEA, Universit\'{e} Paris Sud XI, IRFU/Service d'Astrophysique, Bat. 709, 91191 Gif-sur-Yvette, France, \\
	      \email{aurelie.remy@cea.fr} \\
                \and UK ALMA Regional Centre Node, Jodrell Bank Centre for Astrophysics, School of Physics \& Astronomy, University of Manchester, Oxford Road, Manchester M13 9PL, UK \\
                \and Institut d'Astrophysique de Paris, UMR7095 CNRS, Universit\'e Pierre \& Marie Curie, 98 bis Boulevard Arago, 75014 Paris, France \\
                \and School of Physics \& Astronomy, Cardiff University, The Parade, Cardiff, CF24 3AA, UK \\
                \and Institute of Astronomy, University of Cambridge, Madingley Road, Cambridge CB3 0HA, UK \\
                \and Sterrenkundig Observatorium, Universiteit Gent, Krijgslaan 281 S9, B-9000 Gent, Belgium \\                
                \and Department of Physics and Astronomy, University College London, Gower St, London WC1E 6BT, UK \\
                \and Laboratoire dÕAstrophysique de Marseille - LAM, Universit\'e dÕAix-Marseille \& CNRS, UMR7326, 38 rue F. Joliot-Curie, 13388 Marseille Cedex 13, France \\
                \and Department of Physics, University of Crete, GR-71003, Heraklion, Greece \\
                \and GEPI, Observatoire de Paris, CNRS, Univ. Paris Diderot, Place Jules Janssen 92190 Meudon, France \\
                \and Instituto di Astrofisica e Planetologia Spaziali, INAF-IAPS, Via Fosso del Cavaliere 100, I-00133 Roma, Italy \\
                \and Physics Department, University of the Western Cape, Private Bag X17, 7535, Bellville, Cape Town, South Africa \\
                \and Department of Physics \& Astronomy, McMaster University, Hamilton Ontario L8S 4M1 Canada \\
        }

\date{Received date/Accepted date}

%===============================================================
% Abstract
%===============================================================

 \abstract
{We present new photometric data from our \hersc\ Guaranteed Time Key Programme, the Dwarf Galaxy Survey (DGS), dedicated to the observation of the gas and dust in low-metallicity environments. A total of 48 dwarf galaxies were observed with the PACS and SPIRE instruments onboard the \hersc\ Space Observatory at 70, 100, 160, 250, 350, and 500 \mic.}
 {The goal of this paper is to provide reliable far infrared (FIR) photometry for the DGS sample and to analyse the FIR/submillimetre (submm) behaviour of the DGS galaxies. We focus on a systematic comparison of the derived FIR properties (FIR luminosity, {\it L$_{FIR}$}, dust mass, {\it M$_{dust}$}, dust temperature, {\it T}, emissivity index, $\beta$) with more metal-rich galaxies and investigate the detection of a potential submm excess.}
 {The data reduction method is adapted for each galaxy in order to derive the most reliable photometry from the final maps. The derived PACS flux densities are compared with the \spitz\ MIPS 70 and 160 \mic\ bands. We use colour-colour diagrams to analyse the FIR/submm behaviour of the DGS galaxies and modified blackbody fitting procedures to determine their dust properties. To study the variation in these dust properties with metallicity, we also include galaxies from the \hersc\ KINGFISH sample, which contains more metal-rich environments, totalling 109 galaxies.}
 {The location of the DGS galaxies on \hersc\ colour-colour diagrams highlights the differences in dust grain properties and/or global environments of low-metallicity dwarf galaxies. The dust in DGS galaxies is generally warmer than in KINGFISH galaxies ({\it T$_{DGS}$} $\sim$ 32 K and {\it T}$_{KINGFISH}$ $\sim$ 23 K). The emissivity index, $\beta$, is $\sim$ 1.7 in the DGS, however metallicity does not make a strong effect on $\beta$. The proportion of dust mass relative to stellar mass is lower in low-metallicity galaxies: {\it M$_{dust}$}/{\it M$_{star}$} $\sim$ 0.02\% for the DGS versus 0.1\% for KINGFISH. However, per unit dust mass, dwarf galaxies emit about six times more in the FIR/submm than higher metallicity galaxies. Out of the 22 DGS galaxies detected at 500 \mic, about 41\% present an excess in the submm beyond the explanation of our dust SED model, and this excess can go up to 150\% above the prediction from the model. The excess mainly appears in lower metallicity galaxies (12+log(O/H) $\lesssim$ 8.3), and the strongest excesses are detected in the most metal-poor galaxies. However, we also stress the need for observations longwards of the \hersc\ wavelengths to detect any submm excess appearing beyond 500 \mic.}
 {}


     \authorrunning{R\'emy-Ruyer et al.}
     \titlerunning{The Dwarf Galaxy Survey with \hersc}

 \maketitle

%===============================================================
%Introduction
%===============================================================

\section{Introduction}\label{intro}

The continuous interplay between stars and the interstellar medium (ISM) is one of the major drivers of galaxy evolution. The ISM is primarily composed of gas and dust, and it plays a key role in this evolution, as the repository of stellar ejecta and the site of stellar birth. It thus contains the imprint of the astrophysical processes occurring in a galaxy. Interstellar dust is present in most phases of the ISM, from warm ionized regions around young stars to the cores of dense molecular clouds. Because dust is mainly formed from the available metals in the ISM, the dust content traces its internal evolution through metal enrichment. Dust thus influences the subsequent star formation and has a significant impact on the total spectral energy distribution (SED) of a galaxy: the absorbed stellar light by dust in the ultraviolet (UV) and visible wavelengths is re-emitted in the infrared (IR) domain by the dust grains. In our Galaxy, dust reprocesses about 30\% of the stellar power, and it can grow to as large as $\sim$ 99\% in a starburst galaxy. Studying the IR emission of galaxies thus provides valuable information on the dust properties of the galaxies and on their overall star formation activity. 

Our Galaxy, as well as other well studied local Universe galaxies, provide various observational benchmarks to calibrate the physical dust properties around solar metallicity. However, for galaxies of the high-redshift Universe, dust properties are still poorly known, due to observational constraints and to the unsure variations in dust properties as the metallicity decreases. Because of their low metal abundance and active star formation, dwarf galaxies of the local Universe are ideal laboratories for studying star formation and its feedback on the ISM in conditions that may be representative of different stages in early Universe environments. 

From \iras\ to \spitz, many studies have been dedicated to dwarf galaxies over the past decades, and have uncovered peculiar ISM properties compared to their metal-rich counterparts. Among these, are the following: 

%- overall warmer dust: 
{\it Overall warmer dust}: the SEDs in some low-metallicity star-forming dwarf galaxies often peak at shorter wavelengths, sometimes well below 100 \mic, whereas for more metal-rich galaxies, the peak of the SED is around 100 - 200 \mic\ \citep{Galliano2003, Galliano2005, Walter2007, Engelbracht2008, Galametz2009}. This is a consequence of the harder interstellar radiation field (ISRF) interacting with the porous ISM of dwarf galaxies \citep[e.g.][]{Madden2006}.

%- weak/absent PAH features :
{\it Weak mid infrared (MIR) aromatic features}: the polycyclic aromatic hydrocarbons (PAHs) are often barely detected, if at all, in these galaxies \citep[e.g.][]{Sauvage1990, Madden2000, Boselli2004, Engelbracht2005}. The combination of young star clusters and metal-poor ISM creates a harder galaxy-wide radiation field compared to that of our Galaxy. The paucity of dust allows the harder UV photons to travel deeper into the ISM and destroy PAH molecules by photoevaporation or photodissociation \citep{Galliano2003, Galliano2005, Madden2006}. The dearth of PAH features in dwarf galaxies has also been explained by the destruction of the molecules by supernovae (SN) shocks \citep{OHalloran2006} or by a delayed carbon injection in the ISM by asymptotic giant branch (AGB) stars \citep{Galliano2008}.

%- submm excess: 
{\it The submillimetre (submm) excess}: an excess emission, unaccountable by usual SED models, is appearing in the FIR to submm/millimetre (mm) domain for some dwarf galaxies \citep{Galliano2003, Galliano2005, Galametz2009, Bot2010, Grossi2010}. An excess emission has also been observed in our Galaxy with COBE \citep{Reach1995} but with an intensity less pronounced compared to that found in low-metallicity systems. \cite{Dumke2004, Bendo2006, Zhu2009} found a submm excess in some low-metallicity spiral galaxies as well. The discovery of this excess renders even more uncertain the determination of a quantity as fundamental as the dust mass.

%- faint CO emission: 
{\it The faint CO emission}: CO is difficult to observe in dwarf galaxies \citep[i.e.][]{Leroy2009, Schruba2012}, and the determination of the molecular gas reservoir at low metallicities through the usual CO-to-H$_2$ conversion factor is still very uncertain. The dependence of the CO-to-H$_2$ conversion factor on metallicity has been studied extensively \citep{Wilson1995, Boselli2002, Leroy2011, Schruba2012} but have been limited to metallicities greater than $\sim$ 1/5 \zsun\footnote{Throughout, we assume (O/H)$_\odot$ = 4.90 $\times$ 10$^4$, i.e., 12+log(O/H)$_\odot$ = 8.69 \citep{Asplund2009}} due to the difficulty of detecting CO at lower metallicities. This renders accurate determinations of gas-to-dust mass ratios (G/D) very difficult, as H$_2$ may account for a significant fraction of the total (atomic HI and molecular H$_2$) gas mass. We now believe that the structure of molecular clouds in dwarf galaxies is very different from that of metal-rich systems, and that CO does not trace the full molecular gas reservoir. A potentially large reservoir of CO-dark molecular gas could exist in low-metallicity galaxies, traceable by the FIR cooling line [CII] \citep{Poglitsch1995, Israel1996, Madden1997, Madden2012}, or by neutral carbon [CI] \citep{Papadopoulos2004, Wilson2005}.

The wavelength ranges and sensitivities covered by \spitz, {\it Infrared Space Observatory} (\iso) and \iras\ do not sample the cold dust component of the dust SED beyond 160 \mic. Some ground-based telescopes such as \JCMT, \APEX, {\it SEST}, {\it IRAM} could detect the cold dust beyond 160 \mic, but because of sensitivity limitations, accurate measures of the photometry could only be obtained for the brightest and highest metallicity dwarf galaxies. The \hersc\ Space Observatory \citep{Pilbratt2010}, launched in 2009, is helping to fill this gap and complete our view of dust in galaxies by constraining the cold dust contribution. \hersc\ covers a wide range of wavelengths in the FIR and submm, with unprecedented resolution: its 3.5 m diameter mirror is the largest ever launched in space so far for this wavelength range. \hersc\ carries three instruments {\revised among which are} the Photodetector Array Camera and Spectrometer \citep[PACS -][]{Poglitsch2010} {\revised and} the Spectral and Photometric Imaging REceiver \citep[SPIRE -][]{Griffin2010}, both imaging photometres and medium resolution spectrometres. 
%\removed{and the Heterodyne Instrument for the Far Infrared %\citep[HIFI -][]{deGraauw2010},
 %a very high resolution heterodyne spectrometre.} 
 The PACS and SPIRE photometres in combination cover a 70 to 500 \mic\ range, and the spectrometres together cover 55 to 670 \mic. 

We focus here on local dwarf galaxies by presenting new results of the \hersc\ Guaranteed Time Key Progam, the Dwarf Galaxy Survey \citep[DGS - P.I. Madden ; ][]{Madden2013}. Dwarf galaxies are studied here in a systematic way, enabling us to derive general properties that are representative of these systems. We will focus our study on overall dust properties and look at the submm excess. We present the observed sample and the data reduction processes in Section \ref{datared}. We then present the flux extraction method and the flux catalogues for the whole sample in Section \ref{photo}. Section \ref{FIRbehaviour} is dedicated to the comparison of the dwarf galaxies with more metal-rich environments, first qualitatively with colour-colour diagrams, and then quantitatively with modified blackbody fits. We also inspect a sub-sample of galaxies presenting a submm excess. Throughout this last Section we compare our results with those from another \hersc\ sample, KINGFISH \citep{Kennicutt2011}, which is probing predominantely more metal-rich environments, in order to study the various overall effects of metallicity on the derived dust properties.

%===============================================================
% Observations and Data reduction
%===============================================================

\section{Observations and Data reduction}\label{datared}

 \subsection{The Dwarf Galaxy Survey with \hersc}\label{obs}

\subsubsection{Sample}
The DGS aims at studying the gas and dust properties in low-metallicity ISMs with the \hersc\ Space Observatory. It is a photometric and spectroscopic survey of 50 dwarf galaxies at FIR and submm wavelengths \citep{Madden2013}. For a more detailed description of the general goals of the survey and the source selection process, see the Dwarf Galaxy Survey Overview by \cite{Madden2013}. Here, we focus on the 48 targets for which complete photometry was obtained. The names, positions, distances and metallicities of the DGS galaxies are listed in Table \ref{sample} \citep[from][]{Madden2013}.

\begin{figure}
\begin{center}
\includegraphics[width=8.8cm]{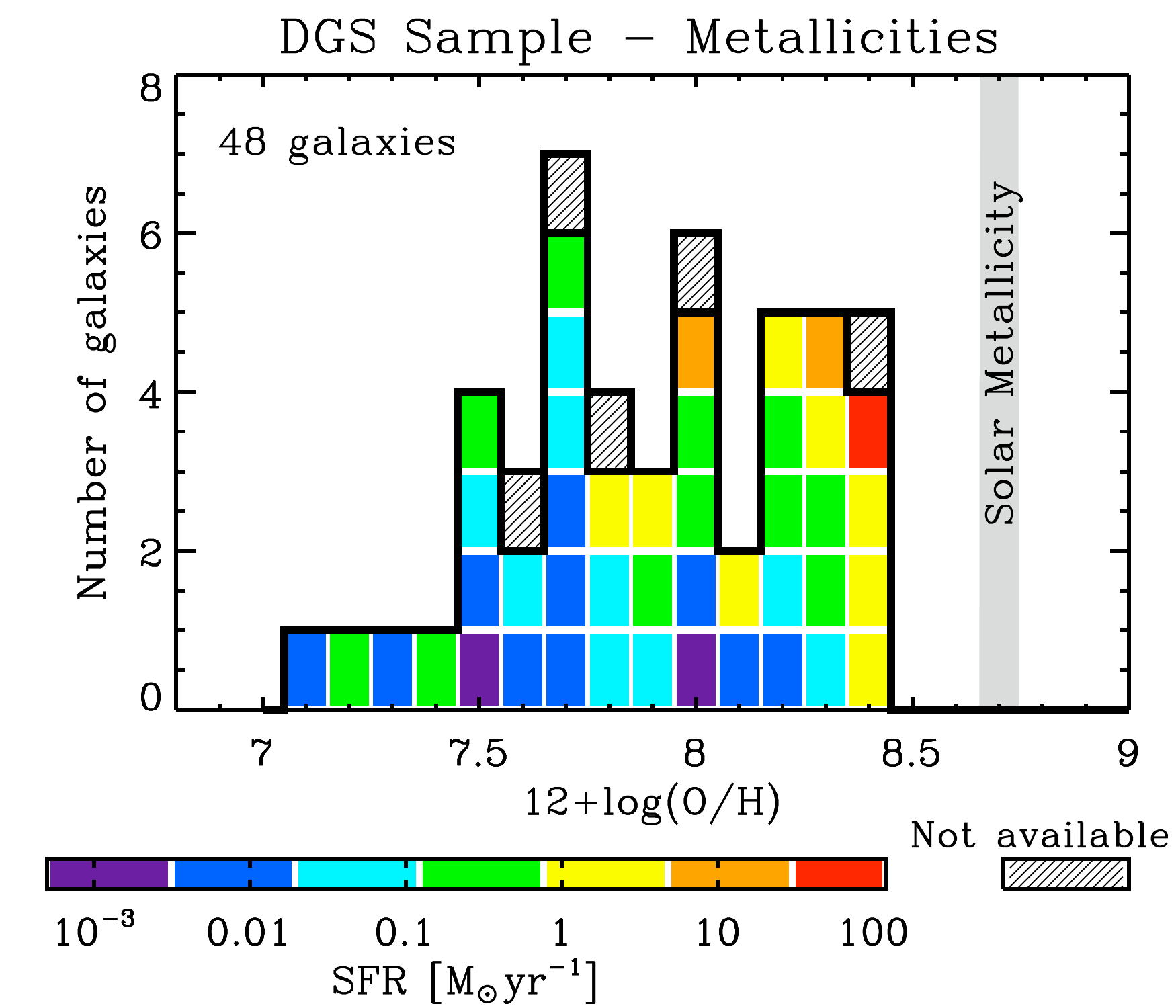}
\caption{Metallicity distribution of the DGS sample from 12+log(O/H) = 7.14 to 8.43. Solar metallicity is indicated here as a guide to the eye. The pre-\hersc\ star formation rate ({\it SFR}) distribution is also indicated by the colour code. They have been converted from {\it L}$_{TIR}$(\spitz) with the \cite{Kennicutt1998} law{\revised, and are given in \cite{Madden2013}}. When no IR data was available, {\it H$\alpha$} or {\it H$\beta$} emission lines were used and converted to {\it SFR} \citep{Kennicutt1998}. The dashed cells indicate that none of these data were available for the galaxy. The most actively star-forming galaxy (in red) corresponds to the starburst luminous infrared galaxy (LIRG) Haro 11.}
\label{metallicities}
\end{center}
\end{figure}

These targets span a wide range in metallicity from 12+log(O/H) = 7.14 to 8.43, including I~Zw~18 with Z $\sim$ 1/40 \zsun\ \citep{Lequeux1979, Izotov1999} which is one of the most metal-poor galaxies in the local Universe known to date (see Figure \ref{metallicities} for the metallicity distribution of the DGS targets). 

%Figure: Histogramme metallicity + couleurs pour SFR

%---------------------------- Sample -------------------------------------------------------------------------------------------
\begin{table*}
\begin{center}
\caption{Characteristics of the sample.}
\label{sample}
 \centering
 \begin{tabular}{lccccccc}
\hline
\hline
Source & RA (J2000) & DEC (J2000) & Distance (Mpc) & Ref & 12 + log(O/H) & Ref \\
\hline
&&&&&&&\\
Haro 11   	  	  	&00h36m52.7s		&-33d33m17.0s  	&92.1	&1 		&8.36 $\pm$ 0.01		&1 \\
Haro 2  	  	  	&10h32m32.0s 	&+54d24m02.0s 	&21.7 	&2		&8.23 $\pm$ 0.03	&2 \\
Haro 3 	  	  	&10h45m22.4s 	&+55d57m37.0s 	&19.3 	&3		&8.28 $\pm$ 0.01	&3 \\
He 2-10 			&08h36m15.1s 	& -26d24m34.0s 	&8.7	 	&21		&8.43 $\pm$ 0.01	&4 \\
HS 0017+1055  	&00h20m21.4s		&+11d12m21.0s 	&79.1 	&3		&7.63 $\pm$ 0.10 &5 \\
HS 0052+2536  	&00h54m56.4s		&+25d53m08.0s 	&191.0	&3		&8.04 $\pm$ 0.10$^a$	&5 \\
HS 0822+3542 	&08h25m55.5s 	&+35d32m32.0s	&11.0	&4		&7.32 $\pm$ 0.	03	&6 \\
HS 1222+3741	  	&12h24m36.7s 	& +37d24m37.0s	&181.7 	&3		&7.79 $\pm$ 0.01	&7 \\
HS 1236+3937  	&12h39m20.2s 	&+39d21m05.0s 	&86.3 	&3		&7.72 $\pm$ 0.10	&8 \\
HS 1304+3529  	&13h06m24.2s 	&+35d13m43.0s 	&78.7 	&3		&7.93 $\pm$ 0.10	&8 \\
HS 1319+3224  	&13h21m19.7s 	&+32d08m25.0s	&86.3 	&3		&7.81 $\pm$ 0.10	&8 \\
HS 1330+3651  	&13h33m08.3s 	&+36d36m33.0s 	&79.7 	&3		&7.98 $\pm$ 0.10	&8 \\
HS 1442+4250  	&14h44m12.8s 	&+42d37m44.0s 	&14.4	&3		&7.60 $\pm$ 0.01	&9 \\	
HS 2352+2733  	&23h54m56.7s		&+27d49m59.0s 	&116.7 	&3		&8.40 $\pm$ 0.10		&5 \\
I Zw 18  			&09h34m02.0s 	&+55d14m28.0s	&18.2 	&5		&7.14 $\pm$ 0.01		&10 \\
II Zw 40  			&05h55m42.6s	 	&+03d23m32.0s	&12.1 	&20		&8.23 $\pm$ 0.01 		&12 \\
IC 10  	 		&00h20m17.3s		&+59d18m14.0s	&0.7 		&6		&8.17 $\pm$ 0.03	&11 \\
Mrk 1089 			&05h01m37.7s 	&-04d15m28.0s 	&56.6 	&3		&8.10 $\pm$ 0.08$^a$		&13 \\
Mrk 1450  		&11h38m35.7s 	&+57d52m27.0s 	&19.8 	&3		&7.84 $\pm$ 0.01		&14 \\
Mrk 153  			&10h49m05.0s 	&+52d20m08.0s 	&40.3 	&3		&7.86 $\pm$ 0.04		&15 \\
Mrk 209  			&12h26m15.9s 	&+48d29m37.0s 	&5.8 		&7		&7.74 $\pm$ 0.01	&16 \\
Mrk 930  	  		&23h31m58.3s 	&+28h56m50.0s	&77.8 	&3		&8.03 $\pm$ 0.01	&17 \\
NGC 1140  		&02h54m33.6s		&-10d01m40.0s 	&20.0 	&8		&8.38 $\pm$ 0.01 	&3 \\
NGC 1569  		&04h30m49.0s		&+64d50m53.0s	&3.1 		&9		&8.02 $\pm$ 0.02	&18 \\
NGC 1705  		&04h54m13.5s		&-53d21m40.0s 	&5.1 		&10		&8.27 $\pm$ 0.11	&19 \\
NGC 2366  		&07h28m54.6s		&+69d12m57.0s 	&3.2 		&11		&7.70 $\pm$ 0.01		&20 \\
NGC 4214  		&12h15m39.2s		&+36d19m37.0s 	&2.9 		&12		&8.26 $\pm$ 0.01	&4 \\
NGC 4449  		&12h28m11.1s		&+44d05m37.0s 	&4.2 		&13		&8.20 $\pm$ 0.11	&21 \\
NGC 4861  		&12h59m02.3s		&+34d51m34.0s 	&7.5	 	&14		&7.89 $\pm$ 0.01	&16 \\
NGC 5253  		&13h39m55.9s		&-31d38m24.0s 	&4.0 		&12		&8.25 $\pm$ 0.02 	&4 \\
NGC 625  		&01h35m04.6s		&-41d26m10.0s 	&3.9		&15		&8.22 $\pm$ 0.02	&22\\
NGC 6822  		&19h44m57.7s		&-14d48m12.0s 	&0.5 		&16		&7.96 $\pm$ 0.01	&23 \\
Pox 186  			&13h25m48.6s 	&-11d36m38.0s 	&18.3 	&3		&7.70 $\pm$ 0.01	&24 \\
SBS 0335-052 		&03h37m44.0s		&-05d02m40.0s 	&56.0 	&3		&7.25 $\pm$ 0.01	&17\\
SBS 1159+545 	&12h02m02.4s 	&+54d15m50.0s 	&57.0 	&3		&7.44 $\pm$ 0.01	&14 \\
SBS 1211+540	  	&12h14m02.5s 	&+53d45m17.0s 	&19.3 	&3		&7.58 $\pm$ 0.01	&14 \\
SBS 1249+493 	&12h51m52.5s 	&+49d03m28.0s 	&110.8 	&3		&7.68 $\pm$ 0.02	&25 \\
SBS 1415+437	   	&14h17m01.4s 	&+43d30m05.0s	&13.6	&17		&7.55 $\pm$ 0.01	&26 \\
SBS 1533+574  	&15h34m13.8s 	&+57d17m06.0s	&54.2 	&3		&8.05 $\pm$ 0.01	&16 \\
Tol 0618-402  		&06h20m02.5s 	&-40d18m09.0s 	&150.8 	&3		&8.09 $\pm$ 0.01	&27 \\
Tol 1214-277  		&12h17m17.1s 	&-28d02m33.0s 	&120.5 	&3		&7.52 $\pm$ 0.01	&4 \\
UGC 4483  		&08h37m03.0s 	&+69d46m31.0s 	&3.2 		&11		&7.46 $\pm$ 0.02	&28 \\
UGCA 20  	 	&01h43m14.7s		&+19d58m32.0s 	&11.0 	&18		&7.50 $\pm$ 0.02	&29 \\
UM 133  			&01h44m41.3s		&+04d53m26.0s 	&22.7 	&3		&7.82 $\pm$ 0.01	&4 \\
UM 311  			&01h15m34.4s		&-00d51m46.0s 	&23.5 	&3		&8.36 $\pm$ 0.01$^a$	&17 \\
UM 448  			&11h42m12.4s 	&+00d20m03.0s 	&87.8 	&3		&8.32 $\pm$ 0.01	&17 \\
UM 461  			&11h51m33.3s 	&-02d22m22.0s 	&13.2 	&3		&7.73 $\pm$ 0.01	&15 \\
VII Zw 403  		&11h27m59.9s 	&+78d59m39.0s 	&4.5		&19		&7.66 $\pm$ 0.01	&16 \\
 \hline
\end{tabular}
\end{center}
\scriptsize{
{\bf References for positions}:  The positions have been taken from the Nasa/Ipac Extragalactic Database (NED). \\

{\bf References for distances}: (1) \cite{Bergvall2006} ; (2) \cite{Kennicutt2003} ; (3) this work, calculated from the redshifts available in NED, the Hubble flow model from \cite{Mould2000} and assuming H$_0$ = 70 km.s$^{-1}$.Mpc$^{-1}$ ; (4) \cite{Pustilnik2003} ; (5) \cite{Aloisi2007} ; (6) \cite{Kim2009} ; (7) \cite{Schulte-Ladbeck2001} ; (8) \cite{Moll2007} ; (9) \cite{Grocholski2012} ; (10) \cite{Tosi2001} ; (11) \cite{Karachentsev2002} ; (12) \cite{Karachentsev2004} ; (13) \cite{Karachentsev2003} ; (14) \cite{DeVaucouleurs1991} ; (15) \cite{Cannon2003} ; (16) \cite{Gieren2006} ; (17) \cite{Aloisi2005} ; (18) \cite{Sharina1996} ; (19) \cite{Lynds1998} ; (20) \cite{Bordalo2009} ; (21) \cite{Tully1988} \\

{\bf References for metallicities}: (1) \cite{Guseva2012} ; (2) \cite{Kong2002} ; (3) \cite{Izotov2004} ; (4) \cite{Kobulnicky1999} ; (5) \cite{Ugryumov2003} ; (6) \cite{Pustilnik2003} ; (7) \cite{Izotov2007} ; (8) \cite{Popescu2000} ; (9) \cite{Guseva2003a} ; (10) \cite{Izotov1999} ; (11) \cite{Magrini2009} ; (12) \cite{Guseva2000} ; (13) \cite{LopezSanchez2004} ; (14) \cite{Izotov1994} ; (15) \cite{Izotov2006} ; (16) \cite{Izotov1997} ; (17) \cite{Izotov1998} ; (18) \cite{Kobulnicky1997} ; (19) \cite{LeeSkillman2004} ; (20) \cite{Saviane2008} ; (21) \cite{McCall1985} ; (22) \cite{Skillman2003} ; (23) \cite{Lee2006} ; (24) \cite{Guseva2007} ; (25) \cite{Thuan1995} ; (26) \cite{Guseva2003b} ; (27) \cite{Masegosa1994} ; (28) \cite{VanZeeHaynes2006} ; (29) \cite{VanZee1996} \\

$^a$: These objects are galaxies within compact groups of galaxies or are parts of other galaxies. The metallicity quoted here is the mean value for the group of objects. For Mrk1089, the group is composed of regions A-C, B, E, F1, F2, H from \cite{LopezSanchez2004}. For UM311, the group is composed of regions 1-2-3 of \cite{Moles1994} plus NGC450 and UGC807. For HS0052+2536 the group is composed of HS0052+2536 and HS0052+2537. For all of the objects, the ``group" corresponds to the objects included in the aperture used for the photometry (see Section \ref{pacsphotofluxes}). For the metallicity of the object only, see \cite{Madden2013}.
}
 \end{table*} 

\subsubsection{Observations}

The Dwarf Galaxy Survey was granted $\sim$ 230 hours of observations, part of which were used to observe the sample with the two \hersc\ imaging photometres: all of them (48) with PACS at 70, 100 and 160 \mic\ and 41 with SPIRE at 250, 350 and 500 \mic.  Seven sources were not observed with SPIRE because they were predicted to be too faint for SPIRE. The full width at half maximum (FWHM) of the beam in each band is 5.6, 6.8, 11.4\footnote{{\revised The PACS Observers' Manual is available at http://herschel.esac.esa.int/
Docs/PACS/pdf/pacs\_om.pdf.}},
 18.2, 24.9, 36.3\arcsec\footnote{{\revised The SPIRE Observers' Manual is available at http://herschel.esac.esa.int/Docs/SPIRE/pdf/spire\_om.pdf.}} at 70, 100, 160, 250, 350 and 500 \mic\ respectively. 
Most of the sources have also been observed by the PACS spectrometre in order to complement the photometry \citep[e.g.][Cormier et al. 2013, in prep.]{Cormier2011, Cormier2012, Lebouteiller2012, Madden2013}. 

For all of our galaxies, the PACS photometry observations have been done in the PACS scan-map mode at a medium scan speed (20\arcsec/s). The SPIRE observations have been made using the SPIRE large and small scan-map modes, depending on the source sizes, at the nominal scan speed (30\arcsec/s).

Substantial ancillary data are available over a large wavelength range, from UV to radio wavelengths. A summary of all the available ancillary data for these galaxies is presented in \cite{Madden2013}. 

\subsection{Data reduction process}

In this section we describe the data reduction process followed to produce the final \hersc\ maps. 

\subsubsection{PACS data reduction}

For the PACS data reduction we use the \hersc\ Interactive Processing Environment \citep[HIPE,][]{Ott2010}, with version 7 of the photometric calibration\footnote{{\revised The version 7 cited here corresponds to the value of the {\sc calFileVersion} metadata of the Responsitivity Calibration Product in HIPE.}}, and a modified version of the available pipeline which we describe here.\\

The pipeline begins with the Level 0 Products, at a purely instrumental level. All the auxiliary data (such as ``housekeeping'' parameters, pointings, etc) is stored as Products. Level 0 also contains the Calibration Tree, needed for flux conversion. Then we perform the usual steps such as flagging the ``bad'' saturated pixels, converting the signal into Jy$\cdot$pix$^{-1}$ and applying flatfield correction. We systematically mask the column 0 of all the matrices (the PACS array is composed of groups of 16$\times$16 bolometres) to avoid electrical crosstalk issues. We perform second level deglitching to remove all the glitches, which represent on average $\sim$ 0.3\% of the data.\\

After performing all of the above steps we reach the Level 1 stage of data reduction. Note that we still have the bolometre drifts (the so-called ``1/{\it f}'' noise) at this stage of the data reduction. This low-frequency noise is originating from two sources: thermal noise, strongly correlated between the bolometres, and uncorrelated non-thermal noise. The method employed to remove the drifts will greatly affect the final reconstructed map (also called Level 2 data). We thus analyse three different map making methods in order to systematically compare the maps and extracted flux densities, to determine if there is an optimized method for each galaxy. The first two map making methods are provided in HIPE: the {\sc PhotProject} and the {\sc MADmap} method. The last method is the {\sc Scanamorphos} method \citep{Roussel2012}.

The first technique we use for the final reconstruction of the map is {\sc PhotProject}. We first remove the 1/{\it f} noise (corresponding to data with low spatial frequencies or large scale structures in the map) using a high-pass filter. We then use {\sc PhotProject} to reproject the data on the sky. The high-pass filtering step is optimum for compact sources but can lead to suppression of extended features (corresponding to low spatial frequencies) in extended sources. 

{\sc MADmap} (Microwave Anisotropy Dataset mapper) produces maximum likelihood maps from the time ordered data \citep{Cantalupo2010}. The main assumption here is that the noise is uncorrelated from pixel to pixel. However, one component of the 1/{\it f} noise is strongly correlated from pixel to pixel, as it is due to the thermal drift of the bolometres, and thus not treated by {\sc MADmap}. Nevertheless, {\sc MADmap} is more efficient than {\sc PhotProject} in reconstructing the extended structures within a map. 

{\sc Scanamorphos} is another technique specially developed to process scan observations \citep{Roussel2012}. The particularity of {\sc Scanamorphos}, compared to {\sc MADmap}, is that no particular noise model is assumed to deal with the low-frequency noise (the 1/{\it f} noise). Indeed {\sc Scanamorphos} takes advantage of the redundancy in the observations, i.e., of the fact that a portion of the sky is observed more than once and by more than one bolometre. The two noise sources contributing to the low-frequency noise are inferred from the redundancy of the data and removed \citep{Roussel2012}. The maps are made using the default parameters. We add the {\sc {\sc minimap}} option when reducing data with a field size of the order of 8.4 arcmin.
For consistency in the following flux computation, we produce maps with the same pixel sizes for all of the methods: 2, 2 and 4\arcsec\ for 70, 100 and 160 \mic\ respectively. 

\subsubsection{PACS data reduction: choosing between {\sc PhotProject}, {\sc MADmap} and {\sc Scanamorphos}}

To determine the best mapmaking method for each galaxy (summarized in Table \ref{fluxesPACS}), we compute the flux densities (see Section \ref{pacsphotofluxes} for PACS flux extraction) for the three bands for the three methods for each galaxy and compare the photometry for the three different methods. For consistency, we use the same apertures for the three different types of maps.

\begin{figure}
\begin{center}
\includegraphics[width=8.8cm]{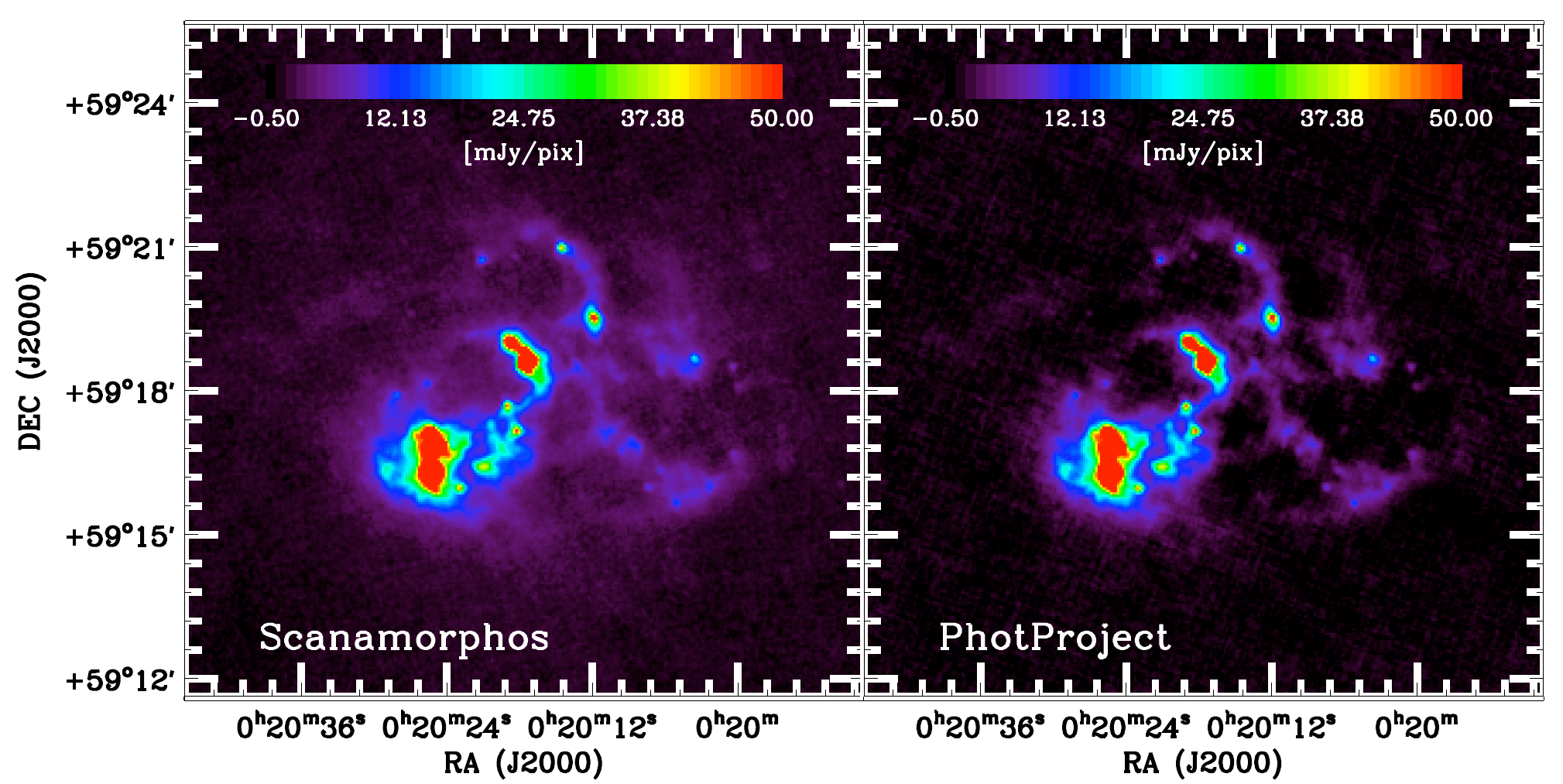}
\caption{{\sc Scanamorphos} ({\it left}) and {\sc PhotProject} ({\it right}) images of IC 10 at 70 \mic\ to illustrate how {\sc PhotProject} tends to clip out the extended features. The colours and spatial scales are the same on both images. Here the diffuse extended emission is best visible on the {\sc Scanamorphos} map. The comparison of the total flux densities coming from the 2 methods confirms that {\sc PhotProject} misses the extended emission: in this case, {\it F}$_{70}$({\sc Scanamorphos})/{\it F}$_{70}$({\sc PhotProject}) = 1.5.}
\label{ic10}
\end{center}
\end{figure}

As mentioned above, the {\sc PhotProject} method is optimized for compact sources. Indeed, the filtering step partly removes large scale structures in the map. It is not adapted for extended sources as this filtering step can sometimes also remove the large scale structures of our sources such as diffuse extended emission (Figure \ref{ic10}), also noted by \cite{Aniano2012} for two extended KINGFISH galaxies. {\revised Moreover the source is automatically masked before the high-pass filtering step, and this mask may be too small for extended sources with peculiar morphology, leading to suppression of extended features during the filtering step.}
Therefore, we decided to take as final, the maps produced by {\sc PhotProject} for compact sources only.

Some galaxies are not detected in one or several bands. When deriving upper limits on the flux densities for these galaxies, the three methods give very different results. As the ``non-detection'' criterion is directly linked to the background determination through its contribution to the total flux density and the corresponding uncertainty, we need to choose the method that gives the most reliable background structure. {\sc MADmap} and {\sc Scanamorphos} do not have any constraints on the background values, whereas {\sc PhotProject} is constrained to an average statistically-null background. Because {\sc Scanamorphos} does not make assumptions on the background, sometimes positive residual noise structures can remain in the maps. {\sc MADmap} presents features, such as a curved background for some maps, due to a too-simple treatment of missing data. Again, the {\sc PhotProject} maps here are used because they are the most constrained as far as the background is concerned. Moreover, when the galaxy is not detected at 160 \mic\, it is usually a compact point source at the other PACS wavelengths. So this choice is consistent with the previous choice for compact sources.

For more extended sources, we only consider {\sc MADmap} and {\sc Scanamorphos}. As mentioned before, {\sc MADmap} maps sometimes present a curved background: the source in the map centre is surrounded by lower background levels than those used in the background aperture for the photometry. This therefore results in a high background leading to an underestimation of the source flux density. Moreover this is not consistent with the assumption of a flat background made for the photometry (see Section \ref{pacsphotofluxes}). To avoid this problem, we decide to use the {\sc Scanamorphos} maps for the extended sources. 

%Figure: extended emission PhP vs Scanam: IC10

\subsubsection{SPIRE data reduction}\label{dataredspire}
 
Following the same method as in \cite{Ciesla2012} for the Herschel Reference Survey, or in \cite{Auld2013} for the HErschel VIrgo Cluster Survey, the SPIRE maps are processed through HIPE\footnote{{\revised The version 5 of HIPE was used for producing the SPIRE maps.}} using a modified version of the available SPIRE pipeline. The steps from the Level 0 to Level 1 are basically the same as in the official version provided by the SPIRE Instrument Control Centre (ICC). The pipeline starts with a first deglitching step, then a time response correction is applied to match the detector timelines to the astronomical pointing timelines. A second deglitching step is then performed as it improves the removal of residual glitches. After, an additional time response correction, the flux calibration step is performed, where non-linearity corrections are taken into account. An additional correction is applied to the bolometre timelines to account for the fact that there is a delay in the response of the bolometres to the incoming signal. The temperature drifts of the bolometres are then removed. For this step, the pipeline temperature drift removal is not run, instead a custom temperature drift correction ({\sc BriGAdE}, Smith et al. in prep) is applied to the whole observation timeline (rather than an individual scan-leg). Finally, the {\sc Naive mapper} is used to construct the final map with pixel sizes of 6, 8, 12\arcsec\ for the 250, 350, 500 \mic\ band respectively. For galaxies with heavy cirrus contamination an additional destriping step is performed. 
A complete description of the data processing step will be given in Smith et al. in prep.
  
%===============================================================
% Photometry on the sample
%===============================================================

\section{Photometry measurements for the DGS sample}\label{photo}

In this section, we describe how we obtain the different PACS and SPIRE flux densities, together with their uncertainties, for the DGS sample (Tables \ref{fluxesPACS} and \ref{fluxesSPIRE}, Sections \ref{pacsphoto} and \ref{spirephoto}). The PACS flux densities are then compared with the existing MIPS flux densities (Section \ref{compspitzer}).

\subsection{PACS photometry}\label{pacsphoto}

\subsubsection{Extracting the fluxes}\label{pacsphotofluxes}

For PACS measurements, we perform aperture photometry, placing an aperture on the source and a background region to estimate the sky level. Using the {\revised version 7 of the PACS} photometric calibration {\revised available in} HIPE, the point spread functions (PSFs) have been measured out to 1000\arcsec. Most of our maps are smaller than this, which means, in principle, that some contribution from the PSF of the source can basically be found everywhere on the map, and, any emission from the source falling in the background region must be taken into account when estimating the total source flux density. 

Taking into account this aperture correction, aperture photometry is performed, using circular apertures of 1.5 times the optical radius whenever possible. 
For cases where it is not, we adjust our apertures to be sure to encompass all the FIR emission of the galaxy (Table \ref{fluxesPACS}). There are three special cases. For HS0052+2536 the chosen aperture also encompasses the neighbouring very faint galaxy HS0052+2537. Mrk1089 is a  galaxy within a compact group of galaxies and UM311 is part of another galaxy and the chosen apertures encompass the whole group of objects. For these galaxies, the spatial resolution of the SPIRE bands makes it very difficult, if not impossible, to separate them from the other objects in their respective groups. For these few cases, the entire group is considered and is noted in Tables \ref{sample}, \ref{fluxesPACS} and \ref{fluxesSPIRE}. The background region is a circular annulus around the source. In most cases, the inner radius of the background region is the same as that of the source aperture and the outer radius is about two times the source aperture radius.

The maps are assumed to consist of the sum of a constant, flat background plus the contribution from the source. Flux densities are measured in the aperture ({\it f$_{ap}$}) and in the background annulus ({\it f$_{bg}$}) by summing the pixels in both regions. The contribution to the measured flux densities ({\it f$_{ap}$} and {\it f$_{bg}$}) from the total flux density of the galaxy ({\it f$_{tot}$}) and from the background ({\it b}) is determined for each aperture using the encircled energy fraction ({\it eef}) tables. These tables, given by HIPE, are measurements of the fraction of the total flux density contained in a given aperture on the PSFs (inverse of the aperture correction). This gives us a simple linear system of two equations with two unknowns: the total flux density from the galaxy ({\it f$_{tot}$}) and the background level ({\it b}) :

\begin{equation}
\left  \{
 \begin{array}{rcl}
f_{ap} & = & f_{tot} \cdot {\it eef}_{r_0} + N_{ap} \cdot b \\
f_{bg} & = & f_{tot} \cdot ({\it eef}_{r_2} - {\it eef}_{r_1}) + N_{bg} \cdot b \end{array} 
 \right .
\end{equation}

where {\it r$_0$}, {\it r$_1$}, {\it r$_2$} are the source aperture radius and the background annulus radii respectively, and {\it eef}$_{r_0}$, {\it eef}$_{r_1}$ and {\it eef}$_{r_2}$ are the encircled energy fractions at radii {\it r$_0$}, {\it r$_1$}, {\it r$_2$}. {\it N$_{ap}$} (resp. {\it N$_{bg}$}) is the number of pixels in the source (resp. background) aperture.
Inverting this system gives us the values for {\it f$_{tot}$} and {\it b}.

{\revised If one considers that there is no contribution from the source outside the source aperture, i.e. setting {\it eef}$_{r_0}$=1 and {\it eef}$_{r_1}$={\it eef}$_{r_2}$=0, the flux density will be underestimated, as some contribution from the source will have been removed during the background subtraction. This underestimation depends on the source aperture size {\it r$_0$} and can be important for small apertures. The error made on the flux density becomes greater than the calibration error, which is the dominant source of uncertainty ($\sim$ 5\%, see Section \ref{pacsphotoerror}), when {\it r$_0$} $\lesssim$ 1\arcmin. Given that the median {\it r$_0$} in the DGS sample is $\sim$ 0.6\arcmin, it is thus important to take into account the contribution from the source falling outside the source aperture.}

\subsubsection{Computing the uncertainties}\label{pacsphotoerror}
The uncertainties on the flux density arise from the non-systematic errors due to the measurement of the flux density on the maps, ({\it unc$_{f_{tot}}$}), and the systematic errors due to calibration, ({\it unc$_{calib}$}).

For the measurement on the maps, the system of equations being linear, the uncertainties arising from the two measurements on the map ({\it unc$_{ap}$} and {\it unc$_{bg}$}) can be linearly propagated to the total flux density and the background level, giving us the uncertainty on the total flux density ({\it unc$_{f_{tot}}$}) and the uncertainty on the background level ({\it unc$_{b}$}). The determination of {\it unc$_{ap}$} and {\it unc$_{bg}$} is the same for both errors as the measure is the same: summing pixels in a given region of the map. Thus we detail the calculation for {\it unc$_{ap}$} only.

There are two sources of errors to {\it unc$_{ap}$}: one coming from the sum of the pixels, {\it unc$_{sum}$}, one coming from the intrinsic error on the flux density value in each pixel, {\it unc$_{int}$}. 

\noindent \textit{Determination of {\it unc$_{sum}$} :}
For each pixel there is a contribution from the background noise to the total measured flux density. This error, $\sigma_{sky}$, is the same for a pixel in the source aperture as well as in the background aperture, repeated {\it N$_{ap}$} times here. The error, $\sigma_{sky}$, is the standard deviation of all pixels in the background aperture. The final uncertainty, {\it unc$_{sum}$}, is then: 

\begin{equation}
unc_{sum} = \sqrt{N_{ap}} \sigma_{sky}
\end{equation}

\noindent \textit{Determination of {\it unc$_{int}$} :}
For each pixel there is an underlying uncertainty for the flux density value in the pixel, $\sigma_{int,i}$, and is independent from pixel to pixel. This uncertainty arises from the data reduction step when the flux density for each pixel is computed. A map of these uncertainties is produced during the data reduction process. The uncertainty, {\it unc$_{int}$}, is then derived by adding quadratically all of the errors in the considered pixels :

\begin{equation}
unc_{int} = \sqrt{\sum_{i=0}^{N_{ap}} \sigma_{int,i}^2}
\label{unc_intP}
\end{equation}

Note that the assumption of pixel-to-pixel independent uncertainty is not applicable for PACS maps and this can result in an underestimation of {\it unc$_{int}$}.
\\

The total error on the source aperture measurement is then :

\begin{equation}
unc_{ap} = \sqrt{unc_{sum}^2 + unc_{int}^2}
\end{equation}

The {\it unc$_{bg}$} is derived the same way and we can then compute {\it unc$_{f_{tot}}$} and {\it unc$_{b}$}. The quantity {\it unc$_{f_{tot}}$} is thus the total error on the flux density due to measurement on the map. To this uncertainty, we add in quadrature the systematic calibration uncertainty, {\it unc$_{calib}$}, of 5\% for the three PACS bands (M. Sauvage \& T. M\"uller, priv. com.), giving, in the end, the $\sigma_{70-100-160}$ reported in Table \ref{fluxesPACS}: 

\begin{equation}\label{finalUnc}
\sigma_{\lambda} = \sqrt{unc_{f_{tot}}^2 + unc_{calib}^2}
\end{equation}

{\revised Note that in {\it unc$_{sum}$}, we have a combination of uncertainties from small scale astronomical noise and instrumental uncertainties. These instrumental uncertainties can be redundant with part of the instrumental uncertainties taken into account in {\it unc$_{int}$}, leading to an overestimate of {\it unc$_{ap}$} and thus {\it unc$_{f_{tot}}$}. However, it has a minor impact on the final uncertainties, $\sigma_{70-100-160}$, as the calibration uncertainty is dominant. 
}

\subsubsection{Case of upper limits}\label{pacsphotoupplim}

\begin{figure}
\begin{center}
\includegraphics[width=8.8cm]{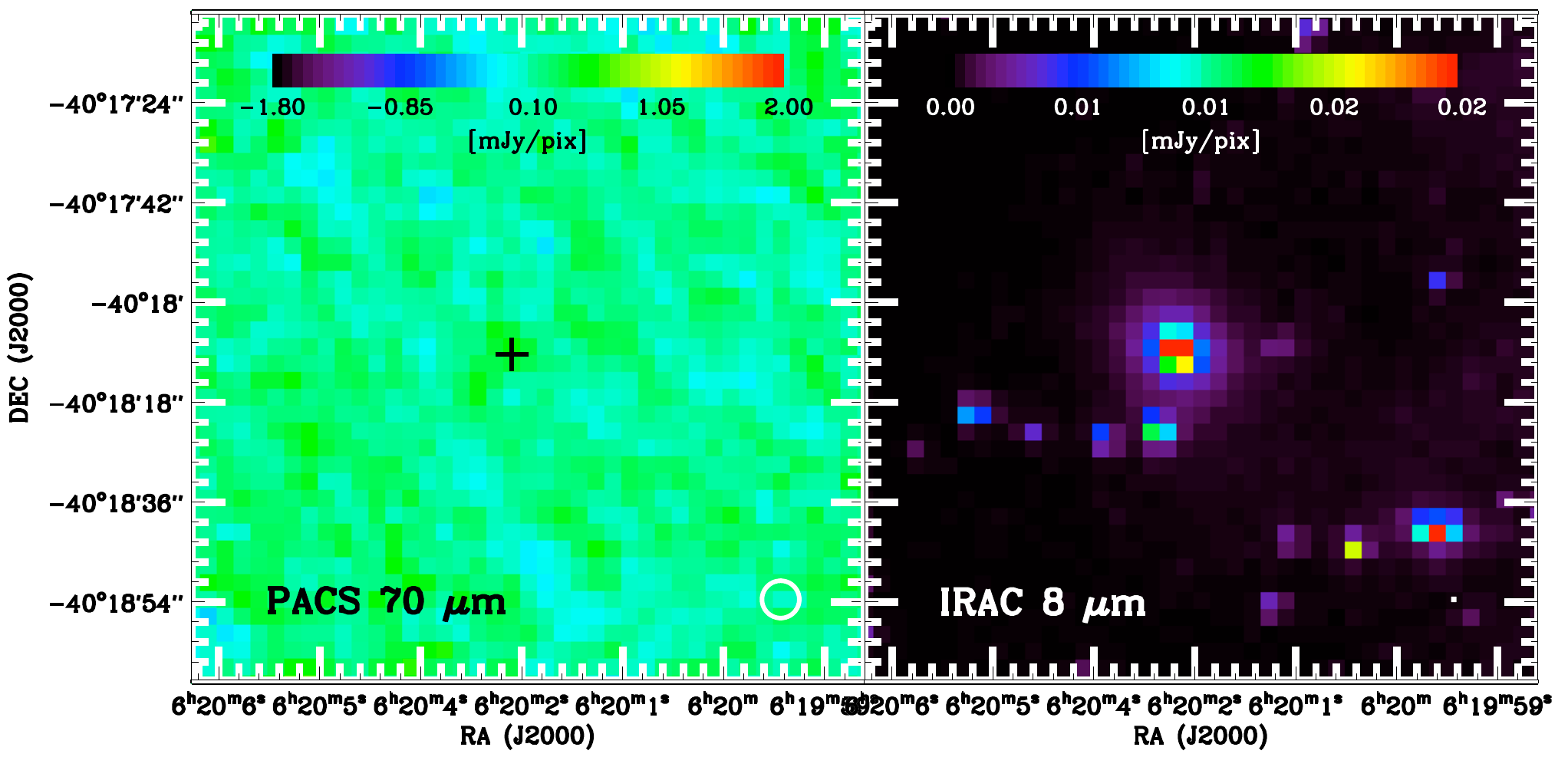}
\caption{Example of a PACS non-detection: (left) PACS 70 \mic\ image of Tol 0618-402. The position of the galaxy is marked with a black cross. The IRAC 8 \mic\ image has been added on the right for comparison. The PACS 70 \mic\ (FWHM=5.6\arcsec) and the IRAC 8 \mic\ (FWHM=2.0\arcsec) beams are indicated as white circles on the bottom right of the images.}
\label{tol0618}
\end{center}
\end{figure}

Some galaxies in our sample are not detected in some or all of the PACS bands. We classify these galaxies as ``upper limits'' when the computed flux density is lower than five times the corresponding uncertainty on the flux density (e.g. Tol 0618-402, Figure \ref{tol0618}). We take as the final upper limit, five times the uncertainty on the flux density value in order to have a 5$\sigma$ upper limit (reported in Table \ref{fluxesPACS}). 

%Figure: Example PACS upper limit: Tol 0618 

%----------------------- PACS flux table-----------------------------------------------
\begin{table*}
 \begin{center}
\caption{Table of PACS flux densities for the DGS galaxies. The map making method is indicated for each galaxy as well as the radius of the circular aperture used for the flux extraction. When an upper limit is given, it is the 5$\sigma$ upper limit computed in \ref{pacsphotoupplim}.}
\label{fluxesPACS}
\begin{tabular}{lcccccccc}
\hline
\hline
Source & F$_{70}$ (Jy) & $\sigma_{70}$ (Jy) & F$_{100}$ (Jy) & $\sigma_{100}$ (Jy) & F$_{160}$ (Jy) & $\sigma_{160}$ (Jy) & Method & Circular aperture radius (\arcsec) \\
\hline
&&&&&&&& \\
Haro11       &        6.14 & 0.31 &        4.96 & 0.25 &        2.42 & 0.12 &    {\sc Scanamorphos} &  45$^b$ \\
Haro2        &        4.99 & 0.25 &        5.33 & 0.27 &        3.95 & 0.20 &    {\sc Scanamorphos} &  50$^a$ \\
Haro3        &        5.30 & 0.26 &        6.41 & 0.32 &        4.83 & 0.24 &    {\sc Scanamorphos} &  60$^a$ \\
He2-10       &        25.6 &  1.3 &        26.6 &  1.3 &        18.8 &  0.9 &    {\sc Scanamorphos} & 108$^a$ \\
HS0017+1055  &        0.046 & 0.005 &        0.033 & 0.004 &        0.019 & 0.004 &     {\sc PhotProject} &  16$^b$ \\
HS0052+2536  &        0.22 & 0.01 &        0.21 & 0.01 &        0.139 & 0.008 &     {\sc PhotProject} &  23$^{a,f}$ \\
HS0822+3542  &        $\leq$ 0.014$^j$ &   - & $\leq$ 0.013$^j$ &   - &        0.034 & 0.003 &     {\sc PhotProject} &  12$^a$ \\
HS1222+3741  &        0.025 & 0.004 & $\leq$ 0.036 &   - & $\leq$ 0.022 &   - &     {\sc PhotProject} &  14$^b$ \\
HS1236+3937  & $\leq$ 0.029 &   - & $\leq$ 0.035 &   - & $\leq$ 0.028 &   - &     {\sc PhotProject} &  15$^a$ \\
HS1304+3529  &        0.121 & 0.007 &        0.150 & 0.009 &        0.069 & 0.005 &     {\sc PhotProject} &  18$^b$ \\
HS1319+3224  &        0.012 & 0.003 &        0.013 & 0.002 & $\leq$ 0.015 &   - &     {\sc PhotProject} &   8$^b$ \\
HS1330+3651  &        0.093 & 0.006 &        0.112 & 0.007 &        0.091 & 0.005 &     {\sc PhotProject} &  20$^b$ \\
HS1442+4250  &        0.09 & 0.01 & $\leq$ 0.016 &   - & $\leq$ 0.047 &   - &     {\sc PhotProject} &  51$^a$ \\
HS2352+2733  &        0.039 & 0.003 &        0.016 & 0.002 & $\leq$ 0.016 &   - &     {\sc PhotProject} &  15$^a$ \\
IZw18        &        0.045 & 0.003 &        0.018 & 0.002 & $\leq$ 0.011 &   - &     {\sc PhotProject} &  14$^a$ \\
IC10         &        140. &   7. &        207. &  10. &        225. &  11. &    {\sc Scanamorphos} & 306$^a$ \\
IIZw40       &        6.39 & 0.32 &        5.79 & 0.29 &        3.53 & 0.18 &    {\sc Scanamorphos} &  66$^b$ \\
Mrk1089      &        4.27 & 0.21 &        4.97 & 0.25 &        4.68 & 0.23 &    {\sc Scanamorphos} &  75$^{b,f}$ \\
Mrk1450      &        0.30 & 0.02 &        0.25 & 0.01 &        0.127 & 0.007 &     {\sc PhotProject} &  20$^a$ \\
Mrk153       &        0.28 & 0.02 &        0.30 & 0.02 &        0.137 & 0.009 &     {\sc PhotProject} &  35$^b$ \\
Mrk209       &        0.32 & 0.02 &        0.35 & 0.02 &        0.16 & 0.01 &    {\sc Scanamorphos} &  24$^c$ \\
Mrk930       &        1.19 & 0.06 &        1.40 & 0.07 &        0.98 & 0.05 &    {\sc Scanamorphos} &  60$^b$ \\
NGC1140      &        4.04 & 0.20 &        4.62 & 0.23 &        4.58 & 0.23 &    {\sc Scanamorphos} & 118$^b$ \\
NGC1569      &        60.4 &  3.0 &        57.3 &  2.9 &        39.7 &  2.0 &    {\sc Scanamorphos} & 150$^d$ \\
NGC1705      &        1.37 & 0.07 &        1.46 & 0.07 &        1.10 & 0.06 &     {\sc PhotProject}$^g$ &  72$^{d}$ \\
NGC2366      &        5.30 & 0.26 &        6.23 & 0.31 &        4.08 & 0.20 &    {\sc Scanamorphos} & 150$^{d,e}$ \\
NGC4214      &        24.5 &  1.2 &        32.2 &  1.6 &        33.7 &  1.7 &    {\sc Scanamorphos} & 300$^d$ \\
NGC4449      &        49.3 &  2.5 &        75.9 &  3.8 &        79.5 &  4.0 &    {\sc Scanamorphos} & 250$^d$ \\
NGC4861      &        2.31 & 0.12 &        2.17 & 0.11 &        1.99 & 0.10 &    {\sc Scanamorphos} & 120$^d$ \\
NGC5253      &        32.9 &  1.6 &        32.3 &  1.6 &        23.2 &  1.2 &    {\sc Scanamorphos} & 120$^d$ \\
NGC625       &        6.49 & 0.32 &        9.47 & 0.47 &        8.20 & 0.41 &    {\sc Scanamorphos} & 170$^d$ \\
NGC6822      &        54.9 &  2.8 &        63.6 &  3.2 &        77.1 &  3.9 &    {\sc Scanamorphos} & 440$^d$ \\
Pox186       &        0.038$^j$ & 0.005 &        0.052$^j$ & 0.005 &        0.047$^j$ & 0.004 &     {\sc PhotProject} &  16$^a$ \\
SBS0335-052  &        0.056 & 0.004 &        0.024$^h$ & 0.001 &        0.007$^h$ & 0.001 &     {\sc PhotProject} &  10$^{a,h}$ \\
SBS1159+545  &        0.019 & 0.003 &        0.019 & 0.003 & $\leq$ 0.018 &   - &     {\sc PhotProject} &   8$^a$ \\
SBS1211+540  &        0.034 & 0.003 &        0.018 & 0.002 &        0.013 & 0.002 &     {\sc PhotProject} &  15$^b$ \\
SBS1249+493  &        0.032 & 0.005 & $\leq$ 0.034 &   - & $\leq$ 0.042 &   - &     {\sc PhotProject} &  12$^b$ \\
SBS1415+437  &        0.18 & 0.01 &        0.16 & 0.01 &        0.065 & 0.007 &     {\sc PhotProject} &  34$^a$ \\
SBS1533+574  &        0.19 & 0.01 &        0.24 & 0.01 &        0.19 & 0.01 &    {\sc Scanamorphos} &  30$^a$ \\
Tol0618-402  & $\leq$ 0.014 &   - & $\leq$ 0.005 &   - & $\leq$ 0.013 &   - &     {\sc PhotProject} &  18$^a$ \\
Tol1214-277  &        0.017 & 0.003 &        0.018 & 0.002 & $\leq$ 0.018 &   - &     {\sc PhotProject} &  12$^b$ \\
UGC4483      &        0.16 & 0.02 &        -$^i$& -$^i$ & $\leq$ 0.037 &   - &     {\sc PhotProject} &  63$^{a}$ \\
UGCA20       & $\leq$ 0.052 &   - & $\leq$ 0.057 &   - & $\leq$ 0.048 &   - &     {\sc PhotProject} &  20$^c$ \\
UM133        &        0.15$^j$ & 0.01 &        0.066 & 0.010 &        0.053 & 0.009 &    {\sc Scanamorphos} &  26$^c$ \\
UM311        &        2.94 & 0.15 &        5.63 & 0.28 &        6.10 & 0.31 &    {\sc Scanamorphos} & 140$^{b,f}$ \\
UM448        &        5.17 & 0.26 &   - &   - &        3.22 & 0.17 &    {\sc Scanamorphos} &  64$^b$ \\
UM461        &        0.21$^j$ & 0.01 &        0.145 & 0.009 &        0.113 & 0.007 &     {\sc PhotProject} &  17$^b$ \\
VIIZw403     &        0.47 & 0.03 &        0.56 & 0.03 &        0.34 & 0.02 &     {\sc PhotProject} &  40$^c$ \\
\hline
\end{tabular}
\end{center}
\scriptsize{
$^{a,b,c}$: The radius is: $^a$: 1.5 times the optical radius. $^b$: larger than 1.5 times the optical radius. $^c$: smaller than 1.5 times the optical radius.

$^{d,e}$: The aperture is: $^d$: adapted from elliptical shape. $^e$: off-centred to match the particular shape of the galaxy.

$^f$: These objects are galaxies within compact groups of galaxies or are parts of other galaxies and the photometry given here is for the whole group (see Section \ref{pacsphotofluxes} for details).

$^g$: The {\sc Scanamorphos} maps of NGC1705 are not satisfactory because of a non-uniform background. Therefore we use the {\sc PhotProject} maps. To preserve as much of the diffuse extended emission as possible we manually mask out the galaxy before performing the high-pass filtering step. 

$^h$: New observations were obtained for SBS0335-052 at 100 and 160 \mic\ with longer integration times (Sauvage et al. in prep). We chose to quote the flux density values from the newest observations for 100 and 160 \mic.\

$^i$: Interferences on the detector are strongly polluting the map for the 100 \mic\ observation of UGC4483. We thus do not report any flux density nor give any 100 \mic\ map for this galaxy.

$^j$: These flux densities might present some discrepancies with other FIR measurements (i.e. MIPS, other PACS and/or SPIRE wavelengths).
}
\end{table*}

\subsection{SPIRE photometry}\label{spirephoto}

For the SPIRE photometre, the relative spectral response function (RSRF) is different for a point source or for an extended source. During the treatment by the pipeline, the measured RSRF-weighted flux density is converted to a monochromatic flux density {\revised for a source where $\nu$*F$_{\nu}$ is constant}, via the ``{\it K$_4$}'' correction defined in the SPIRE Observers' Manual %\footnote{http://herschel.esac.esa.int/Docs/SPIRE/html/spire\_om.html}
(Section 5.2.7), assuming a {\it point-like} source. The output of the pipeline will be, by definition, a monochromatic flux density of a point source. To obtain monochromatic flux densities of extended sources we apply the ratio of {\it K$_4$} corrections for extended and point-like sources, {\it K$_{4e}/K_{4p}$}, defined in the SPIRE Observers' Manual ({\revised Section} 5.2.7). In order to determine which sources need this extra-correction, we have to distinguish between extended and point-like (unresolved) sources in our sample, as well as non-detected sources. Extended sources are defined as galaxies whose spatial extension is larger than the FWHM of the SPIRE beam, and non-detected sources are galaxies that are not visible at SPIRE wavelengths. 

\subsubsection{Extracting the fluxes}\label{spirephotofluxes}

The photometry method is adapted for each type of galaxy. However, as the data reduction has been performed with HIPE v5, the 350 \mic\ maps are first scaled by a factor of 1.0067 to update the maps to the latest version of the 350 \mic\ flux calibration (SPIRE Observers' Manual ({\revised Section} 5.2.8)).\\

\noindent \textit{Point source photometry} \\

To determine the flux densities of point sources, we fit a Gaussian function (which is representative of the shape of the PSF) to the timeline data from the bolometres, using a timeline-based source fitter that is used for deriving the flux calibration for the individual bolometres\footnote{{\revised The last version of this source fitter is incorporated into HIPE v10 (Bendo et al. in prep.).}}. We then check a posteriori that our ``unresolved'' classification was correct: if the FWHM of the fitted Gaussian is {\revised $<$} 20\arcsec, 29\arcsec\ and 37\arcsec\ at 250, 350 and 500 \mic\ respectively, then the source can be considered as truly point-like. As the timeline data is in Jy$\cdot$beam$^{-1}$, the flux density will simply be the amplitude of the fitted Gaussian. This is the most accurate way of computing flux densities for point-like sources as it matches the measurement techniques used for the SPIRE calibration. Moreover we avoid all pixelization issues when using the timeline data rather than the map. On top of that, applying any mapmaking process would also smear the PSFs, causing the peak signal values to decrease by $\sim$ 5\% for point sources. \\

\noindent \textit{Extended source photometry} \\

For the extended sources, we perform aperture photometry on the maps, using the same source and background apertures as those used for the PACS photometry, and check that the PACS apertures do fully encompass the SPIRE emission from the entire galaxy. The maps are converted from Jy$\cdot$beam$^{-1}$ to Jy$\cdot$pix$^{-1}$ considering that the beam area values are 465, 822 and 1768 square arcseconds\footnote{SPIRE photometre reference spectrum values: http://herschel.esac. \\ esa.int/twiki/bin/view/Public/SpirePhotometerBeamProfileAnalysis, September 2012 values.} at 250, 350, 500 \mic\ respectively and the pixel sizes are given in Section \ref{dataredspire}. 

The background level is determined by the median of all of the pixels in the background aperture. The median is preferred rather than the mean because the SPIRE background is contaminated by prolific background sources due to some observations reaching the confusion limit. The background level is then subtracted from our maps and the total flux density is the sum of all of the pixels encompassed in the source aperture{\revised , corrected for {\it K$_{4e}/K_{4p}$}.} %\removed{For the extended sources, the monochromatic intensities of point sources given by the pipeline need to be converted to monochromatic flux densities of extended sources, using the ratio of the {\it K$_4$} corrections for extended sources to point sources.} 
These {\it K$_{4e}/K_{4p}$} correction factors, given in the SPIRE Observers' Manual ({\revised Section} 5.2.8), are 0.98279, 0.98344 and 0.97099 at 250, 350, 500 \mic\ respectively.

However there are also ``marginally'' extended sources (e.g. IIZw40) that do not require this {\it K$_{4e}/K_{4p}$} correction. To identify these sources, we first check that the source is truly resolved by applying the point source method on the timeline data. We verify that the FWHM is indeed greater than the chosen threshold values for the ``unresolved'' classification. As an additional check, the fitted Gaussian is subtracted from the map and the resulting map is visually checked for any remaining emission from the source. If this condition is satisfied, then the source is truly resolved. If the FWHM of the fitted Gaussian is lower than 24\arcsec, 34\arcsec\ and 45\arcsec\ at 250, 350, 500 \mic\ respectively then the source is considered to be ``marginally'' extended only, and thus to not require the {\it K$_{4e}/K_{4p}$} correction.

\subsubsection{Computing the uncertainties}\label{spirephotoerrors}
As for the PACS photometry, there are two types of uncertainties for SPIRE photometry: the errors arising from the determination of the flux density, {\it unc$_{flux}$}, and the calibration errors, {\it unc$_{calib}$}.
As we used different methods for flux extraction depending on the type of the source, the errors contributing to {\it unc$_{flux}$} are determined differently. The method described here has been adapted from the method described in \cite{Ciesla2012}.\\

\noindent \textit{Point source photometry} \\

The uncertainty on the flux density for a point source is determined through a test in which we add 100 artificial sources with the same flux density as the original source. They are added at random locations in the map, within a 0.3 deg box centred on the original source. The same photometry procedure was applied to the artificial sources and the final uncertainty is the standard deviation in the flux densities derived for the artificial sources. We quote the following uncertainties ({\it unc$_{flux}$}) for point-like sources:  
\begin{itemize}
\item{6 mJy at 250 \mic ;}
\item{
7 mJy (for flux densities $>$ 50 mJy) and 10 mJy (for flux densities $\lesssim$ 50 mJy) at 350 \mic ;
}
\item{9 mJy at 500 \mic.}
\end{itemize}

\noindent \textit{Extended source photometry} \\

For the aperture photometry performed on the extended sources, we have three types of uncertainties contributing to {\it unc$_{flux}$}: the uncertainty arising from the background level determination, {\it unc$_{bg}$}, the uncertainty due to background noise in the source aperture, {\it unc$_{source}$}, the underlying uncertainty for the flux density value in the pixel coming from the data reduction, {\it unc$_{int}$}{\revised , and the uncertainty in the beam area value:  {\it unc$_{beam}$}, which is given to be 4\%\footnote{This value is given in: http://herschel.esac.esa.int/twiki/bin/view/ \\ÊPublic/SpirePhotometerBeamProfileAnalysis.}.}

The determination of the background level generates an uncertainty which will affect each pixel in the source aperture when subtracting the background level from the map. The uncertainty on the background level is {\it unc$_{bg_{level}}$} = $\sigma_{sky}/ \sqrt{N_{bg}}$, with $\sigma_{sky}$ being here again the standard deviation of all of the pixels in the background aperture. This will affect the determination of the flux density for each pixel summed in the aperture :

\begin{equation}
unc_{bg} =  N_{ap} unc_{bg_{level}}
\end{equation}

The uncertainty due to background noise in the source aperture, {\it unc$_{source}$}, is determined the same way as the PACS {\it unc$_{ap}$} since it is the uncertainty arising from summing the pixels in a given aperture :

\begin{equation}
unc_{source} = \sqrt{N_{ap}} \sigma_{sky}
\end{equation}

The uncertainty arising from the underlying uncertainties of the flux density value in each pixel is computed the same way as for PACS. Here again, this uncertainty arises from the data reduction step when the flux density for each pixel is computed, and the pipeline produces the corresponding error map :

\begin{equation}
unc_{int} = \sqrt{\sum_{i=0}^{N_{ap}} \sigma_{int,i}^2}
\end{equation}

The total uncertainty coming from the determination of the flux density for an extended source, is then :

\begin{equation}
unc_{flux} = \sqrt{unc_{bg}^2 + unc_{source}^2 +unc_{int}^2 + unc_{beam}^2}
\end{equation}

For both types of sources, we also add calibration uncertainties to {\it unc$_{flux}$} to get the final total uncertainty. There are two different SPIRE calibration uncertainties: a systematic uncertainty of $\sim$ 5\% coming from the models used for Neptune, the primary calibrator, which is correlated between the three bands, and a random uncertainty of $\sim$ 2\% coming from the repetitive measurement of the flux densities of Neptune. These two uncertainties were added linearly instead of in quadrature as advised in the SPIRE Observer's Manual, giving an overall 7\% calibration uncertainty {\it unc$_{calib}$}.
The final total uncertainty, $\sigma_{250-350-500}$ reported in Table \ref{fluxesSPIRE}, is obtained by adding {\it unc$_{flux}$} and {\it unc$_{calib}$} in quadrature.

{\revised As for PACS, with SPIRE we also have a redundancy in the error estimation in {\it unc$_{source}$} and {\it unc$_{int}$}, again with only a minor impact on the final uncertainties, $\sigma_{250-350-500}$, as the calibration uncertainty dominates.} %\removed{Note that in {\it unc$_{source}$}, we have a combination of uncertainties from small scale astronomical noise and instrumental uncertainties. These instrumental uncertainties can be redundant with part of the instrumental uncertainties taken into account in {\it unc$_{int}$}, leading to an overestimate of {\it unc$_{flux}$}. However, it has a minor impact on the final uncertainties, $\sigma_{250-350-500}$, as the calibration uncertainty is dominating.}

\subsubsection{Case of upper limits}\label{spirephotoupplim}

\begin{figure}
\begin{center}
\includegraphics[width=8.8cm]{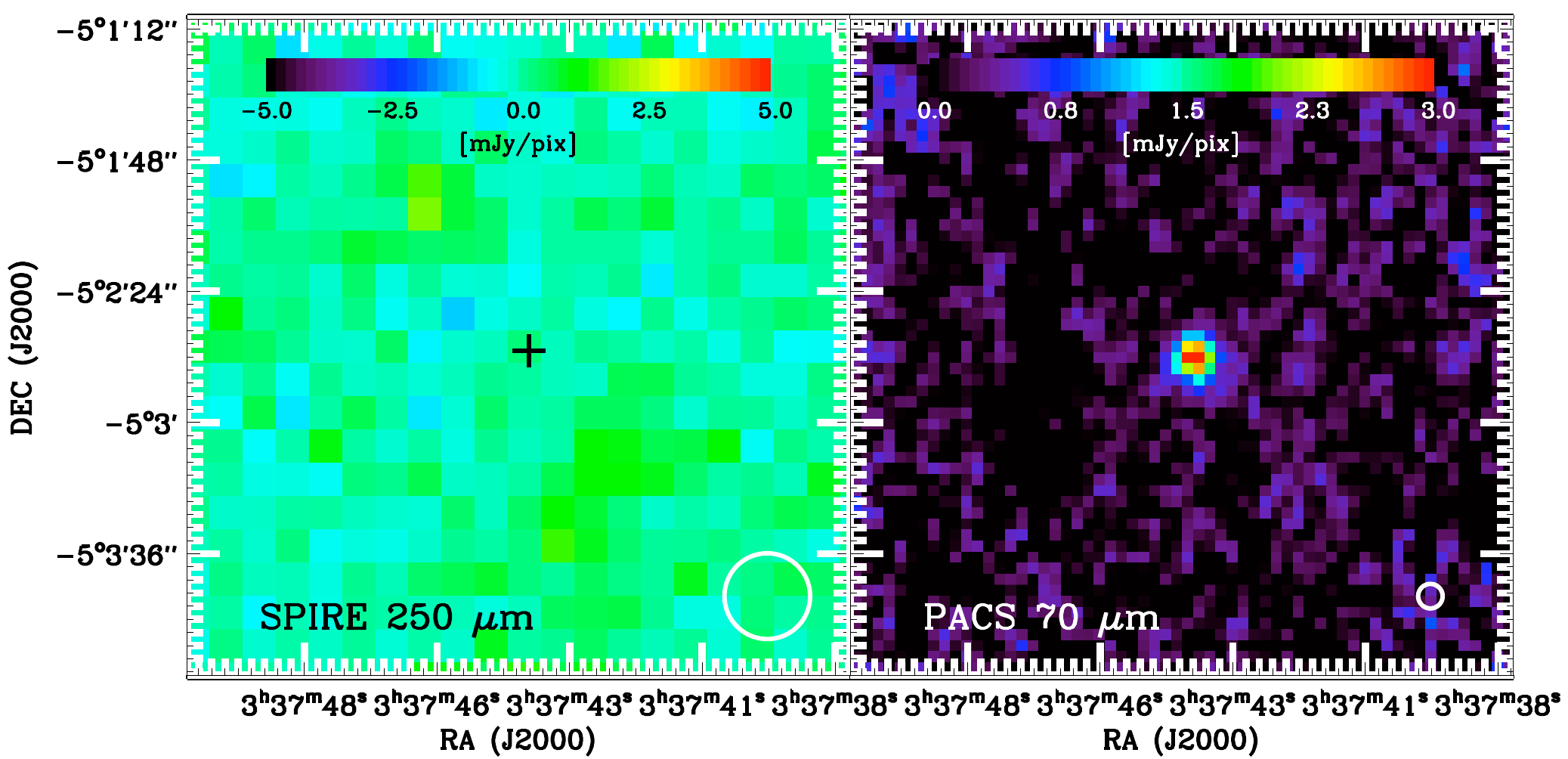}
\caption{Example of a SPIRE non-detection: (left) SPIRE 250 \mic\ and (right) PACS 70 \mic\ image of SBS 0335-052. The position of the galaxy is indicated by a black cross on the SPIRE image. The SPIRE 250 \mic\ (FWHM=18.2\arcsec) and the PACS 70 \mic\ (FWHM=5.6\arcsec) beams are indicated as white circles on the bottom right of the images.}
\label{sbs0335}
\end{center}
\end{figure}

\begin{figure}
\begin{center}
\includegraphics[width=8.8cm]{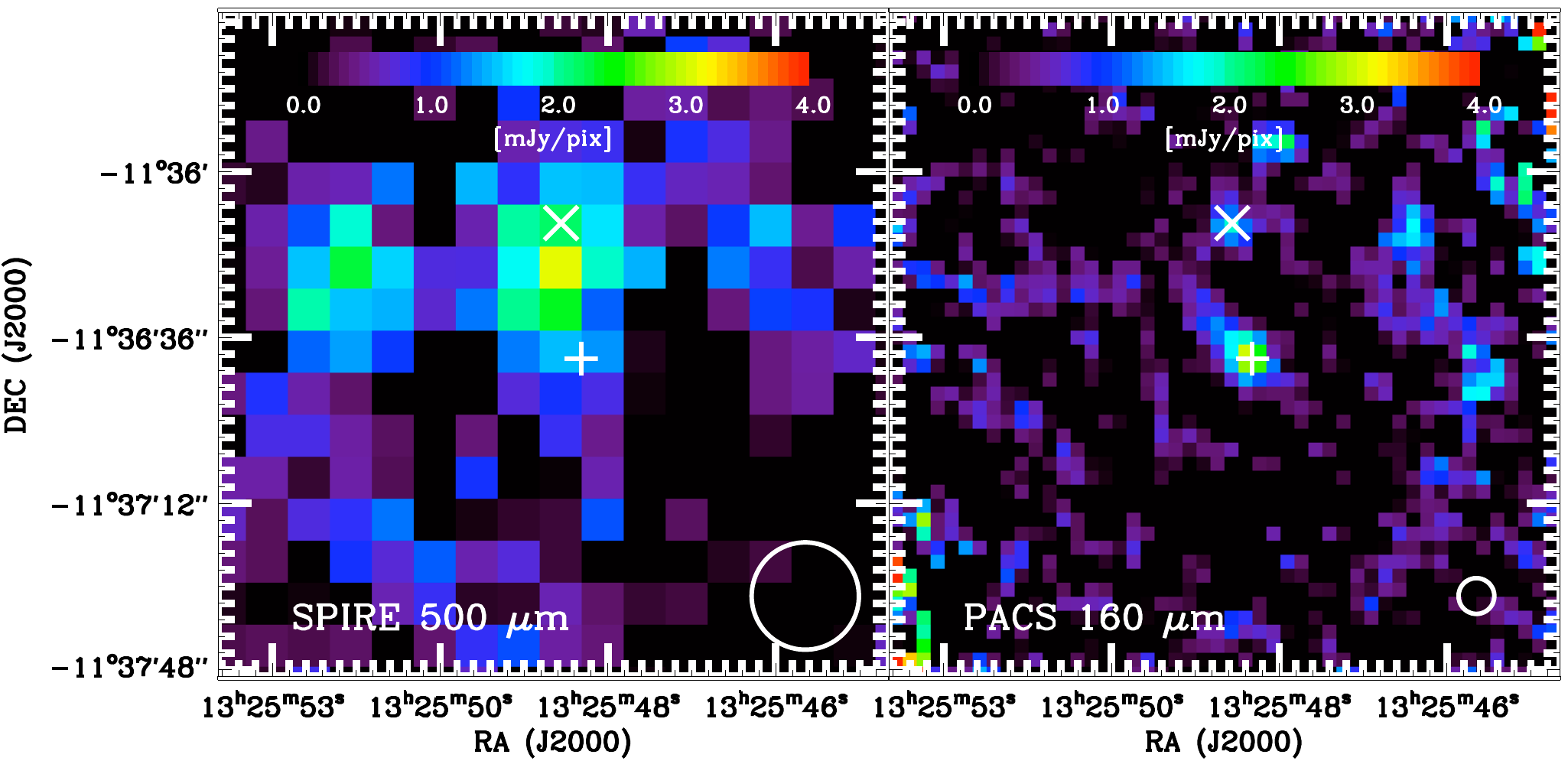}
\caption{Example of a ``mixed'' source. SPIRE 500 \mic\ (left) and PACS 160 (right) images of Pox186 and a contaminating background source. The sources are 20\arcsec\ apart, and are well separated at 160 \mic, but are completely blended at SPIRE 500 \mic\ resolution. Pox186 corresponds to the bottom cross, whereas the contaminating background source is the X. The SPIRE 500 \mic\ (FWHM=36.3\arcsec) and the PACS 160 \mic\ (FWHM=11.3\arcsec) beams are indicated as white circles on the bottom right of the images.
}
\label{pox186}
\end{center}
\end{figure}

When the galaxy is not detected in the SPIRE bands (e.g. SBS 0335-052, Figure \ref{sbs0335}), we can only derive upper limits on the flux density. Also, when the source is blended with another source in the beam and we are unable to confidently separate them (e.g. Pox186 and a background galaxy separated by 20\arcsec, Figure \ref{pox186}), upper limits are reported.
Since the undetected sources are point sources, we use five times the uncertainties reported for point sources in \ref{spirephotoerrors}. 
The only exception is SBS1533+574 which is blended with another source and slightly extended at 250 \mic. The method described above gives an upper limit too low. The extended source photometry method is thus used to derive a 5$\sigma$ upper limit at this wavelength.

%Figure: Example SPIRE upper limit: sbs0335 

%Figure: Example SPIRE mixed sources: pox186 

\subsubsection{Special cases: heavy cirrus contamination}

For NGC 6822 and IC 10, the cirrus contamination from our Galaxy is important in the SPIRE bands. 

\textit{NGC 6822} - \cite{Galametz2010} determined that the contribution from the cirrus to the total emission of the galaxy is of the order of 30\% for all SPIRE bands. To determine the cirrus contribution here, we assume that the entire galaxy is in a homogeneous and flat cirrus region. We determine this cirrus level by considering regions at the same cirrus level outside of the galaxy. This level is used as the background level for the flux determination. We then compare this flux density with the flux density obtained when we consider an uncontaminated background region and get the contamination from the cirrus. We also find that the contribution of the cirrus to the total flux densities is about 30\%, which is coherent with the results from \cite{Galametz2010}. Thus for this galaxy, the flux densities cited in Table \ref{fluxesSPIRE} are flux densities where the cirrus contribution has been subtracted. We also include a conservative 30\% uncertainty in the error for these flux densities to account for the estimation of the cirrus contribution, and for the fact that the cirrus emission may not be flat.

\textit{IC 10} - We apply the same method here. Again, we find that the cirrus contributes $\sim$ 30\% on average, to each SPIRE band. We took this contribution into account by adding this cirrus uncertainty to the other sources of uncertainties for this galaxy. 

This method can be improved, by using the HI maps to better determine the cirrus emission and the background level and thus reducing the uncertainties on the measurements for these two galaxies.

%----------------------- SPIRE flux table----------------------------------------------- 
\begin{table*}
\begin{center}
\caption{Table of SPIRE flux densities for the DGS galaxies. When an upper limit is given, it is the 5$\sigma$ upper limit computed in \ref{spirephotoupplim}.}
\label{fluxesSPIRE}
\begin{tabular}{lcccccc}
\hline
\hline
Source & F$_{250}$ (Jy) & $\sigma_{250}$ (Jy) & F$_{350}$ (Jy) & $\sigma_{350}$ (Jy) & F$_{500}$ (Jy) & $\sigma_{500}$ (Jy) \\
\hline
&&&&&& \\
Haro11       & 0.63 & 0.05 & 0.23 & 0.02 & 0.09 & 0.01 \\
Haro2        & 1.28$^a$ & 0.10 & 0.53$^a$ & 0.04 & 0.15$^a$ & 0.01 \\
Haro3        & 1.79$^a$ & 0.15 & 0.77$^a$ & 0.07 & 0.23$^a$ & 0.02 \\
He2-10       & 6.67$^a$ & 0.54 & 2.64$^a$ & 0.22 & 0.79$^a$ & 0.07 \\
HS0017+1055  & $\leq$0.030 &   - & $\leq$0.050 &   - & $\leq$0.045 &   - \\
HS0052+2536$^b$  & 0.058 & 0.007 & 0.03 & 0.01 & 0.018 & 0.009 \\
HS0822+3542$^c$  &   - &   - &   - &   - &   - &   - \\
HS1222+3741$^c$  &   - &   - &   - &   - &   - &   - \\
HS1236+3937  & $\leq$0.030 &   - & $\leq$0.050 &   - & $\leq$0.045 &   - \\
HS1304+3529  & 0.038 & 0.007 & $\leq$0.050 &   - & $\leq$0.045 &   - \\
HS1319+3224$^c$  &   - &   - &   - &   - &   - &   - \\
HS1330+3651$^c$  &   - &   - &   - &   - &   - &   - \\
HS1442+4250  & $\leq$0.030 &   - & $\leq$0.050 &   - & $\leq$0.045 &   - \\
HS2352+2733  & $\leq$0.030 &   - & $\leq$0.050 &   - & $\leq$0.045 &   - \\
IZw18        & $\leq$0.030 &   - & $\leq$0.050 &   - & $\leq$0.045 &   - \\
IC10$^d$         & 101.$^a$ &  31. & 47.6$^a$ & 14.8 & 16.3$^a$ &  5.1 \\
IIZw40       & 1.33$^a$ & 0.12 & 0.58$^a$ & 0.06 & 0.18 & 0.02 \\
Mrk1089$^b$      & 1.75$^a$ & 0.15 & 0.78$^a$ & 0.07 & 0.24$^a$ & 0.03 \\
Mrk1450      & 0.049 & 0.007 & $\leq$0.050 &   - & $\leq$0.045 &   - \\
Mrk153       & 0.048$^a$ & 0.008 & $\leq$0.050 &   - & $\leq$0.045 &   - \\
Mrk209       & 0.062 & 0.007 & 0.03 & 0.01 & $\leq$0.045 &   - \\
Mrk930       & 0.40$^a$ & 0.04 & 0.13$^a$ & 0.01 & 0.049$^a$ & 0.007 \\
NGC1140      & 1.97$^a$ & 0.17 & 0.94$^a$ & 0.08 & 0.28$^a$ & 0.03 \\
NGC1569      & 12.0$^a$ &  1.0 & 5.02$^a$ & 0.41 & 1.55$^a$ & 0.13 \\
NGC1705      & 0.60$^a$ & 0.05 & 0.29$^a$ & 0.03 & 0.10$^a$ & 0.01 \\
NGC2366      & 2.04$^a$ & 0.17 & 1.01$^a$ & 0.09 & 0.39$^a$ & 0.04 \\
NGC4214      & 18.6$^a$ &  1.5 & 9.92$^a$ & 0.80 & 3.79$^a$ & 0.31 \\
NGC4449      & 32.4$^a$ &  2.6 & 14.8$^a$ &  1.2 & 5.01$^a$ & 0.41 \\
NGC4861      & 1.10$^a$ & 0.10 & 0.54$^a$ & 0.05 & 0.20$^a$ & 0.03 \\
NGC5253      & 7.82$^a$ & 0.63 & 3.64$^a$ & 0.29 & 1.18$^a$ & 0.10 \\
NGC625       & 4.33$^a$ & 0.35 & 2.18$^a$ & 0.18 & 0.80$^a$ & 0.07 \\
NGC6822$^d$      & 48.4$^a$ & 15.0 & 29.7$^a$ &  9.2 & 13.6$^a$ &  4.2 \\
Pox186       & 0.045 & 0.007 & $\leq$0.050 &   - & $\leq$0.045 &   - \\
SBS0335-052  & $\leq$0.030 &   - & $\leq$0.050 &   - & $\leq$0.045 &   - \\
SBS1159+545  & $\leq$0.030 &   - & $\leq$0.050 &   - & $\leq$0.045 &   - \\
SBS1211+540  & $\leq$0.030 &   - & $\leq$0.050 &   - & $\leq$0.045 &   - \\
SBS1249+493  & $\leq$0.030 &   - & $\leq$0.050 &   - & $\leq$0.045 &   - \\
SBS1415+437$^c$  &   - &   - &   - &   - &   - &   - \\
SBS1533+574  & $\leq$0.122$^a$ &   - & $\leq$0.050 &   - & $\leq$0.045 &   - \\
Tol0618-402$^c$  &   - &   - &   - &   - &   - &   - \\
Tol1214-277  & $\leq$0.030 &   - & $\leq$0.050 &   - & $\leq$0.045 &   - \\
UGC4483      & 0.024 & 0.006 & $\leq$0.050 &   - & $\leq$0.045 &   - \\
UGCA20$^c$       &   - &   - &   - &   - &   - &   - \\
UM133        & 0.032 & 0.006 & $\leq$0.050 &   - & $\leq$0.045 &   - \\
UM311$^b$        & 3.84$^a$ & 0.31 & 1.87$^a$ & 0.16 & 0.66$^a$ & 0.06 \\
UM448        & 0.99$^a$ & 0.08 & 0.38$^a$ & 0.03 & 0.13 & 0.01 \\
UM461        & 0.027 & 0.006 & 0.03 & 0.01 & $\leq$0.045 &   - \\
VIIZw403     & 0.14$^a$ & 0.01 & 0.053 & 0.008 & 0.028 & 0.009 \\
\hline
\end{tabular}
\end{center}
\scriptsize{
$^a$: The flux densities are derived from aperture photometry, with the same aperture used for PACS.

$^b$: These objects are galaxies within compact groups of galaxies or are parts of other galaxies and the photometry given here is for the whole group (see Section \ref{pacsphotofluxes} for details).

$^c$: These sources were not observed at all by SPIRE.

$^d$: The quoted flux densities for these sources have been corrected for cirrus contamination.}
\end{table*}

\subsection{Comparison of PACS and MIPS existing flux densities}\label{compspitzer}
We compare our PACS flux densities to the flux densities at 70 and 160 \mic\ from MIPS onboard the \spitz\ Space Telescope from \cite{Bendo2012} to assess the reliability of our measurements.

\subsubsection{MIPS photometry}\label{mipsphoto}
The table of the available MIPS data for the DGS is given in \cite{Madden2013} and \cite{Bendo2012} who give a detailed description of the photometry for total galaxy flux densities. Of the DGS sample, 34 galaxies have been observed by MIPS in the considered bands. \cite{Bendo2012} MIPS flux densities compare well with previously published MIPS samples containing a subset of the DGS galaxies \citep{Dale2007, Engelbracht2008}.
Therefore we are confident about the reliability of these results and will use them to perform the comparison with our PACS flux densities.

\subsubsection{Comparison with PACS}\label{comparison}

The PACS flux densities correspond to monochromatic values for sources with spectra where $\nu f_{\nu}$ is constant, while the MIPS flux densities are monochromatic values for sources with the spectra of a 10$^4$ K blackbody, so colour corrections need to be applied to measurements from both instruments before they are compared to each other. We first fit a blackbody through the three PACS data points and apply the corresponding colour corrections from the available PACS documentation\footnote{{\revised The corresponding documentation for PACS colour corrections is available at http://herschel.esac.esa.int/twiki/pub/Public/Pacs \\ CalibrationWeb/cc\_report\_v1.pdf}}. For the MIPS flux densities, we fit a blackbody through the 70 and 160 \mic\ data points (not using the 24 \mic\ point) and apply the corrections from the MIPS Handbook\footnote{{\revised The MIPS Instrument Handbook is available at} http://irsa.ipac.caltech.edu/data/SPITZER/docs/mips/ \\ mipsinstrumenthandbook/home/}. The typical colour corrections for MIPS are of the order of 10 and 4\% on average at 70 and 160 \mic. However, they are of the order of 1 or 2\% in the 70 and 160 \mic\ PACS bands. 
For non detected galaxies, where we, for PACS, and/or \cite{Bendo2012}, for MIPS, reported upper limits (nine galaxies), we are not able to properly fit a blackbody and therefore derive a proper colour correction. We do not compare PACS and MIPS flux densities for these galaxies for now. 

\begin{figure}[h!tbp]
\begin{center}
\includegraphics[width=8.8cm]{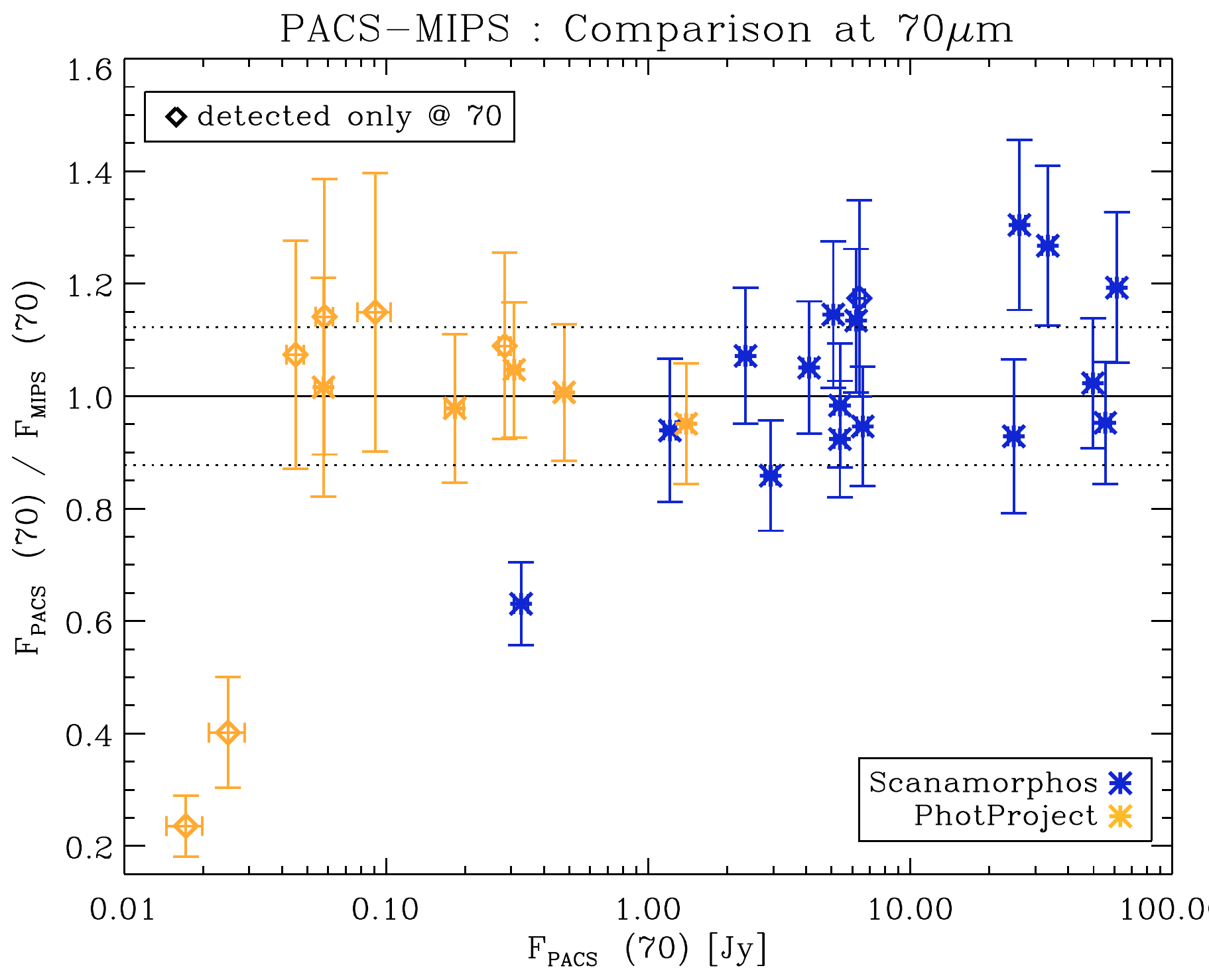}
\includegraphics[width=8.8cm]{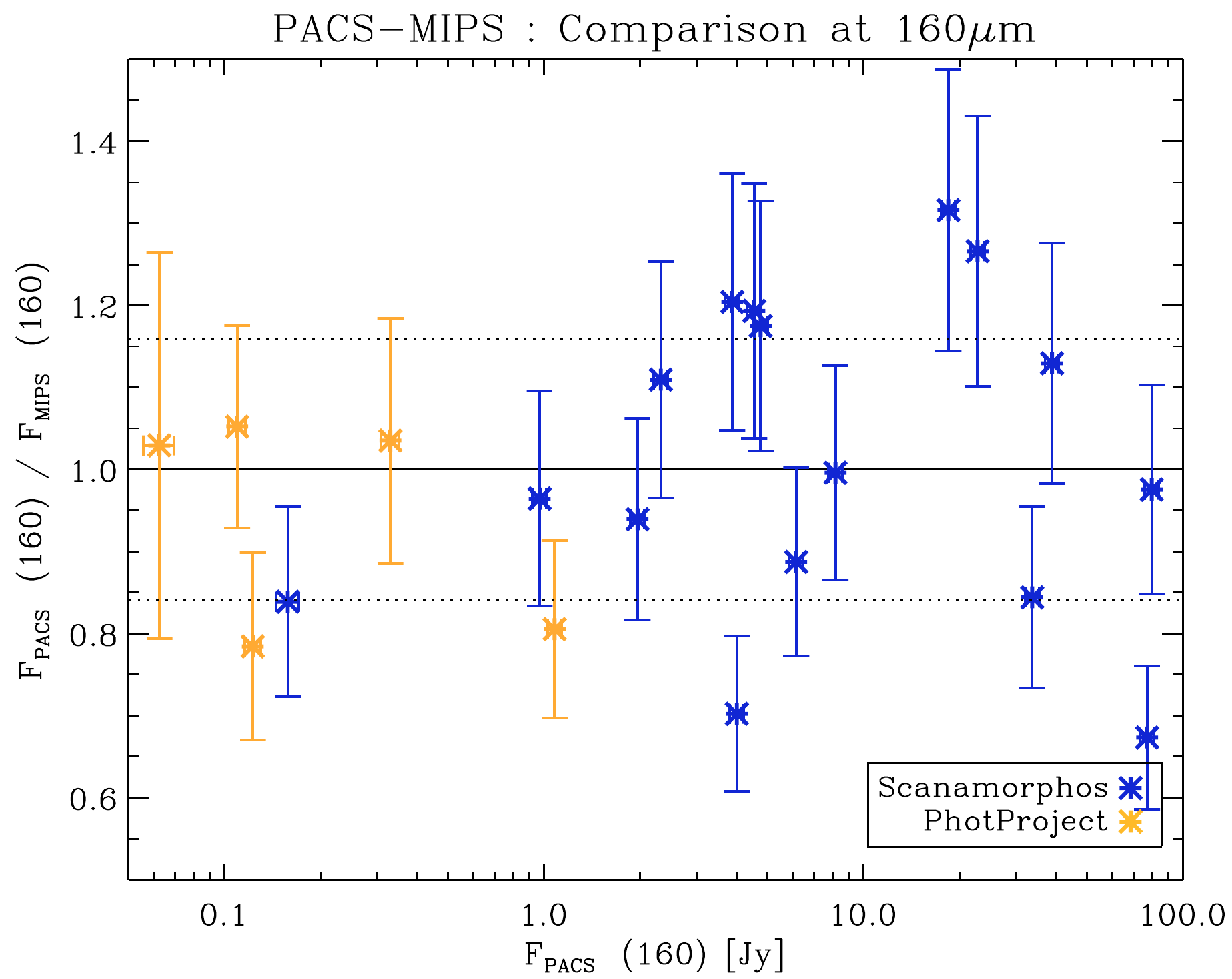}
\caption{Comparison of PACS flux densities and MIPS flux densities: PACS-to-MIPS flux density ratios as a function of PACS flux density at 70 \mic\ {\it(top)} and 160 \mic\ {\it(bottom)}. As a guide to the eye, the unity line is added as a solid line as well as the average uncertainties on the ratio in both bands as dotted lines. These average uncertainties are $\sim$12\% and $\sim$16\% at 70 and 160 \mic. Colours distinguish the selected mapping method.}
\label{pm_comparison}
\end{center}
\end{figure}

We use the ratios of the PACS and MIPS flux densities to assess how well the measurements from the instrument agree with each other; a ratio of one corresponds to a very good agreement. The average PACS/MIPS ratios at 70 and 160 microns are shown in Figure \ref{pm_comparison}, and the correspondence is relatively good. The PACS/MIPS ratio is 1.019 $\pm$ 0.112 at 70 \mic\ and 0.995 $\pm$ 0.153 at 160 \mic. This is to be compared to an average uncertainty of $\sim$12\% ($\sim$11\% from MIPS and $\sim$5\% for PACS, added in quadrature) and $\sim$16\% ($\sim$15\% from MIPS and $\sim$7\% for PACS, added in quadrature) on the ratios at 70 and 160 \mic\ respectively. \cite{Aniano2012} found a slightly less good agreement ($\sim$ 20\%) for integrated fluxes of two KINGFISH galaxies. 
 
If we now consider galaxies detected at 70 \mic\ and not at 160 \mic, indicated by a different symbol on the upper panel of Figure  \ref{pm_comparison}, we are still able to compare, with extra caution, the measurements at 70 \mic. Indeed, as we are not able to derive a proper colour correction for those galaxies, we add to the MIPS 70 \mic\ flux densities a 10\% uncertainty and a 1\% uncertainty to the PACS 70 \mic\ flux densities to account for the colour correction effect. When adding these extra galaxies at 70 \mic, the PACS/MIPS ratio is 0.985 $\pm$ 0.158 at 70 \mic. This is to be compared with an average uncertainty of $\sim$14\% on the 70 \mic\ ratio ($\sim$12\% from MIPS and $\sim$7\% for PACS, added in quadrature, including the extra galaxies). 
The very faint and discrepant galaxies at 70 \mic\ are HS1222+3741 (ratio of 0.40) and Tol1214-277 (ratio of 0.24). For HS1222+3741, the MIPS image contains some bright pixels near the edge of the photometry aperture used for MIPS, and this may have driven the 70 \mic\ MIPS flux density up. For Tol1214-277, a nearby source is present in the MIPS data and, although its contribution has been subtracted when computing the MIPS 70 \mic\ flux, some contribution from this source may still be present. Additionally, measuring accurate flux densities at $\leq$ 50 mJy in both MIPS and PACS data is difficult and may have led to the discrepancies.

The error on the average ratio is comparable to the average uncertainties on the ratio for both bands. 
Thus there is a good photometric agreement between PACS and MIPS photometry for the DGS sample.

% Figure: PACS to MIPS comparison 

%===========================================================================
% Far Infrared and submillimetre behaviour and dust properties of the dwarf galaxies
%===========================================================================

\section{Far Infrared and submillimetre behaviour and dust properties of the dwarf galaxies}\label{FIRbehaviour}
\begin{figure}
\begin{center}
\includegraphics[width=8.8cm]{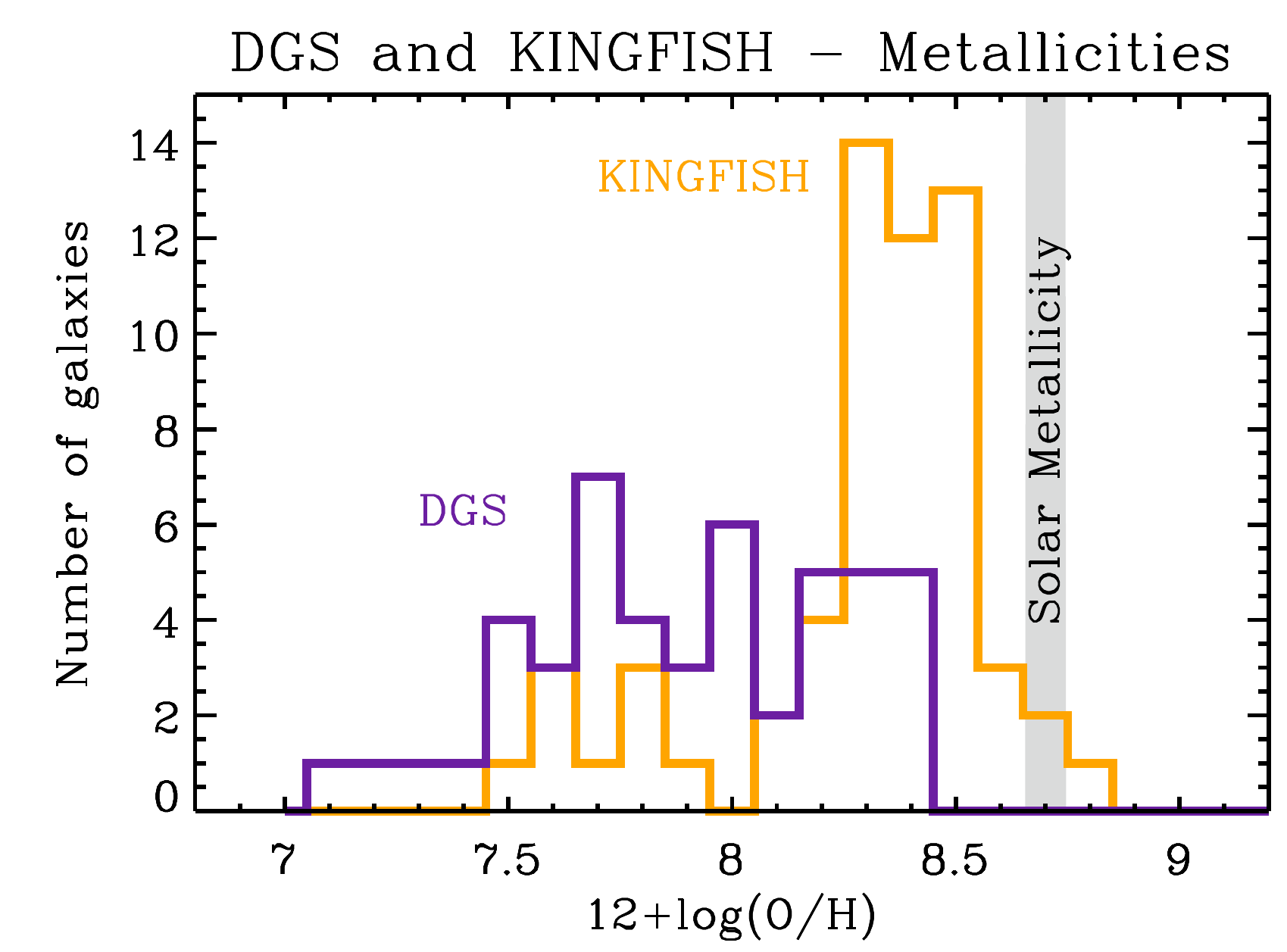}
\caption{Metallicity distributions for both DGS (purple) and KINGFISH (orange) samples. Note how the high-metallicity end is better covered by KINGFISH whereas the low-metallicity end is better covered by the DGS.}
\label{metDGSKF}
\end{center}
\end{figure}

To study the dust properties of the DGS and determine the impact of metallicity, we perform a comparison with galaxies from the KINGFISH sample \citep{Kennicutt2011}. The KINGFISH survey contains 61 galaxies: 41 spiral galaxies, 11 early-type galaxies (E and S0) and nine irregulars \citep{Kennicutt2011}. KINGFISH is a survey including more metal-rich galaxies and enables us to span a wider metallicity range, notably by filling up the high-metallicity end of the  metallicity distribution (Fig. \ref{metDGSKF}). The metallicities adopted here for the KINGFISH sample have been determined the same way as for the DGS in \cite{Kennicutt2011}, using the method of \cite{PilyuginThuan2005}\footnote{See \cite{Madden2013} for the DGS metallicity determination. The KINGFISH metallicities are from Column 9 from Table 1 of \cite{Kennicutt2011}.}. No errors for metallicities are given in \cite{Kennicutt2011} so we adopt a 0.1 dex error for the KINGFISH metallicities. The \hersc\ KINGFISH flux densities are taken from \cite{Dale2012}\footnote{{\revised The KINGFISH SPIRE fluxes and corresponding uncertainties are updated to match the latest SPIRE beam areas. The beam areas used in this paper were released in September 2012, after the publication of \cite{Dale2012} in January 2012.}}.
%The \hersc\ Reference Survey (HRS) \citep{Boselli2010a} has not been considered here as the PACS observations of the HRS are still under way, and thus homogeneous data for the whole sample are not, at present, available.

We use FIR colour-colour diagrams (Section \ref{ccdiagrams}) and modified blackbody models (Section \ref{modelling}) in order to derive some physical dust parameters of the galaxies, such as the temperature ({\it T}), the emissivity index ($\beta$), the dust mass ({\it M$_{dust}$}) and the FIR luminosity ({\it L$_{FIR}$}). 
In Section \ref{submmexcess}, we then investigate the presence of a possible submm excess in the galaxies .

%Figures: Met histo DGS+KF

\subsection{Characterization of the SED shapes}\label{ccdiagrams}

In order to obtain a qualitative view of the FIR-to-submm behaviour of the DGS sample, and to compare with the KINGFISH sample, we inspect the observed \hersc\ SEDs as well as several \hersc\ colour-colour diagrams combining both PACS and SPIRE observations.

\subsubsection{Observed spectral energy distributions}\label{obsSED}

\begin{figure*}[h!tbp]
\begin{center}
\includegraphics[width = 15cm]{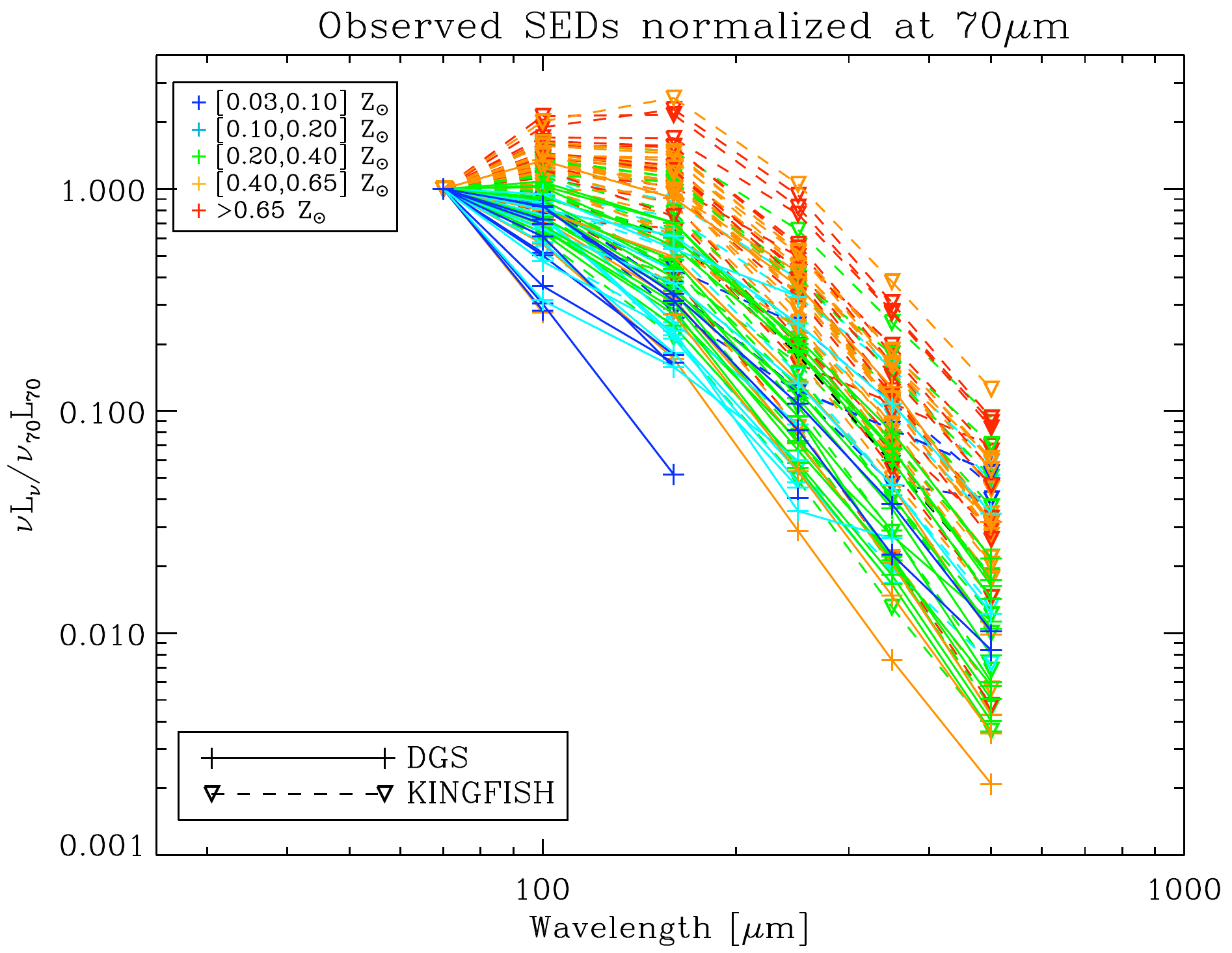}
\caption{Total \hersc\ observed SEDs for both DGS and KINGFISH samples, normalized at 70 \mic. The colours delineate the different metallicity bins, and the lines and symbols differentiate DGS (plain lines and crosses) and KINGFISH galaxies (dashed lines and downward triangles).}
\label{dgs_seds}
\end{center}
\end{figure*}

Total observed SEDs for both samples are computed for a first look at the characteristic SED shapes in the DGS and KINGFISH samples (Figure \ref{dgs_seds}). The upper limits are not indicated here for clarity. The most metal-poor galaxies are also the faintest and therefore not detected with \hersc\ beyond 160 \mic. The observed SEDs are normalized at 70 \mic, and we see here that the peak of the SED shifts towards longer wavelengths as the metallicity increases, reflecting the impact of metallicity on the observed dust properties.

\subsubsection{Dwarf Galaxy Survey colours}

\noindent {\it Constructing the colour-colour diagrams} \\

The \hersc\ colour-colour diagrams are constructed first by computing the observed ratios and the corresponding error bars, for both DGS and KINGFISH, and omitting the galaxies with more than one upper limit in the considered bands. 

We then compute the theoretical \hersc\ flux ratios of simulated modified blackbodies spanning a range in temperature ({\it T} from 0 to 40 K in 2 K bins and from 40 to 100 K in 10 K bins) and emissivity indices ($\beta$ from 0.0 to 2.5). From now on, we define the emissivity index fixed for modelling the simulated \hersc\ flux ratios as ``$\beta_{theo}$'', and ``$\beta_{obs}$'' when we leave the emissivity index as a free parameter in modified blackbody fits (see Section \ref{modelling}). In our simulated modified blackbody, the emitted fluxes are proportional to $ \lambda^{-\beta_{theo}} \times B_\nu(\lambda, T)$, where $B_\nu(\lambda, T)$ is the Planck function.

The pipeline we use for the data reduction gives us monochromatic flux densities for our data points for both PACS and SPIRE.
To mimic the output of the pipeline for our theoretical points we weigh our theoretical flux density estimates by the RSRF of the corresponding bands. For SPIRE simulated measurements, we then convert our RSRF-weighted flux densities into monochromatic flux densities by applying the {\it K$_4$} correction given on the SPIRE Observers' Manual. %\removed{We adapt this correction for PACS simulated measurements.} 
{\revised For PACS, we also colour correct the RSRF-weighted modeled flux densities to a spectrum where $\nu$F$_{\nu}$ is constant (i.e. multiply by the analogous of {\it K$_4$} for PACS).}
These simulated flux ratios from a simple model are useful indicators to interpret the colour-colour diagrams. \\

\noindent {\it FIR/submm colours} \\ 

\begin{figure*}
\begin{center}
\includegraphics[width=15cm]{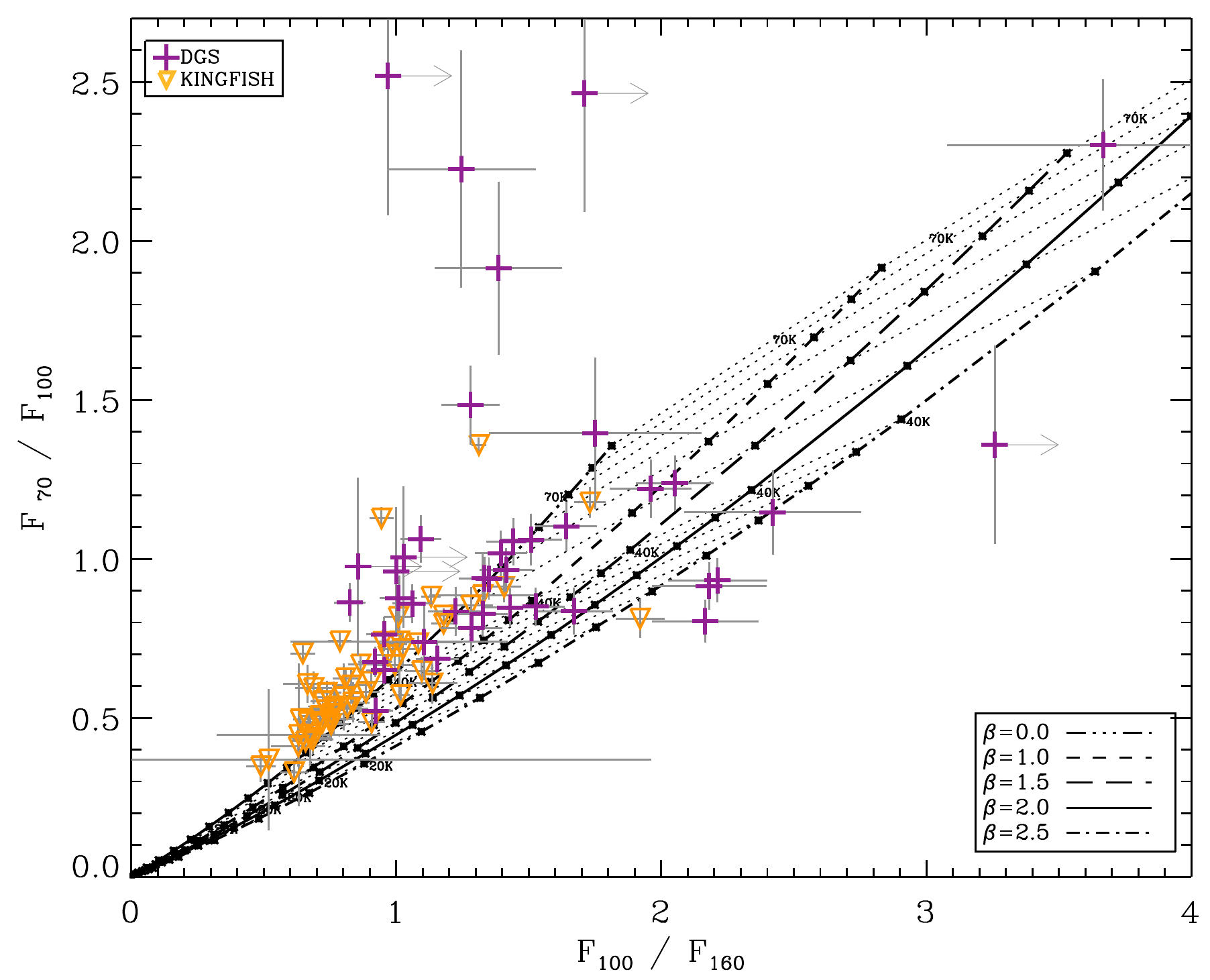}
\includegraphics[width=15cm]{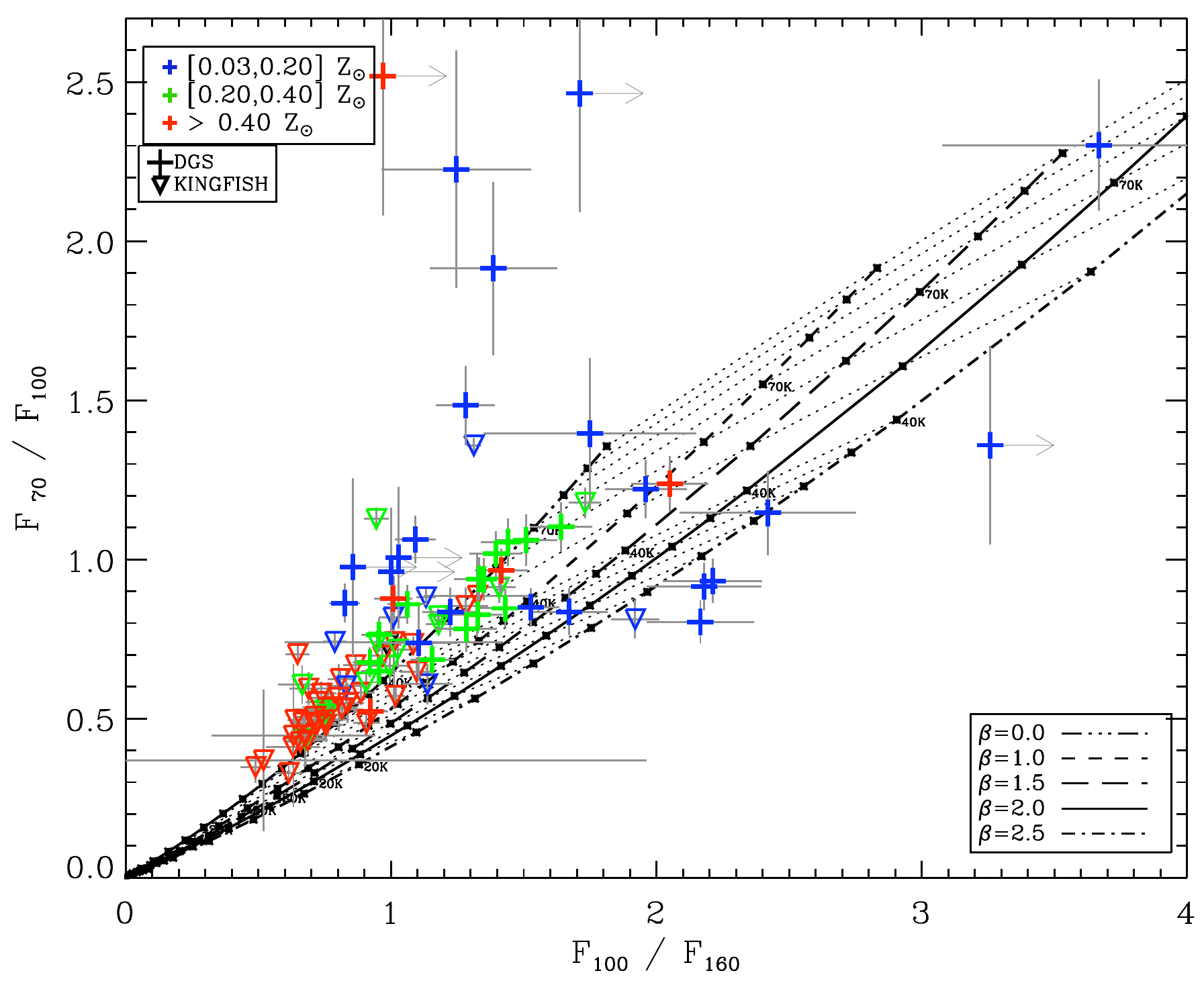}
\caption{Colour-colour diagram: PACS/PACS diagram: {\it F$_{70}$/F$_{100}$} versus {\it F$_{100}$/F$_{160}$}. {\it (top)} The colour and symbol code differentiates DGS (purple crosses) and KINGFISH galaxies (orange downward triangles). {\it (bottom)} The colour code delineates the different metallicity bins this time. Crosses and downward triangles are still representing DGS and KINGFISH galaxies, respectively. For both plots, the curves give theoretical Herschel flux ratios for simulated modified black bodies for $\beta_{theo}$ = 0.0 to 2.5 and {\it T} from 0 to 40 K in 2 K bins and from 40 to 100 K in 10 K bins, as black dots, increasing in {\it T} from left to right. Lines of constant {\it T} are indicated as dotted lines, and a few temperatures have been marked on the plots.}
\label{diagrams}
\end{center}
\end{figure*}

\begin{figure*}
\begin{center}
\includegraphics[width=15cm]{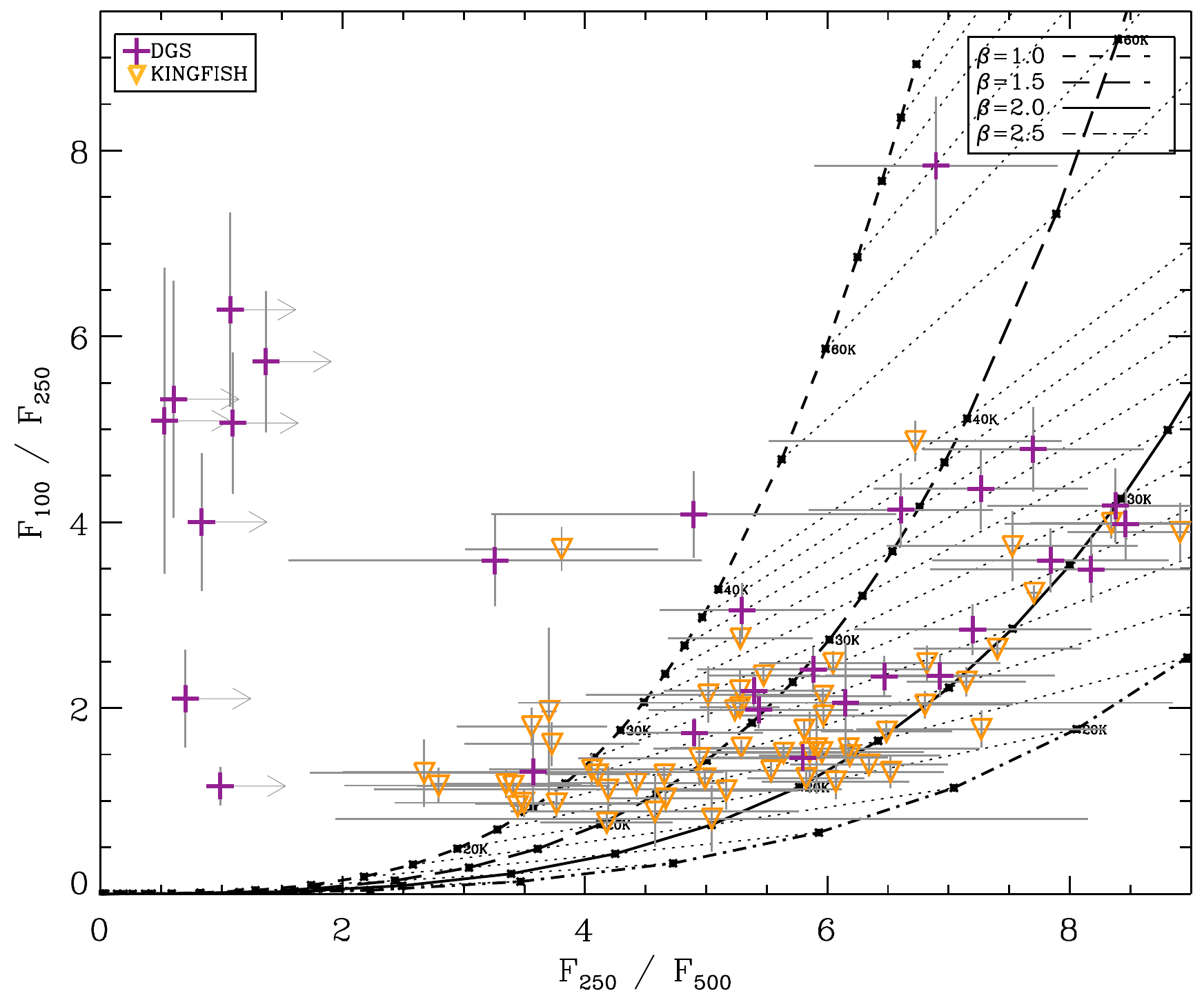}
\includegraphics[width=15cm]{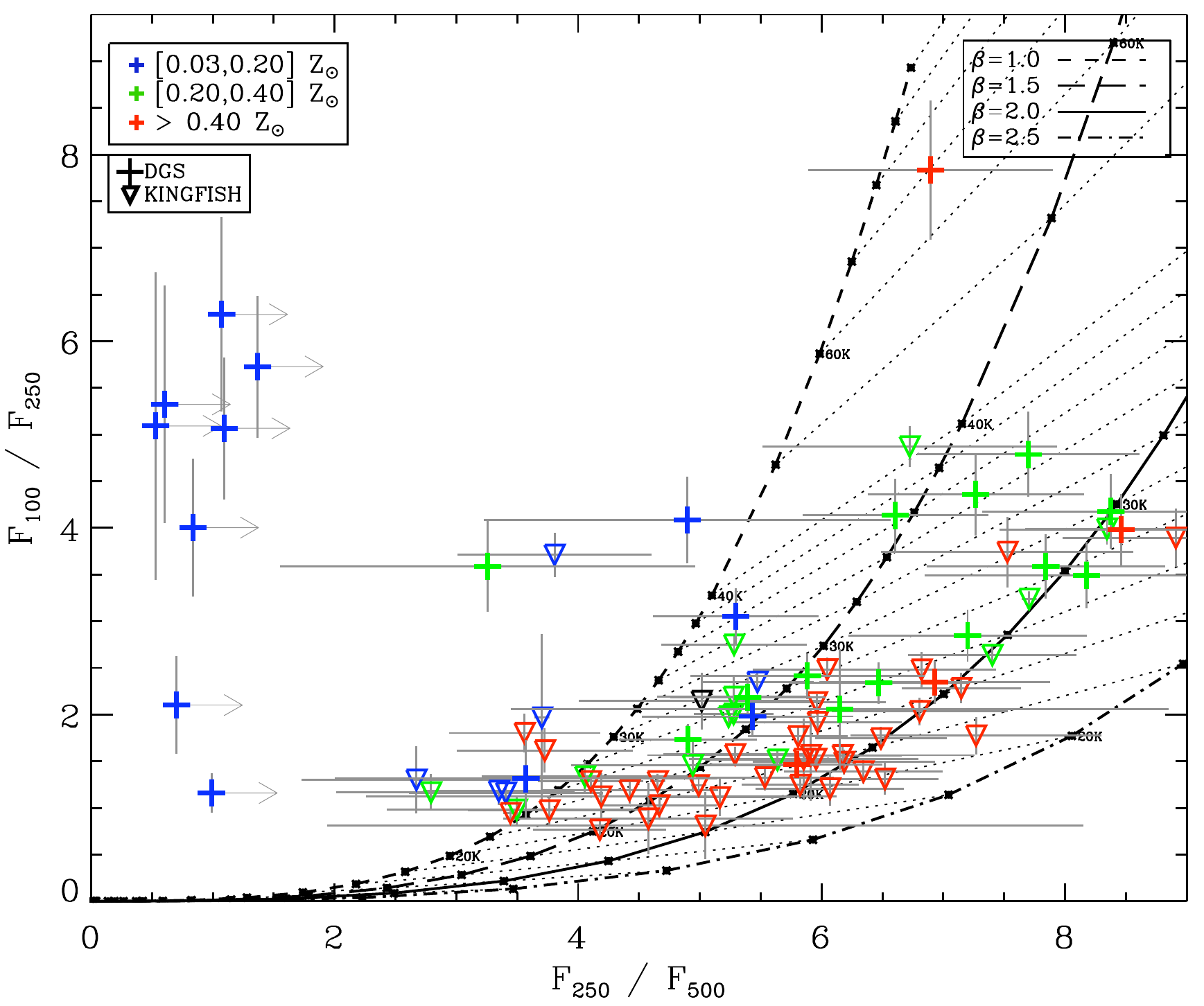}
\caption{Colour-colour diagram: PACS/SPIRE diagram: {\it F$_{100}$/F$_{250}$} versus {\it F$_{250}$/F$_{500}$}. 
The colour and symbol choices are the same as in Figure \ref{diagrams} for both figures. Note that the most metal-poor galaxies (from 0.03 to 0.20 \zsun) are very faint and even not detected anymore at long wavelengths. We were only able to derive upper limits beyond 160 \mic\ for these galaxies and, thus, some galaxies do not appear on this diagram anymore.}
\label{diagramPACS_SPIRE}
\end{center}
\end{figure*}

The spread of galaxies on the colour-colour diagrams (Figures \ref{diagrams} and \ref{diagramPACS_SPIRE}) reflects broad variations in the SED shape and metallicity in our survey. 

Indeed the DGS galaxies show a wider spread in location on the diagrams compared to the KINGFISH galaxies (Figures \ref{diagrams} and \ref{diagramPACS_SPIRE}, top panels), reflecting the differences in the dust properties between dwarf galaxies and the generally more metal-rich environments probed by the KINGFISH survey.

The {\it F$_{70}$/F$_{100}$ vs F$_{100}$/F$_{160}$} diagram (Figure \ref{diagrams}) traces best the peak of the SED. Galaxies usually exhibit a peak in their SED around $\sim$ 100 - 160 \mic. Galaxies presenting FIR flux densities with {\it F$_{70}$ $>$ F$_{100}$ $>$ F$_{160}$} may be quite warm as they peak at wavelengths less than 70 \mic. Colder galaxies would lie in the lower left corner of the plot ({\it F$_{70}$ $<$ F$_{100}$ $<$ F$_{160}$}), as shown by the simulated flux ratio lines. KINGFISH galaxies indeed cluster in the corresponding lower left corner of the plot while DGS galaxies span a wider space (Fig \ref{diagrams}, top). Nonetheless both samples follow the theoretical flux ratio lines from simulated modified blackbodies. There are some outliers, all of them being very faint, extremely metal-poor galaxies (from 0.03 to 0.20 \zsun). 
There is also a metallicity trend in Fig. \ref{diagrams} (bottom), either between KINGFISH and the DGS or within both samples, i.e. low-metallicity (dwarf) galaxies peak at much shorter wavelengths and thus have overally warmer dust (several tens of K), compared to more metal-rich galaxies. 

In dwarf galaxies, the warmer dust is due to the very energetic environment in which the grains reside: the density of young stars causes the ISRF to be much harder on global scales than in normal galaxies \citep{Madden2006}. The low dust extinction enables the FUV photons from the young stars to penetrate deeper into the ISM. The dust grains are thus exposed to harder and more intense ISRF than in a more metal-rich environment. This increases the contribution of hot and warm dust to the total dust emission resulting in overall higher equilibrium dust temperatures.

Note that there is a small excess at 70 \mic\ for most of the galaxies compared to our simulated modified blackbodies, causing them to fall above the lowest $\beta_{theo}$ line. This means that if we were to fit a modified blackbody {\it only} to the FIR flux densities (from 70 \mic\ to 160 \mic) we would get very low $\beta_{obs}$, i.e. a very flat SED in the FIR, which reflects a broad peak in the observed SED. This is due to the crudeness of the isothermal approximation made in the modified blackbody modelling. In a real galaxy, the dust grains are distributed in a range of temperatures, (e.g. hotter dust around star-forming regions vs colder dust in the diffuse ISM). Such a low $\beta_{obs}$ here is only a side effect of the distribution in temperature of the grains in the galaxy. The extremely metal-poor outliers noted before may present an even wider temperature distribution than in more metal-rich galaxies, towards the higher temperatures, causing the broadening of the peak in their dust SED and their peculiar location on the diagrams in Fig. \ref{diagrams}. {\revised Part of this excess at 70 \mic\ could also be due to non-thermal heating, i.e. dust grains whose emission can not be represented by a modified blackbody, such as stochastically heated small grains.}

More accurate values of {\it T} and $\beta_{theo}$ may be illustrated by including submm data in the colour-colour diagrams. At submm wavelengths (beyond $\sim$ 250 \mic), the emissivity index this time represents an intrinsic grain property: the efficiency of the emission from the dust grain. A theoretical emissivity index $\beta_{theo}$~=~2 is commonly used to describe the submm SED for local and distant galaxies in the models as it represents the intrinsic optical properties of Galactic grains (mixture of graphite and silicate grains). More recently $\beta_{theo}$ between 1.5 and 2 have also been used \citep[e.g.][]{Amblard2010, Dunne2011}. 
The {\it F$_{100}$/F$_{250}$ vs F$_{250}$/F$_{500}$} diagram (Figure \ref{diagramPACS_SPIRE}) reflects best the variations in emissivity index $\beta_{theo}$. Here again the DGS galaxies are more wide-spread than the KINGFISH galaxies (Figure \ref{diagramPACS_SPIRE}, top) spanning larger ranges of {\it F$_{100}$/F$_{250}$} and {\it F$_{250}$/F$_{500}$} ratios, that is, wider ranges in temperature and $\beta$ (such as higher {\it T} and lower $\beta$). As far as metallicity is concerned, the trend with temperature already noted in Fig. \ref{diagrams} is still present (Figure \ref{diagramPACS_SPIRE}, bottom). But hardly any trend between $\beta$ and metallicity can be noticed: as the extremely low-metallicity galaxies are not detected at 500 \mic, it is rather difficult to conclude on this point relying only on the FIR/submm colour-colour diagram.

Modelling low-metallicity dwarf galaxies with grain properties derived from the Galaxy (i.e. using $\beta_{theo}$~=2), may thus not be accurate. The galaxies showing a lower $\beta_{obs}$ ($\beta_{obs}$ $<$ 2) will have a flatter submm slope. Smaller {\it F$_{250}$/F$_{500}$} ratios, that can be seen as a sign of lower $\beta_{obs}$, indicative of a flatter submm slope, have already been noted by \cite{Boselli2012} for sub-solar metallicity galaxies.
This flatter slope may be the sign of a contribution from an extra emission in excess of the commonly used $\beta_{theo}$ = 2 models. Thus the flattening of the observed submm slope ($\beta_{obs}$ $<$ 2) could be used as a diagnosis for possible excess emission appearing at 500 \mic\ (see Section \ref{submmexcess}).

%Figure: colour-colour diagrams 

 \subsection{FIR/Submm modelling}\label{modelling}

To complete our observational and qualitative view of the FIR-submm behaviour  of the DGS and KINGFISH galaxies, we use a modified blackbody model to quantitatively determine the parameters already discussed before: {\it L$_{FIR}$}, {\it M$_{dust}$}, {\it T}, $\beta_{obs}$, in the DGS and KINGFISH samples.

 \subsubsection{Modified blackbody fitting}\label{bbfits} 
A single modified blackbody is fitted to the \hersc\ data of each galaxy from the DGS sample where the free parameters are: temperature ({\it T}) and dust mass ({\it M$_{dust}$}) as well as the emissivity index ($\beta_{obs}$), where we leave $\beta_{obs}$ free in the [0.0, 2.5] range. The modeled flux densities are given by: 

\begin{equation}\label{BB}
F_{\nu}(\lambda) = \frac{M_{dust}\kappa(\lambda_0)}{D^2}\left(\frac{\lambda}{\lambda_0}\right)^{-\beta_{obs}}B_{\nu}(\lambda, T)
\end{equation}

where $\kappa(\lambda_0)$ = 4.5 $m^2 kg^{-1}$ is the dust mass absorption opacity at the reference wavelength, $\lambda_0$ = 100 \mic. $\kappa(\lambda_0)$ has been calculated from the grain properties of \cite{Zubko2004}, as in \cite{Galliano2011}\footnote{for their ``Standard Model", see Appendix A of \cite{Galliano2011}.}, {\revised and is consistent with a $\beta_{theo}$ = 2. Leaving $\beta_{obs}$ to vary in our fit can produce lower dust masses for lower $\beta_{obs}$ \citep{Bianchi2013}. This effect is discussed for the two dust masses relations we derive in Section \ref{bbfitsprop}. Moreover, this particular choice for the value $\kappa(\lambda_0)$ will only affect the absolute values of the dust masses. Choosing another model to derive $\kappa(\lambda_0)$ would not affect the intrinsic variations noted in Section \ref{bbfitsprop}.}
{\it D} is the distance to the source (given in Table \ref{sample}) and {\it B}$_{\nu}(\lambda, T)$ is the Planck function. {\revised Colour corrections are included in the fitting procedure.}

\begin{figure}[h!tbp]
\begin{center}
\includegraphics[width=8.8cm]{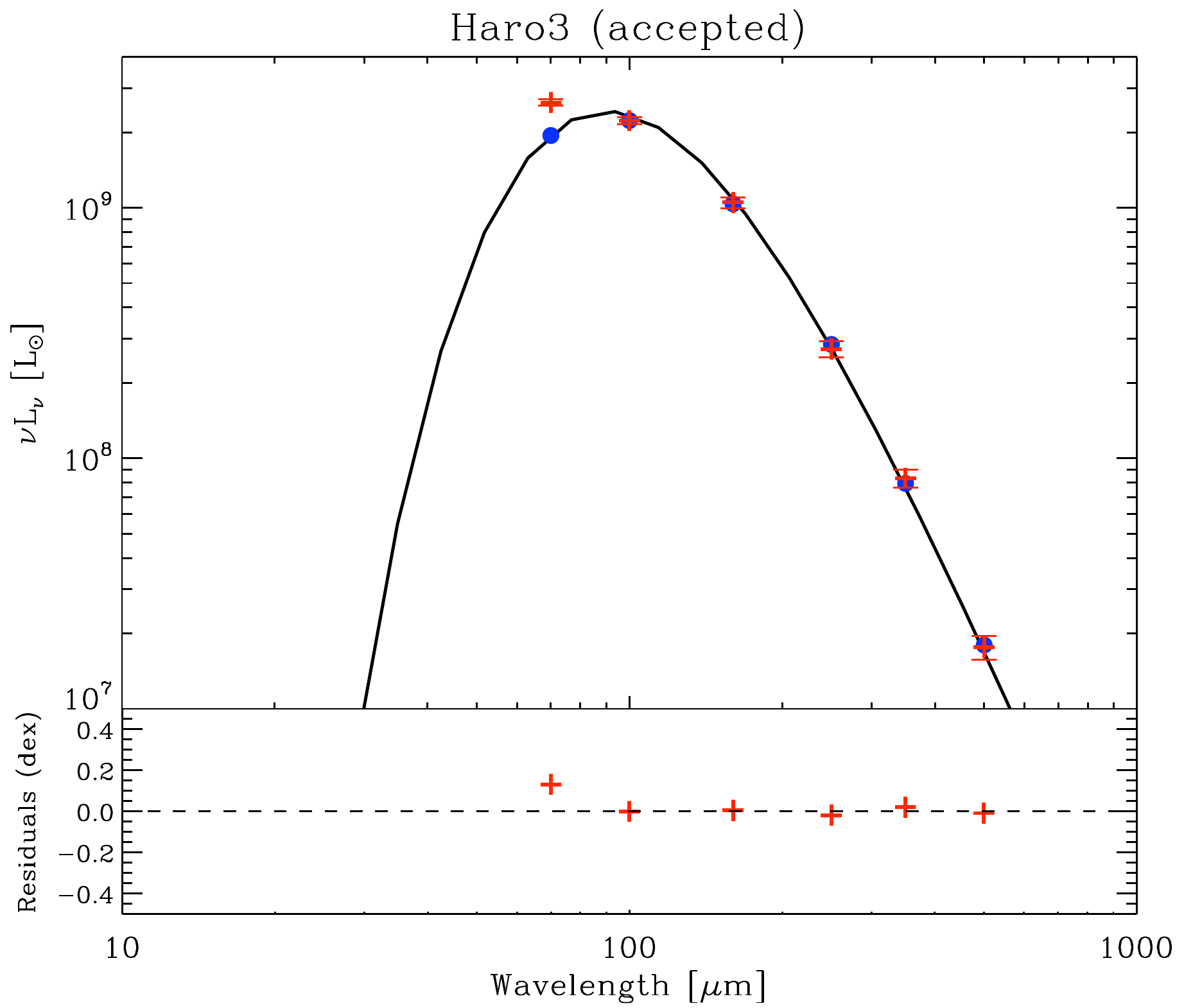}
\includegraphics[width=8.8cm]{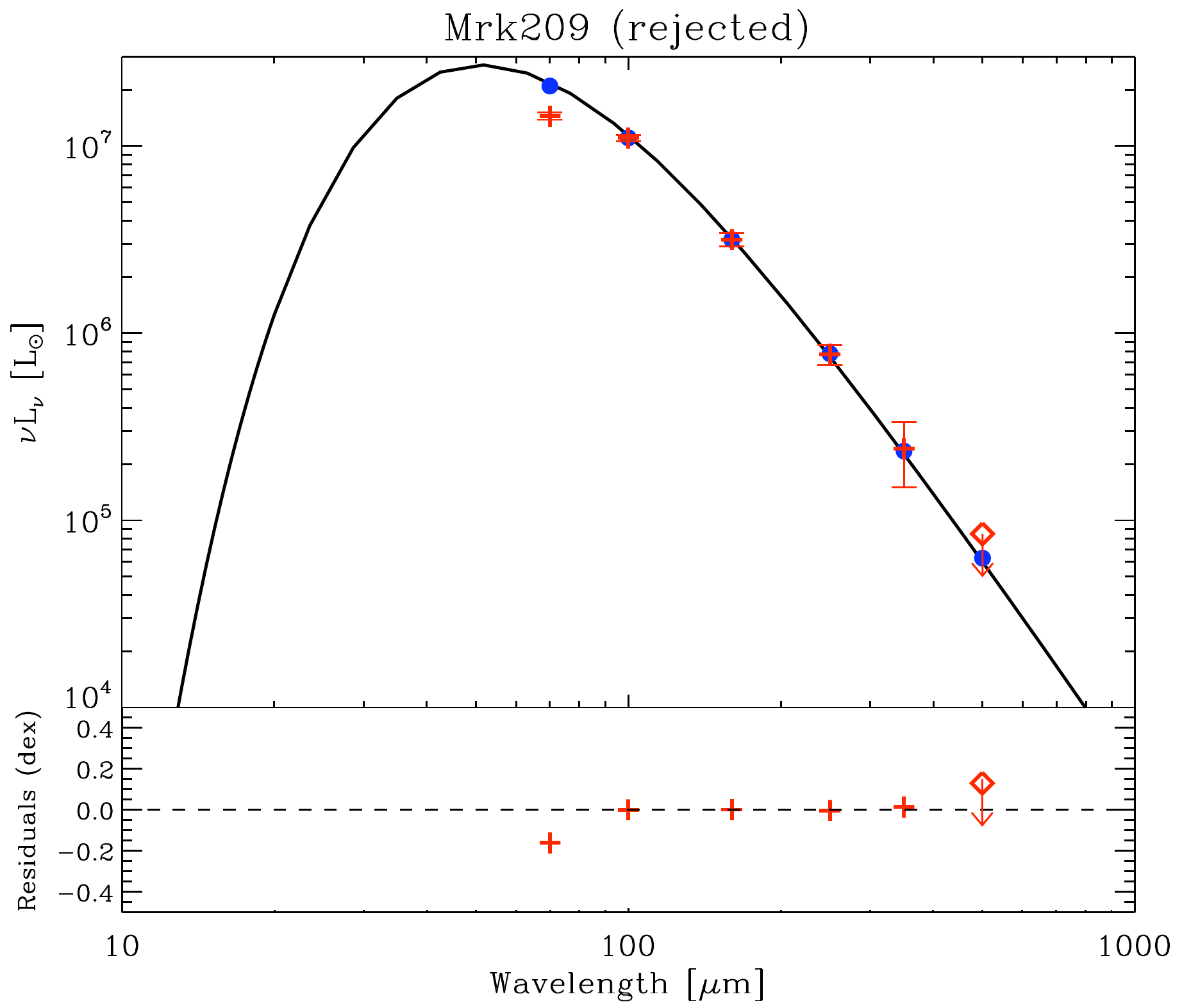}
\includegraphics[width=8.8cm]{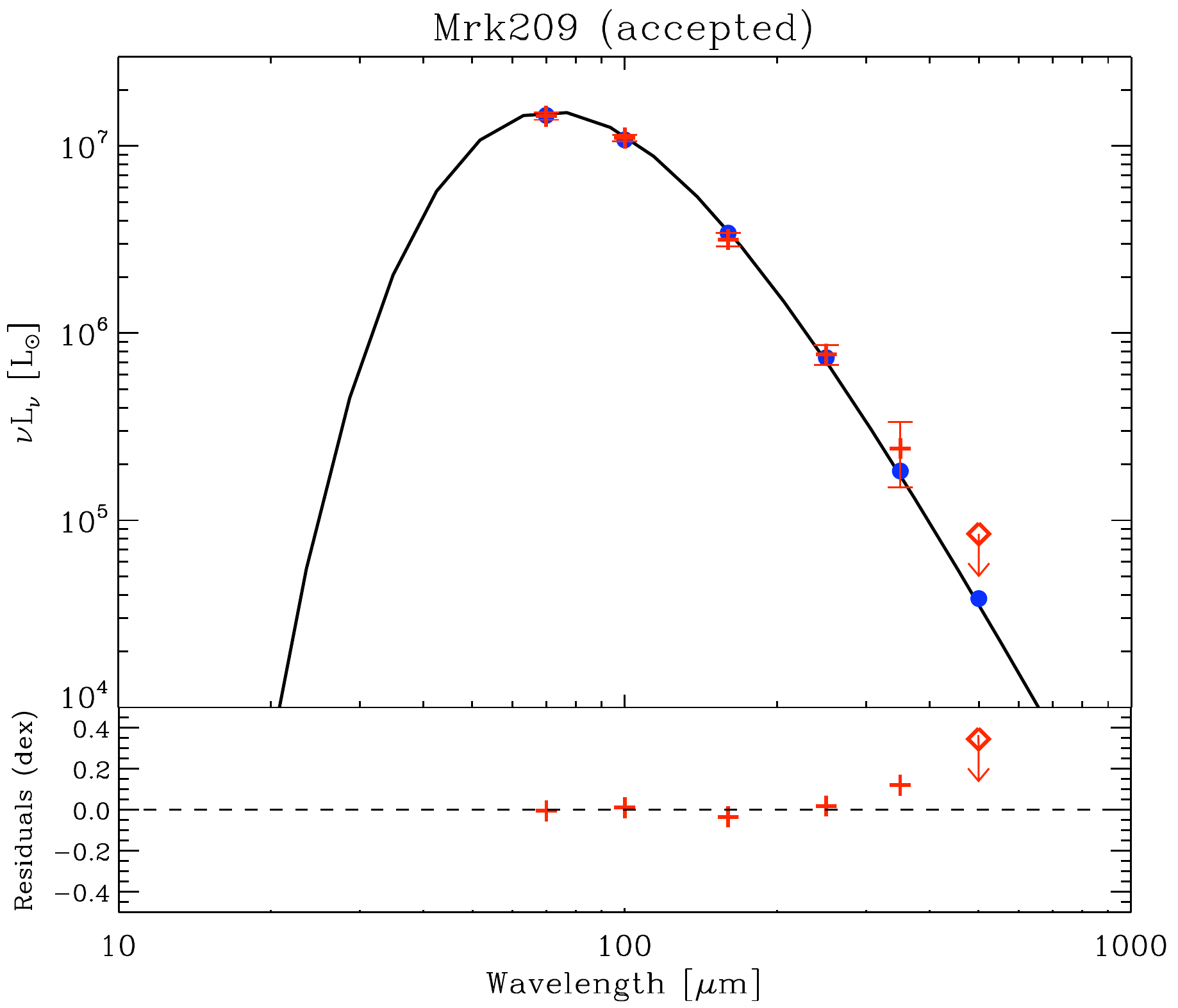}
\caption{Examples of modified blackbody fits: the observed points are the red crosses whereas the modelled points are the filled blue circles. Upper limits are indicated with red diamonds. The bottom panel of each plot indicates the residuals from the fit. ({\it top}) Fit for Haro3, the observed 70 \mic\ point which is not considered at first in our fitting procedure, is above the modelled one. ({\it centre}) Fit for Mrk209. Here the observed 70 \mic\ point is below the modelled one, and the fit should be redone, giving us: ({\it bottom}) Fit for Mrk209 using the 70 \mic\ point. Note how the shape of the modified blackbody varies between the two: for example, the dust temperature for Mrk209 goes from 56~K (without 70 \mic) to 33K (with 70 \mic).}
\label{bb70}
\end{center}
\end{figure}

At 70 \mic, possible contamination by dust grains that are not in thermal equilibrium, and whose emission cannot be represented by a modified blackbody, can occur in galaxies. An excess at 70 \mic\ compared to a modified blackbody model can also appear, as seen in Fig. \ref{diagrams}, because dust grains in a galaxy are more likely to have a temperature distribution rather than a single temperature. For example in spiral galaxies (present in the KINGFISH sample), the dust emitted at 70-500 \mic\ can originate from two components with different heating sources and potentially different temperatures \citep{Bendo2010, Bendo2012, Boquien2011, Smith2012a}. Therefore, we restrict our wavelength fitting range to 100 - 500 \mic. The 70 \mic\ point can be useful as an upper limit for a single temperature dust component. We redo the fit including the 70 \mic\ point only if the modelled point from the fit without 70 \mic\ data violates this upper limit condition, i.e. if it is greater than the observed point (e.g. Mrk 209 in Figure \ref{bb70}).

 %Figures: BB fits

Some of our galaxies are not detected at some wavelengths. To have enough constraints for the fit, at least a detection up to 250 \mic\ is required. If the galaxy is not detected beyond 160 \mic, we fit a modified blackbody including the 70 \mic\ point. Indeed some galaxies peaking at very short wavelengths have their Rayleigh Jeans contribution dropping at FIR and submm wavelengths, and are often not detected by SPIRE. For these galaxies, the 70 \mic\ point is already on the Rayleigh Jeans side of the modified blackbody, and in this case we also include it in our fit.

All of these conditions are matched for 35 galaxies, and we use the 70 \mic\ point for 11 of them (five because of the violation of the upper limit condition at 70 \mic, four because the galaxy is not detected beyond 160 \mic, and two because the galaxy is not observed by SPIRE, see Table \ref{bbfitsparam} for details). 
 
 From the fitted modified blackbodies, we also derive the total FIR luminosity, {\it L$_{FIR}$}, by integrating the modelled curve between 50 and 650 \mic. 
 The resulting parameters from the fits are given in Table \ref{bbfitsparam}. The SEDs are shown in Appendix A for all 35 DGS galaxies.

%----------------------- BB fits parameters table----------------------------------------------- 
\begin{table*}[h!tbp]
\caption{Table of modified blackbody fit parameters for the DGS galaxies.}
\begin{center}
\label{bbfitsparam}
\small{
\begin{tabular}{lcccc}
\hline
\hline
Source & Temperature (K) & $\beta_{obs}$ & M$_{dust-BB}$ (\msun) &  L$_{FIR-BB}$ (\lsun) \\
\hline
&&&& \\
Haro11       & $       38  _{-        6  }^{+       11  } $ & $      1.96  _{-      0.41  }^{+      0.44  } $ & $     5.0  _{-     3.3 }^{+    6.5} \times10^{ 6}$ & $     5.3  _{-     0.5 }^{+    1.1} \times10^{10  }$ \\
Haro2        & $       25  _{-        1  }^{+        3  } $ & $      2.38  _{-      0.38  }^{+      0.09  } $ & $     2.1  _{-     0.9 }^{+    0.5} \times10^{ 6}$ & $     2.5  _{-     0.1 }^{+    0.1} \times10^{ 9  }$ \\
Haro3        & $       26  _{-        2  }^{+        3  } $ & $      2.15  _{-      0.34  }^{+      0.31  } $ & $     1.7  _{-     0.8 }^{+    1.1} \times10^{ 6}$ & $     2.4  _{-     0.1 }^{+    0.2} \times10^{ 9  }$ \\
He2-10       & $       26  _{-        1  }^{+        4  } $ & $      2.24  _{-      0.40  }^{+      0.21  } $ & $     1.3  _{-     0.7 }^{+    0.5} \times10^{ 6}$ & $     2.0  _{-     0.1 }^{+    0.1} \times10^{ 9  }$ \\
HS0017+1055$^{2a}$  & $       98  _{-       36  }^{+       34  } $ & $      0.00  _{-      0.00  }^{+      1.34  } $ & $     1.9  _{-     1.4 }^{+    8.2} \times10^{ 3}$ & $     3.4  _{-     0.4 }^{+    0.3} \times10^{ 8  }$ \\
HS0052+2536  & $       37  _{-        9  }^{+       14  } $ & $      1.20  _{-      0.73  }^{+      0.79  } $ & $     1.1  _{-     0.8 }^{+    3.1} \times10^{ 6}$ & $     8.5  _{-     1.0 }^{+    1.6} \times10^{ 9  }$ \\
HS0822+3542  &  - & - & - & - \\
HS1222+3741  &  - & - & - & - \\
HS1236+3937  &  - & - & - & - \\
HS1304+3529$^1$  & $       32  _{-        3  }^{+        7  } $ & $      2.05  _{-      0.72  }^{+      0.47  } $ & $     2.2  _{-     1.3 }^{+    1.5} \times10^{ 5}$ & $     9.4  _{-     0.4 }^{+    0.5} \times10^{ 8  }$ \\
HS1319+3224  &  - & - & - & - \\
HS1330+3651$^{2b}$  & $       50  _{-        8  }^{+        3  } $ & $      0.00  _{-      0.00  }^{+      0.61  } $ & $     3.3  _{-     0.6 }^{+    3.4} \times10^{ 4}$ & $     8.8  _{-     0.4 }^{+    0.4} \times10^{ 8  }$ \\
HS1442+4250  &  - & - & - & - \\
HS2352+2733  &  - & - & - & - \\
IZw18        &  - & - & - & - \\
IC10         & $       21  _{-        1 }^{+        3  } $ & $      2.25  _{-      0.49  }^{+      0.26  } $ & $     2.6  _{-     1.4 }^{+    1.1} \times10^{ 5}$ & $     1.1  _{-     0.0 }^{+    0.0} \times10^{ 8  }$ \\
IIZw40       & $       33  _{-        4  }^{+        5  } $ & $      1.71  _{-      0.32  }^{+      0.41  } $ & $     1.9  _{-     0.9 }^{+    1.8} \times10^{ 5}$ & $     9.2  _{-     0.7 }^{+    0.9} \times10^{ 8  }$ \\
Mrk1089      & $       23  _{-        1  }^{+        3  } $ & $      2.34  _{-      0.39  }^{+      0.15  } $ & $     2.5  _{-     1.2 }^{+    0.7} \times10^{ 7}$ & $     1.6  _{-     0.1 }^{+    0.1} \times10^{10  }$ \\
Mrk1450      & $       43  _{-       13 }^{+       40  } $ & $      1.35  _{-      0.92  }^{+      1.00  } $ & $     7.6  _{-     6.1 }^{+   24.6} \times10^{ 3}$ & $     1.2  _{-     0.2 }^{+    0.4} \times10^{ 8  }$ \\
Mrk153$^1$       & $       32  _{-        2  }^{+        5  } $ & $      2.33  _{-      0.61  }^{+      0.15  } $ & $     1.2  _{-     0.6 }^{+    0.4} \times10^{ 5}$ & $     5.3  _{-     0.2 }^{+    0.3} \times10^{ 8  }$ \\
Mrk209$^1$       & $       34  _{-        3  }^{+        6  } $ & $      1.95  _{-      0.47  }^{+      0.42  } $ & $     2.1  _{-     1.1 }^{+    1.4} \times10^{ 3}$ & $     1.3  _{-     0.1 }^{+    0.1} \times10^{ 7  }$ \\
Mrk930       & $       26  _{-        2  }^{+        4  } $ & $      2.22  _{-      0.43  }^{+      0.21  } $ & $     5.7  _{-     3.0 }^{+    3.0} \times10^{ 6}$ & $     8.6  _{-     0.5 }^{+    0.6} \times10^{ 9  }$ \\
NGC1140      & $       23  _{-        1  }^{+        2  } $ & $      2.17  _{-      0.36  }^{+      0.31  } $ & $     3.0  _{-     1.5 }^{+    1.7} \times10^{ 6}$ & $     1.9  _{-     0.1 }^{+    0.1} \times10^{ 9  }$ \\
NGC1569      & $       28  _{-        2  }^{+        4  } $ & $      2.20  _{-      0.38  }^{+      0.28  } $ & $     2.8  _{-     1.4 }^{+    1.5} \times10^{ 5}$ & $     5.7  _{-     0.4 }^{+    0.5} \times10^{ 8  }$ \\
NGC1705      & $       33  _{-        4  }^{+        5  } $ & $      1.16  _{-      0.28  }^{+      0.33  } $ & $     8.4  _{-     3.7 }^{+    6.7} \times10^{ 3}$ & $     4.1  _{-     0.3 }^{+    0.3} \times10^{ 7  }$ \\
NGC2366      & $       39  _{-        4  }^{+        4  } $ & $      0.96  _{-      0.23  }^{+      0.22  } $ & $     6.8  _{-     2.1 }^{+    2.9} \times10^{ 3}$ & $     7.3  _{-     0.2 }^{+    0.3} \times10^{ 7  }$ \\
NGC4214      & $       26  _{-        3  }^{+        3  } $ & $      1.39  _{-      0.37  }^{+      0.37  } $ & $     2.0  _{-     1.0 }^{+    2.0} \times10^{ 5}$ & $     2.9  _{-     0.1 }^{+    0.2} \times10^{ 8  }$ \\
NGC4449      & $       22  _{-        1  }^{+        3  } $ & $      2.18  _{-      0.41  }^{+      0.27  } $ & $     2.3  _{-     1.2 }^{+    1.2} \times10^{ 6}$ & $     1.4  _{-     0.1 }^{+    0.1} \times10^{ 9  }$ \\
NGC4861      & $       28  _{-        3  }^{+        3  } $ & $      1.36  _{-      0.32  }^{+      0.34  } $ & $     6.1  _{-     2.7 }^{+    5.2} \times10^{ 4}$ & $     1.3  _{-     0.1 }^{+    0.1} \times10^{ 8  }$ \\
NGC5253      & $       30  _{-        3  }^{+        5  } $ & $      1.86  _{-      0.37  }^{+      0.40  } $ & $     1.9  _{-     1.0 }^{+    1.8} \times10^{ 5}$ & $     5.5  _{-     0.4 }^{+    0.5} \times10^{ 8  }$ \\
NGC625       & $       29  _{-        3 }^{+        4  } $ & $      1.33  _{-      0.30  }^{+      0.35  } $ & $     5.9  _{-     2.7 }^{+    5.9} \times10^{ 4}$ & $     1.5  _{-     0.1 }^{+    0.1} \times10^{ 8  }$ \\
NGC6822      & $       26  _{-        4  }^{+        4  } $ & $      0.99  _{-      0.50  }^{+      0.59  } $ & $     1.1  _{-     0.6 }^{+    1.6} \times10^{ 4}$ & $     1.8  _{-     0.1 }^{+    0.1} \times10^{ 7  }$ \\
Pox186       & $       40  _{-        4  }^{+        4  } $ & $      0.00  _{-      0.00  }^{+      0.00  } $ & $     1.6  _{-     0.5 }^{+    0.7} \times10^{ 3}$ & $     2.0  _{-     0.2 }^{+    0.2} \times10^{ 7  }$ \\
SBS0335-052$^{2a,3}$  & $       89  _{-        8  }^{+       10  } $ & $      1.64  _{-      0.33  }^{+      0.39  } $ & $     8.0  _{-     1.4 }^{+    1.8} \times10^{ 2}$ & $     1.2  _{-     1.2 }^{+    1.4} \times10^{ 7  }$ \\
SBS1159+545  &  - & - & - & - \\
SBS1211+540$^{2a,3}$  & $       71  _{-        7  }^{+        7 } $ & $      0.34  _{-      0.41  }^{+      0.46  } $ & $     1.0  _{-     0.3 }^{+    0.4} \times10^{ 2}$ & $     1.2  _{-     0.1 }^{+    0.1} \times10^{ 7  }$ \\
SBS1249+493  &  - & - & - & - \\
SBS1415+437$^{2b}$  & $       35  _{-        3  }^{+       16  } $ & $      2.37  _{-      1.20  }^{+      0.15  } $ & $     4.9  _{-     3.1 }^{+    2.1} \times10^{ 3}$ & $     3.5  _{-     0.3 }^{+    0.3} \times10^{ 7  }$ \\
SBS1533+574$^{2a}$  & $       42  _{-       10  }^{+        8  } $ & $      0.44  _{-      0.40  }^{+      1.00  } $ & $     5.6  _{-     2.5 }^{+   11.4} \times10^{ 4}$ & $     8.2  _{-     0.4 }^{+    0.4} \times10^{ 8  }$ \\
Tol0618-402  &  - & - & - & - \\
Tol1214-277  &  - & - & - & - \\
UGC4483      &  - & - & - & - \\
UGCA20        & - & - & - & - \\
UM133        & $       41  _{-       15  }^{+       16  } $ & $      0.44  _{-      0.57  }^{+      1.89  } $ & $     3.1  _{-     2.8 }^{+   56.4} \times10^{ 3}$ & $     4.0  _{-     0.5 }^{+    0.9} \times10^{ 7  }$ \\
UM311        & $       24  _{-        2  }^{+        3  } $ & $      1.58  _{-      0.39  }^{+      0.34  } $ & $     3.7  _{-     1.9 }^{+    3.5} \times10^{ 6}$ & $     3.4  _{-     0.1 }^{+    0.2} \times10^{ 9  }$ \\
UM448$^1$        & $       33  _{-        2  }^{+        2  } $ & $      1.99  _{-      0.16  }^{+      0.18  } $ & $     9.9  _{-     2.5 }^{+    3.5} \times10^{ 6}$ & $     4.9  _{-     0.2 }^{+    0.2} \times10^{10  }$ \\
UM461        & $       24  _{-        1  }^{+        1  } $ & $      2.50  _{-      0.00  }^{+      0.00  } $ & $     2.7  _{-     0.5 }^{+    0.5} \times10^{ 4}$ & $     2.5  _{-     0.2 }^{+    0.2} \times10^{ 7  }$ \\
VIIZw403$^1$     & $       34  _{-        3  }^{+        3  } $ & $      1.57  _{-      0.23  }^{+      0.27  } $ & $     2.1  _{-     0.7 }^{+    1.1} \times10^{ 3}$ & $     1.2  _{-     0.0 }^{+    0.1} \times10^{ 7  }$ \\
\hline
\end{tabular}
}
\end{center}
\scriptsize{

$^1$: 70 \mic\ point included in fit: violation of the upper limit condition at 70 \mic.

$^{2a}$: 70 \mic\ point included in fit: no detections beyond 160 \mic.

$^{2b}$: 70 \mic\ point included in fit: no observations beyond 160 \mic.

$^{3}$: For these particular galaxies, we included the 24 \mic\ point in the fit as the 24 \mic\ point fell below the modelled modified blackbody when we just overlaid it on the plot.}

\end{table*}

\subsubsection{Rigorous error estimation}\label{bbfitserrors}
In order to derive conservative errors for our {\it T}, $\beta$, {\it M$_{dust}$} and {\it L$_{FIR}$} parameters we performed Monte Carlo iterations for each fit, following the method in \cite{Galliano2011}. For each galaxy, we randomly perturb our fluxes within the errors bars and perform fits of the perturbed SEDs (300 for each galaxy). To be able to do this we must first carefully identify the various types of error and take special care for errors which are correlated between different bands.

As explained in Sections \ref{pacsphotoerror} and \ref{spirephotoerrors}, we have measurement errors and calibration errors in our error estimates. 
The measurement errors are independent from one band to another and are usually well represented by a Gaussian distribution.
The calibration errors, however, are correlated between different bands as it is the error on the flux conversion factor. It can be summarized for our case as follow :
\\ 

\noindent PACS: Although the total calibration error is 5\% in the three PACS bands it can be decomposed into two components :
 \begin{itemize}
 \item{the uncertainty on the calibration model is 5\% (according to the PACS photometre point-source flux calibration documentation\footnote{http://herschel.esac.esa.int/twiki/bin/view/Public/PacsCalibration\\ Web?template=viewprint}) and is correlated between the three bands.}
 \item{the uncertainties due to noise in the calibration observations are: 1.4, 1.6, 3.5 \% at 70, 100, 160 \mic, respectively (PACS photometre point-source flux calibration). These uncertainties are independent from one band to another.}
\end{itemize}

\noindent SPIRE: The SPIRE ICC recommend using 7\% in each band but here again we can decompose it :
 \begin{itemize}
  \item{the uncertainty on the calibration model is 5\% (SPIRE Observer's Manual) and is correlated between the three bands.}
 \item{the uncertainties due to noise in the calibration observations are 2\% for each band (SPIRE Observer's Manual). These uncertainties are independent.}
 \item{As SPIRE maps are given in Jy$\cdot$beam$^{-1}$, the error on the beam area will also affect the calibration. The uncertainty on the beam area {\revised is given to be 4\%} in each band\footnote{This value is given in: http://herschel.esac.esa.int/twiki/bin/view/ \\ÊPublic/SpirePhotometerBeamProfileAnalysis.} and is independent.}
\end{itemize}

The perturbation of the observed fluxes will then be the sum of two components :
\begin{itemize}
  \item{A normal random independent variable representing the measurement errors.}
 \item{A normal random variable describing the calibration errors that takes into account the correlation between the wavebands as described above, the same for each galaxy.}
\end{itemize}

\begin{figure}[h!tbp]
\begin{center}
\includegraphics[width=8.8cm]{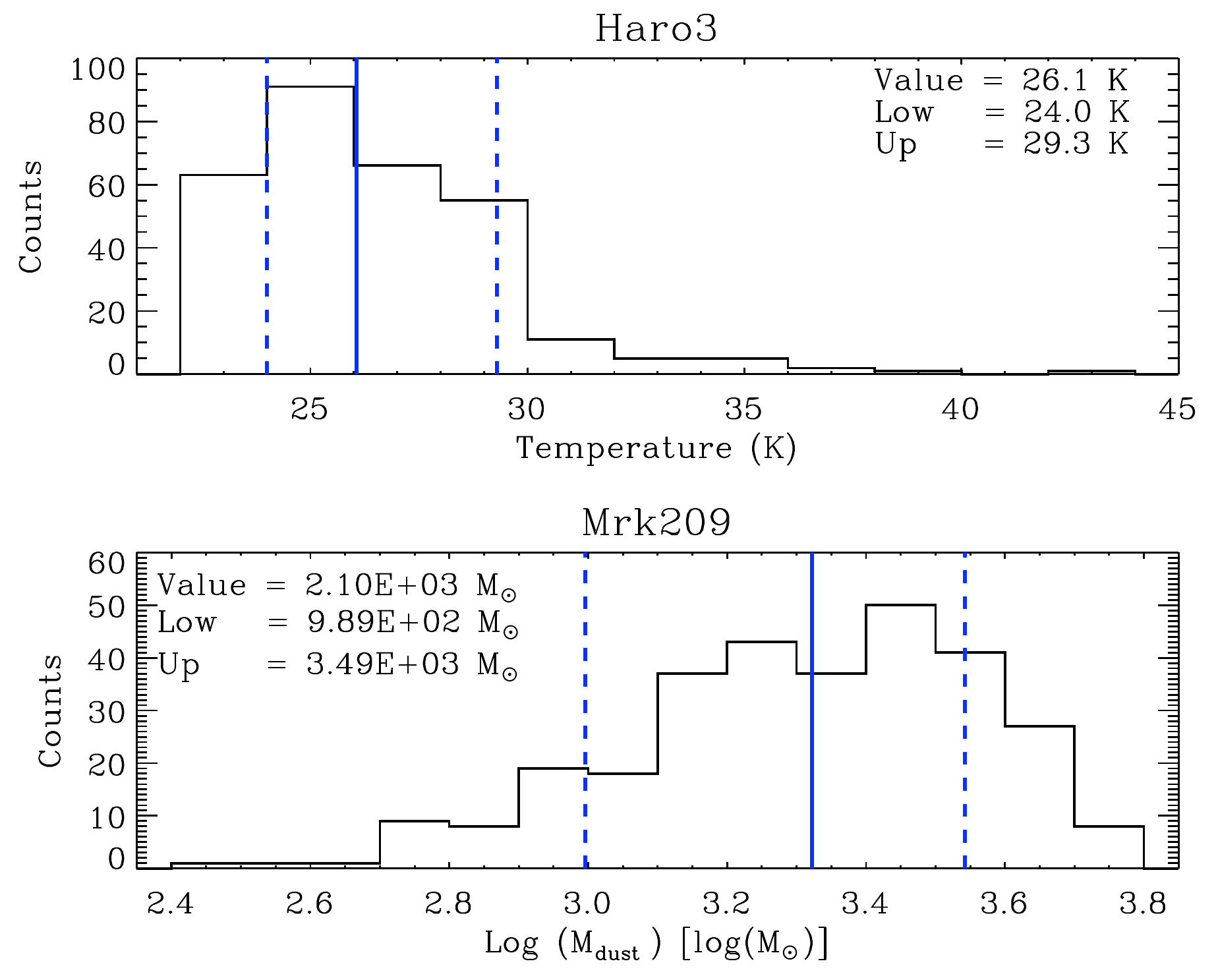}
\caption{Examples of the obtained parameter distributions for the 300 Monte-Carlo iterations for the modified blackbody fits: ({\it top}) distribution of temperature,  {\it T}, for Haro3, ({\it bottom}) distribution of dust mass, {\it M$_{dust}$}, for Mrk209 with the 70 \mic\ point included in the fit. The plain blue line notes the value of the parameter and the dashed blue lines note the 66.67\% confidence level for the parameters.}
\label{bberr}
\end{center}
\end{figure}

After performing 300 Monte-Carlo iterations, a distribution for each of the three model parameters {\it T}, $\beta$, {\it M$_{dust}$} as well as for {\it L$_{FIR}$} is obtained for each galaxy (see example on Fig. \ref{bberr}). We chose to quote the 66.67\% confidence level for our parameters defined by the range of the parameter values between 0.1667 and 0.8333 of the repartition function. As the distributions are often asymmetric we obtain asymmetric error bars on our parameters. These error bars are given in Table \ref{bbfitsparam}. 

 %Figures: BB errors distribution

\subsubsection{FIR properties}\label{bbfitsprop} 

\begin{figure*}[h!tbp]
\begin{center}
\begin{tabular}{ p{8cm}p{8cm}}
    {\LARGE a)} & {\LARGE b)} \\
    \includegraphics[width=8.5cm]{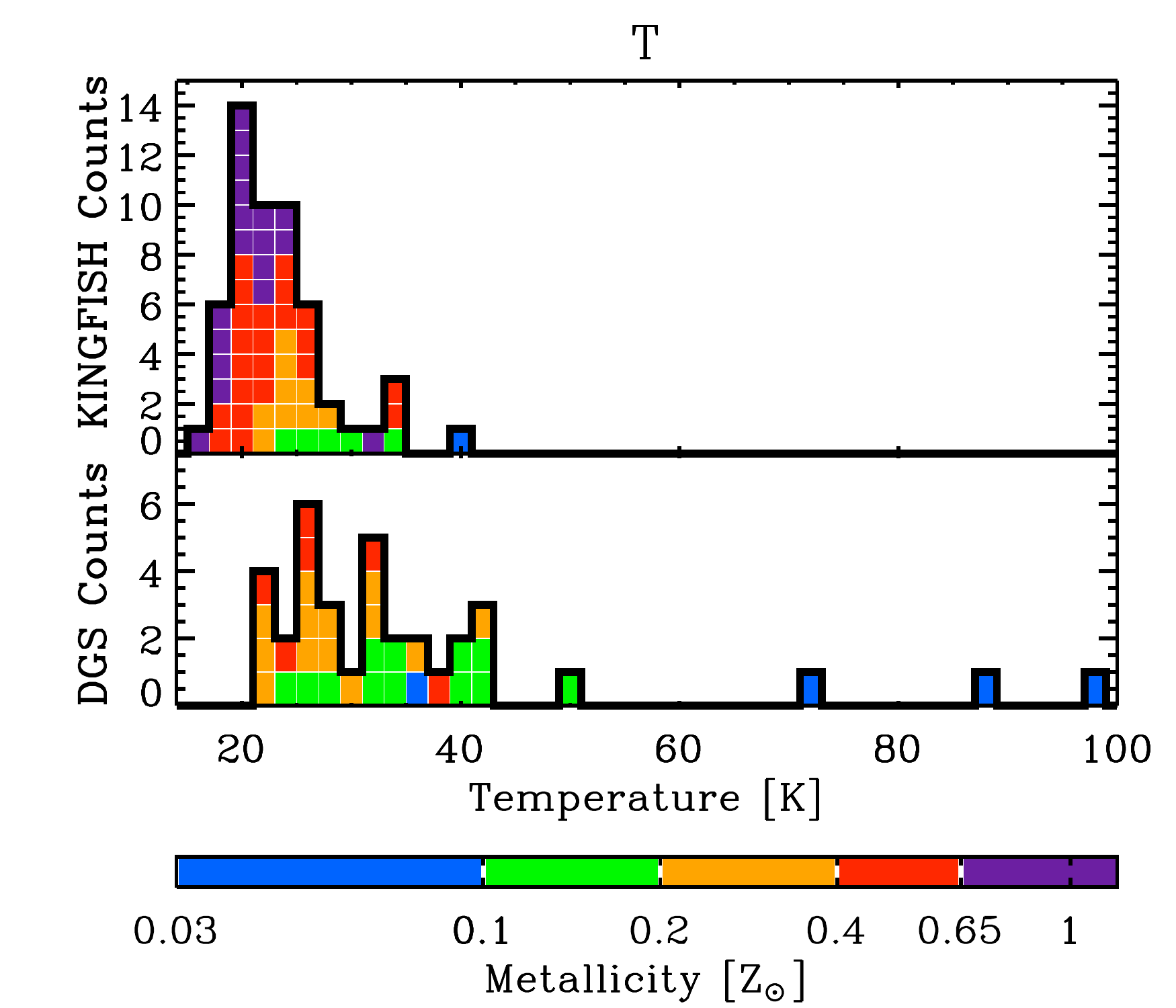} &
   \includegraphics[width=8.5cm]{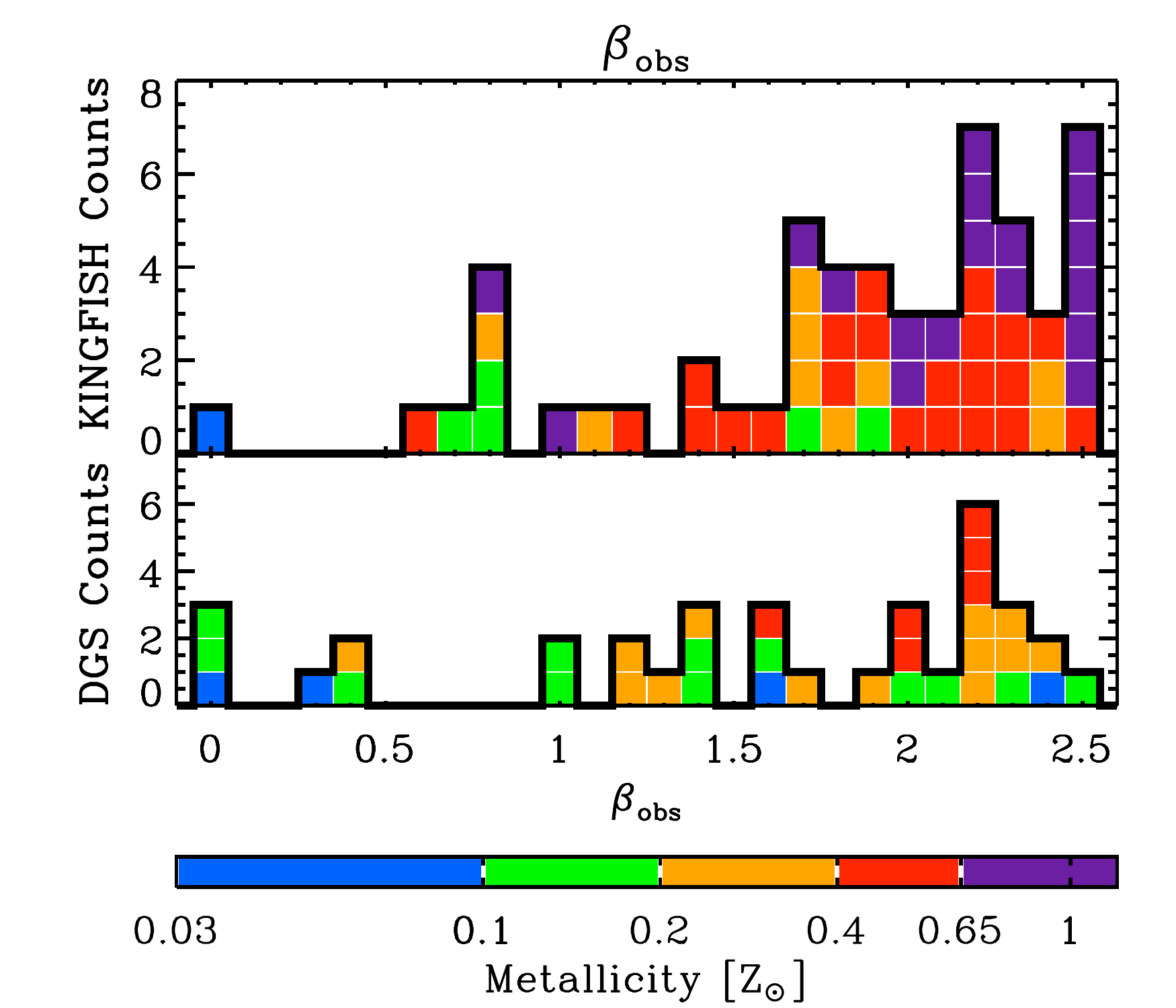} \\
    {\LARGE c)} & {\LARGE d)} \\ 
    \includegraphics[width=8.5cm]{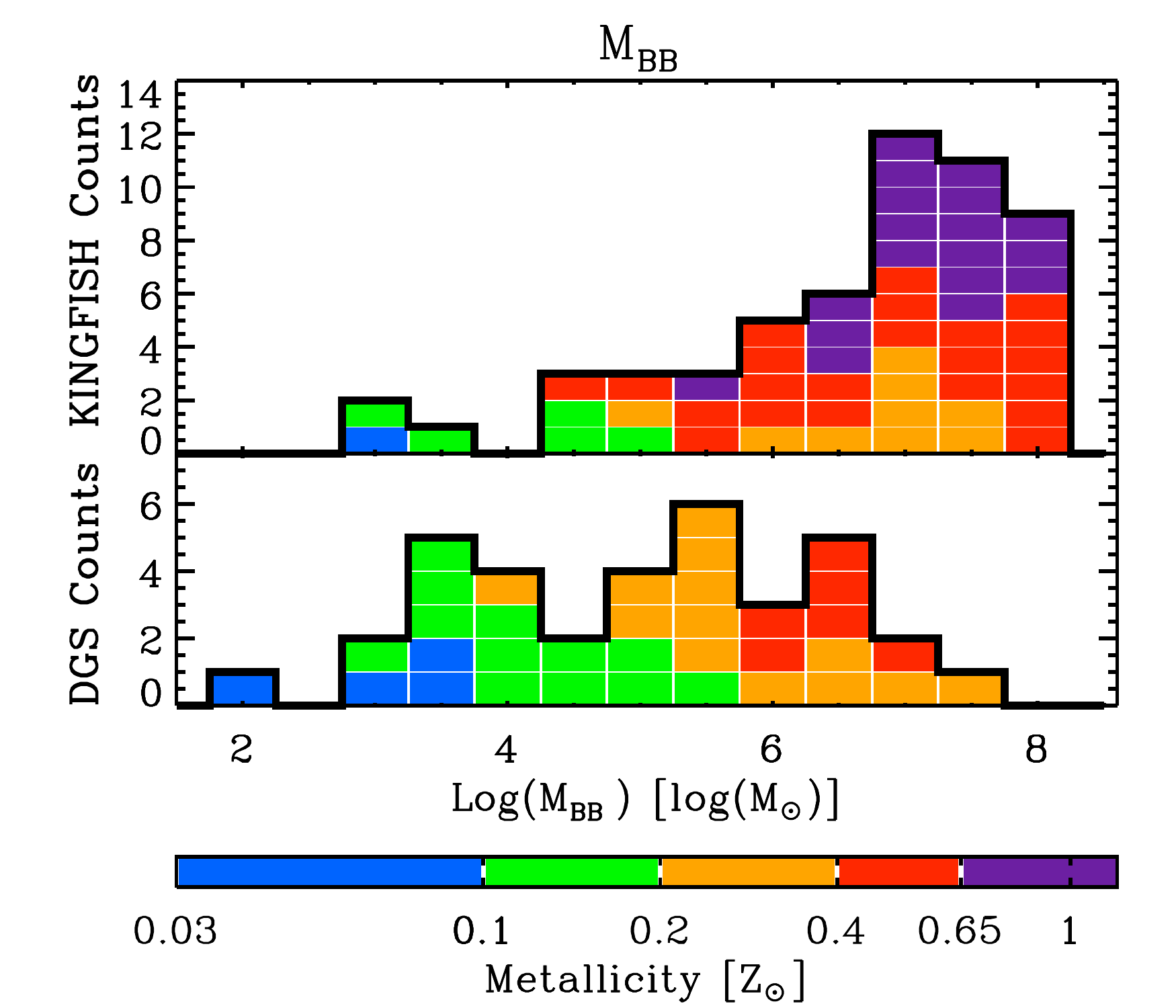} &
    \includegraphics[width=8.5cm]{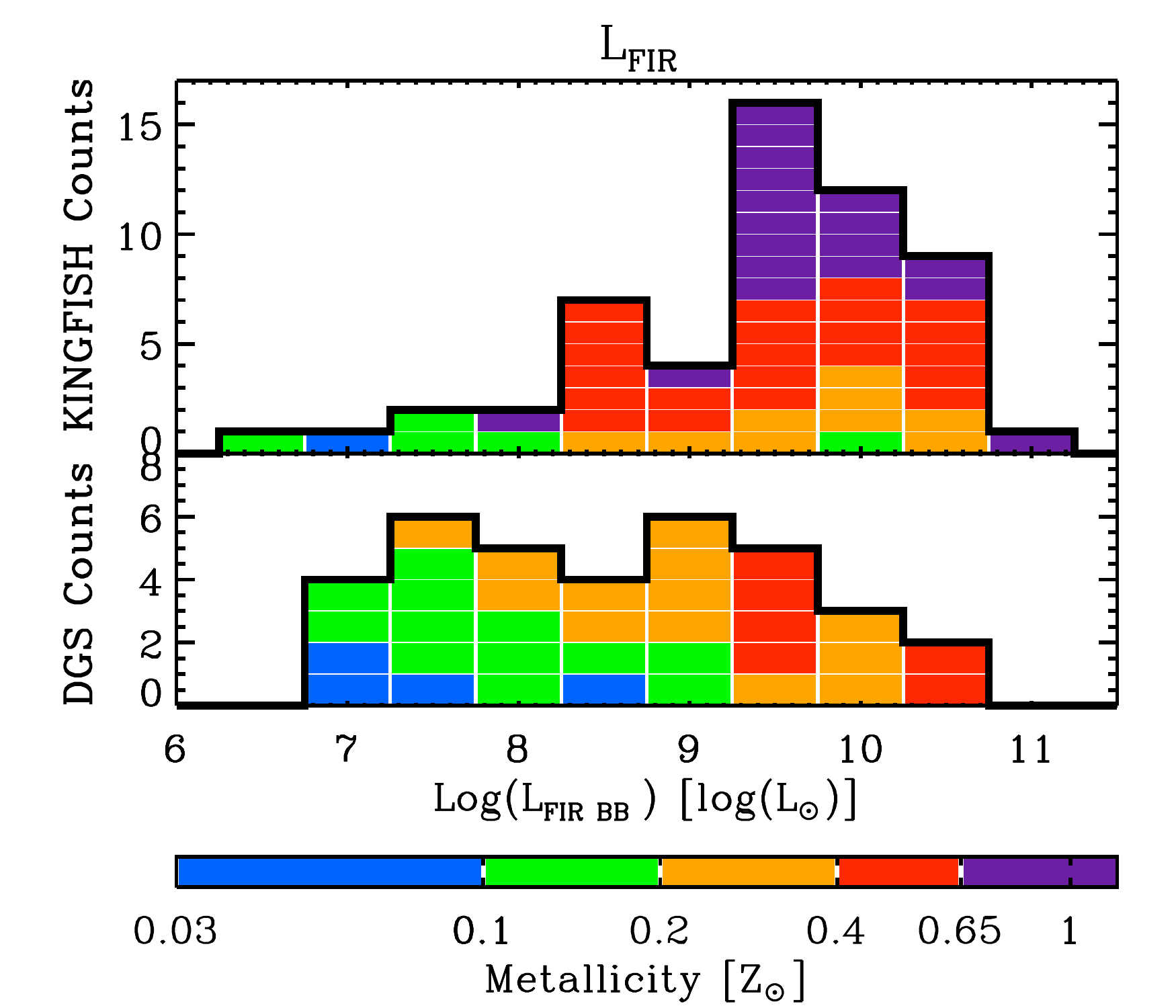} \\ 
 \end{tabular} 
\caption{Distributions of temperature {\it(a)}, emissivity index {\it(b)}, dust mass {\it (c)} and FIR luminosity {\it (d)} from modified blackbody fits for \hersc\ data for the DGS and KINGFISH samples. The colour scale represents the range of metallicity values. On each panel, the upper/lower histogram is the KINGFISH/DGS distribution for the parameter.}
\label{histos}
\end{center}
\end{figure*}

\begin{figure*}[h!tbp]
\begin{center}
\includegraphics[width=15cm]{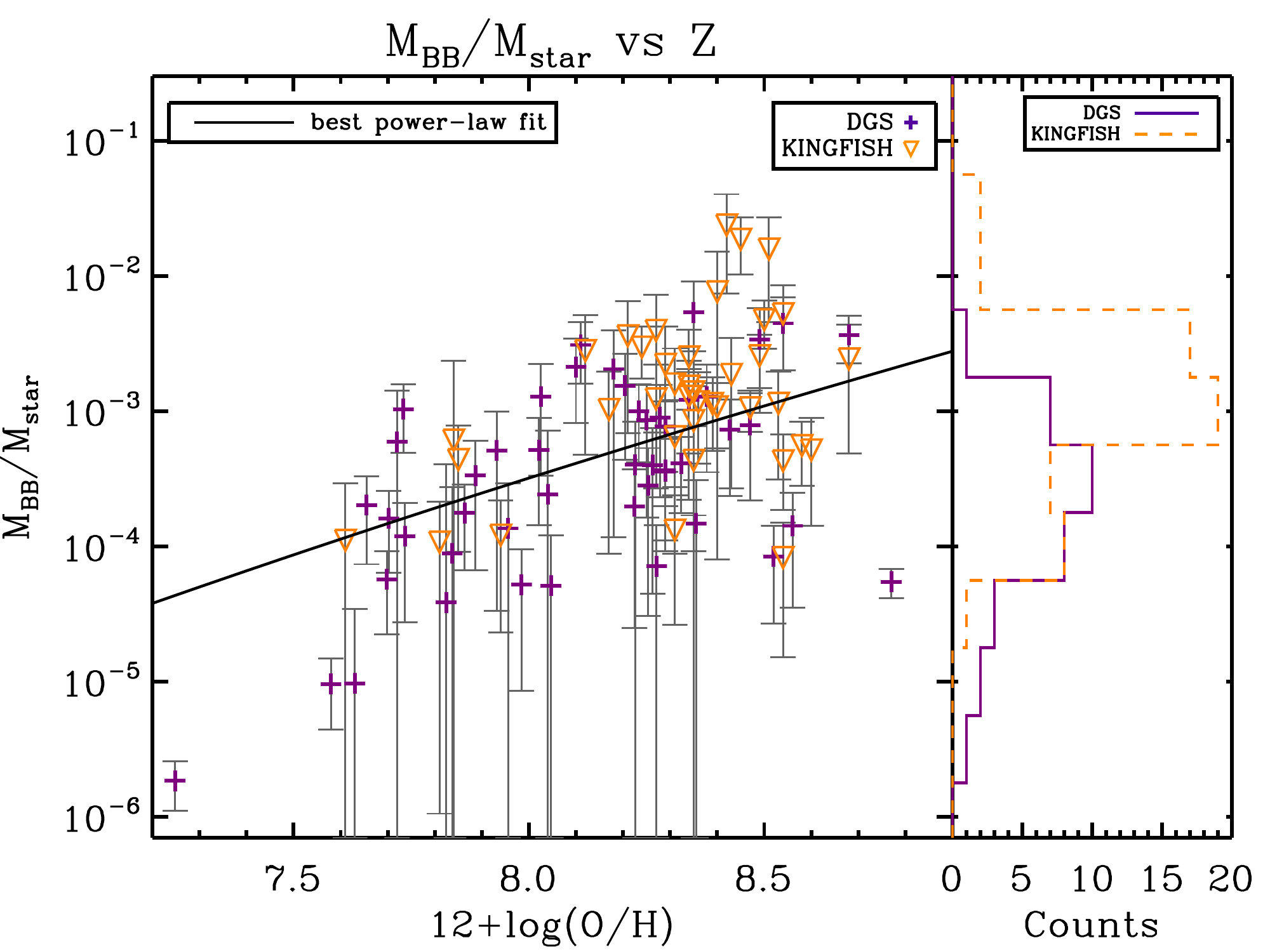}
\includegraphics[width=15cm]{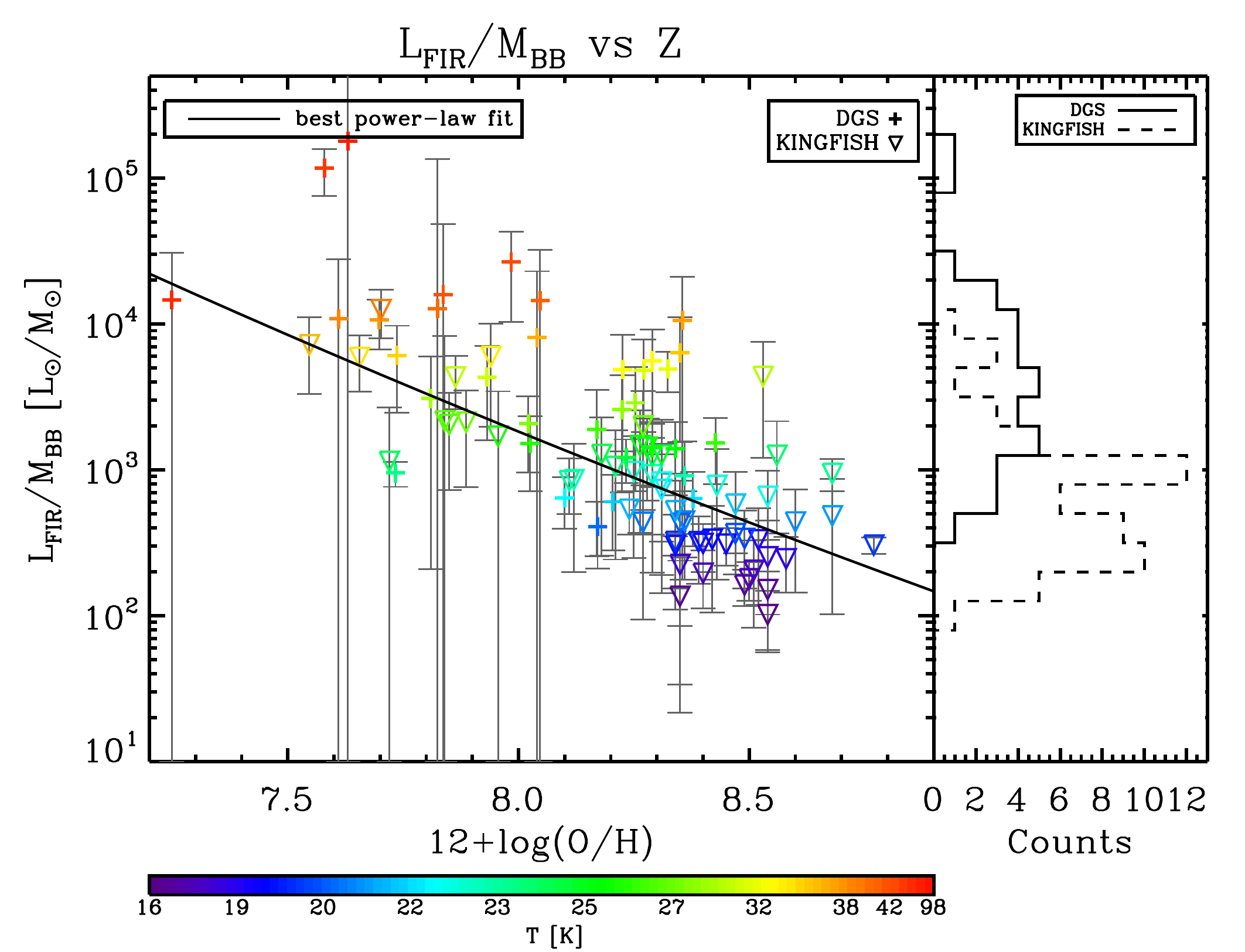}
\caption{{\it(top)} {\it M$_{BB}$/M$_{star}$} as a function of metallicity for DGS (purple crosses) and KINGFISH (orange downward triangles). The best power-law fit is indicated as a black line, and corresponds to: log({\it M$_{BB}$/M$_{star}$}) = (-21.8 $\pm$ 1.5) + (20.3 $\pm$ 1.6) $\times$ log(12+log(O/H)).
The distribution of {\it M$_{BB}$/M$_{star}$} is indicated on the side for both samples: plain purple line for DGS and dashed orange line for KINGFISH. 
{\it(bottom)} {\it L$_{FIR}$/M$_{BB}$} as a function of metallicity for DGS (crosses) and KINGFISH (downward triangles). The colours code the temperature, {\it T}. The best power-law fit line is indicated as a black line, and corresponds to: log({\it L$_{FIR}$}/{\it M$_{BB}$}) = (24.4 $\pm$ 1.1) + (-23.6 $\pm$ 1.2) $\times$ log(12+log(O/H)). The distribution of {\it L$_{FIR}$/M$_{BB}$} is indicated on the side for both samples: plain line for DGS and dashed line for KINGFISH. 
On both plots: the errors on the metallicities are omitted for clarity. They are of about 0.1 dex on average.}
\label{trend}
\end{center}
\end{figure*}

We now have the {\it T}, $\beta$,  {\it M$_{dust}$} and {\it L$_{FIR}$} distributions of the DGS. We perform the same analysis for the KINGFISH sample in order to compare the distribution of parameters of the dwarf galaxies with those of the KINGFISH sample (Figures \ref{histos} and \ref{trend}). Note that KINGFISH is not a volume- or flux-limited sample but a cross-section of galaxies with different properties. Due to the heterogeneity of both samples we thus quote the median rather than the mean to compare the samples. \\

%Figures: Temp, beta, MBB and LFIR

 %Figures: MBB/Mstar, LFIR/MBB

%Temp histo
\noindent {\it Temperature} \\

The range in dust temperature of the DGS galaxies is 21 to 98~K with a median {\it T} $\sim$ 32~K (Figure \ref{histos}a). The most metal-poor galaxies are among the warmer ones. If we compare the KINGFISH to the DGS galaxies, our lowest temperatures are quite comparable (17 vs 21 K), but the DGS galaxies have higher maximal dust temperatures (39 vs 98 K). In Figure \ref{histos}a, we see that the KINGFISH dust temperature distribution has a narrow peak around $\sim$20-25 K whereas the DGS distribution is broader. This difference is due to some galaxies in our sample that peak at extremely short wavelengths, a distinguishing feature of star-forming dwarf galaxies, resulting in very high dust temperatures for a single modified blackbody fit. The dust in DGS galaxies is thus overally warmer than that in more metal-rich galaxies ({\it T}$_{DGS}^{med}$ = 32 K and {\it T}$_{KINGFISH}^{med}$ = 23 K). This is coherent with the temperature trends presented in the previous section. Note that the high temperature tail of the DGS temperature distribution could be even more prominent: some galaxies are not detected beyond 100-160 \mic\ rendering impossible the determination of their dust temperature with a modified blackbody fit (13 galaxies in the DGS). The SEDs for these galaxies likely peak at very short wavelengths giving very warm average dust temperatures. \\

% Beta histo
\noindent {\it Emissivity index} \\

The ``observed'' emissivity index ($\beta_{obs}$, see Eq. \ref{BB}) distribution is shown on Figure \ref{histos}b, spanning a range from 0.0 to 2.5 with a median $\beta_{obs} \sim$ 1.7. There does not appear to be any clear correlation with metallicity here. Nonetheless, even if some DGS galaxies are nicely fitted by an often-presumed $\beta_{obs}$ = 2.0 blackbody, some require a $\beta_{obs} \leq$ 2.0, and those are primarily metal-poor to moderately metal-poor galaxies (0.10 to 0.4 \zsun). Note also that for the KINGFISH sample, all of the galaxies, but two, within this metallicity range have 0.5 $ \leq \beta_{obs} \leq$ 2.0. From SPIRE band ratios, \cite{Boselli2012} also found that low-metallicity galaxies from the HRS sample were presenting submm colours consistent with an emissivity index $\leq$ 2.0. Arbitrarily fixing $\beta$ = 2.0 in blackbody fitting, in order to mimic the emissivity index appropriate for a mixture of amorphous silicate and graphite (reproducing the Milky Way observations), may not always be appropriate for low-metallicity galaxies. However, we note that several DGS galaxies suggest a $\beta_{obs}$ = 0.0-0.5. These six galaxies with $\beta_{obs}$ $<$ 0.5 in the DGS, are not detected beyond 160 \mic\ and such a low $\beta_{obs}$ is probably due to the poorly constrained submm part of the SED.

In summary, there are metal-poor to moderately metal-poor galaxies, with metallicities between 0.1 and 0.4 \zsun, for which 0.5 $ \leq \beta_{obs} \leq$ 2.0. These lower $\beta_{obs}$ values, not necessarily realistic in term of actual grain properties, are representative of a flatter submm slope in the FIR observations, and could perhaps be an indicator of the presence of a submm excess in these sources (see Section \ref{submmexcess}). \\

% Dust mass histo
\noindent {\it Dust mass} \\

The dust masses estimated from our modified blackbody fits range from $1.0\times10^2$ to $2.5\times10^7$ \msun\ (Figure \ref{histos}c), with a median of $\sim 1.2 \times 10^5$ \msun. From Figure \ref{histos}c we see that the most metal-poor galaxies are the least massive galaxies compared to the moderately metal-poor galaxies. The dwarf galaxies are, not surprisingly, less massive in dust than the galaxies from the KINGFISH sample: the median dust mass of the KINGFISH sample is about two orders of magnitude higher than for the DGS: $\sim1.1 \times 10^7$ \msun. In order to determine if this is only an effect due to the smaller sizes of dwarfs, we consider the ratio between the dust and stellar masses. The stellar masses for KINGFISH can be found in \cite{Skibba2011} and the DGS stellar masses in \cite{Madden2013}. Figure \ref{trend} shows that there is a strong decrease (two orders of magnitude) of the proportion of dust mass relative to the stellar mass with decreasing metallicity: we have a Spearman rank coefficient\footnote{The Spearman rank coefficient, $\rho$, indicates how well the relationship between X and Y can be described by a monotonic function: monotonically increasing: $\rho$ $>$ 0, or monotonically decreasing: $\rho$ $<$ 0.} $\rho$=0.58. The median for the ratio {\it M$_{BB}$/M$_{star}$} is 0.02\% for DGS versus 0.1\% for KINGFISH. The best power-law fit gives: 

\begin{equation}
M_{BB}/M_{star} = 1.6 \times 10^{-22} \times (12+log(O/H))^{20.3}
\end{equation}

The stellar masses from the DGS are derived from the formula of \cite{Eskew2012} from the IRAC 3.6 and 4.5 \mic\ broadband flux densities. The scatter in their relation corresponds to a 1$\sigma$ uncertainties for their stellar masses of $\sim$ 30\%, which is within the uncertainties we have for the DGS stellar masses ($\sim$ 50\% on average). The stellar masses for KINGFISH have been derived by \cite{Skibba2011} following \cite{Zibetti2009} from optical and NIR colours. With this estimate, the KINGFISH stellar masses could be biased low by up to 40\% \cite{Zibetti2009}. Even if this could decrease the ratios by a factor of $\sim$ 1.7, this could not explain the order of magnitude difference seen between the dust-to-stellar mass ratios of the two samples.

However the dust masses derived here for both samples are probably lower limits of the real dust masses in many cases \citep[for example, see][for KINGFISH]{Dale2012}. Indeed we allow our $\beta_{obs}$ to go to very low values, giving lower dust masses than if we fixed it to 1.5 or even 2.0: as we allow a greater emission effciency for the grains, we need less mass than if we were using a higher emissivity index, to account for the same amount of luminosity. We perform the test by fixing the emissivity index parameter to 1.5 then 2.0 but find that the dust masses were increasing by only a factor $\sim$ 1.5 - 3, again insufficient to explain the order of magnitude difference between the proportion of dust relative to the stars between the metal-poor and metal-rich galaxies.
Nonetheless, with our modified blackbody fits we are considering only one temperature and grain size. We may be missing here a fraction of the dust mass coming from warmer big grains, and this contribution may be more important in low-metallicity galaxies rather than in more metal-rich ones. Thus, part of the observed trend may just be a side effect of using modified blackbodies.  
The mass corresponding to the stochastically heated grains is, however, negligible. In a follow-up paper (R\'emy-Ruyer et al. 2013, in prep.), we will obtain total dust masses from a full semi-empirical SED model, which will allow us to study this effect in more details. \\ 

%L_FIR histo
\noindent {\it FIR luminosity} \\

The FIR luminosities in the DGS sample range from $1.2\times10^7$ to $5.3\times10^{10}$ \lsun\ (Figure \ref{histos}d), with a median of $\sim 5.3 \times 10^{8}$ \lsun. We see in Figure \ref{histos}d that dwarf galaxies are less luminous in the FIR than the galaxies from the KINGFISH sample. However if we consider {\it L$_{FIR}$}/{\it M$_{BB}$}, which represents the quantity of light emitted by the available amount of dust, there is a strong trend of increasing {\it L$_{FIR}$}/{\it M$_{BB}$} with decreasing metallicity (Figure \ref{trend}): here we have a Spearman rank coefficient $\rho$=-0.72. The best power-law fit gives: 

\begin{equation}
L_{FIR}/M_{BB} = 4.2 \times 10^{24} \times (12+log(O/H))^{-23.6}
\end{equation}

Despite their lower dust masses, dwarf galaxies emit more in the FIR/submm than more metal-rich galaxies, per unit dust mass ($\sim$ six times more for the DGS). {\revised Here again, fixing $\beta_{obs}$=2 would only change the dust masses by a factor of 1.5 - 3, insufficient to explain the difference between the two samples. This difference is rather} a direct consequence of the higher temperature of dust grains in dwarf galaxies, as shown by the colours on Fig. \ref{trend}, due to the stronger and harder ISRF in which the grains are embedded. However, as mentioned above, the total dust mass may be underestimated by the modified blackbody model in lower metallicity galaxies and this trend could be weaker. \\

%T vs B plot
\noindent {\it Temperature - emissivity index relation} \\

\begin{figure*}
\begin{center}
\includegraphics[width=15cm]{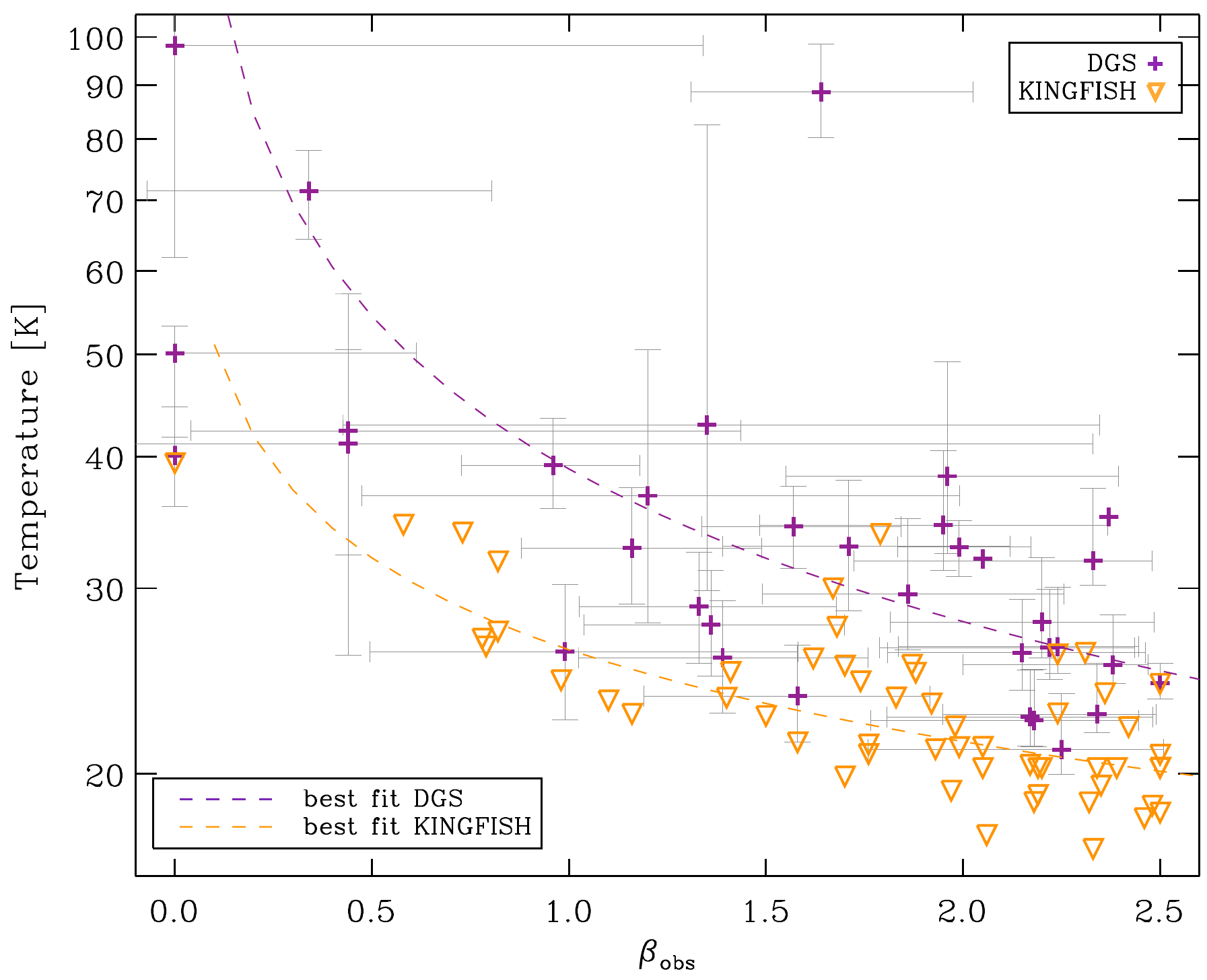}
\caption{Temperature versus $\beta_{obs}$ from the modified blackbody fits for the DGS (purple crosses) and for KINGFISH galaxies (orange downward triangles). The dotted lines correspond to the best power law fit for the DGS (purple) and KINGFISH (orange) galaxies, excluding the galaxies for which $\beta_{obs}$=0.0. They correspond to: log({\it T$_{DGS}$}) = (1.59 $\pm$ 0.01) + (-0.48 $\pm$ 0.04) $\times$ log($\beta_{obs,DGS}$), and log({\it T$_{KINGFISH}$}) = (1.41 $\pm$ 0.02) + (-0.29 $\pm$ 0.05) $\times$ log($\beta_{obs,KINGFISH}$). For clarity the error bars on the parameters have been displayed only for the DGS sample.}
\label{Tvsbeta}
\end{center}
\end{figure*}

Some studies have noted an inverse $\beta$ / temperature correlation in objects from starless cores to galaxies \citep{Dupac2003, YangPhillips2007, Anderson2010, Paradis2010, PlanckCollaboration2011XIX, Galametz2012, Smith2012b}. To investigate this possible effect in the DGS sample, we plot these two parameters from our modified blackbody fits ({\it T} and $\beta_{obs}$) (Figure \ref{Tvsbeta}). We also add the KINGFISH galaxies.
First we note that the DGS galaxies have overall higher dust temperature than the KINGFISH galaxies as already noted in Figures \ref{diagrams}, \ref{diagramPACS_SPIRE} and \ref{histos}. We also have the DGS galaxies where the fit gives $\beta_{obs}$ = 0.0, without detections beyond 160 \mic\ that we believe to be due to a poorly constrained submm SED. If we exclude these galaxies, the KINGFISH and DGS samples present an anticorrelation between {\it T} and $\beta_{obs}$, and this anticorrelation seems to be steeper in the DGS: the best power law fit gives {\it T} $\propto$ $\beta_{obs}^{-0.48}$ for the DGS and {\it T} $\propto$ $\beta_{obs}^{-0.29}$ for the KINGFISH galaxies. However, the anticorrelation seems stronger in KINGFISH than in the DGS sample ($\rho_{KINGFISH}$= -0.69 vs $\rho_{DGS}$= -0.56).

\cite{Shetty2009a, Shetty2009b} and \cite{JuvelaYsard2012b, JuvelaYsard2012a} showed that such an observed anticorrelation comes from the assumption of a constant temperature along the line-of-sight in modified blackbody fits and from noise in the measurements. They advise caution when interpreting this $\beta$ / temperature relationship when derived from $\chi^2$ modified blackbody fits. \cite{Kelly2012} show that a $\chi^2$ fit can artificially produce an anticorrelation between {\it T} and $\beta_{obs}$, where a Bayesian fit does not, and recovers the true parameters more accurately. 

Nonetheless, if we assume that the differences in the observed ({\it T}, $\beta_{obs}$) relations between DGS and KINGFISH can be due to changes in dust optical grain properties in the submm \citep[as suggested in][]{Meny2007, Paradis2010}, this may be the sign that the assumption of a single grain temperature, the presence of noise in the measurements and the use of a $\chi^2$ fitting procedure may only be {\it partially} responsible for the observed trends. However, given the very large errors on the {\it T} and $\beta$ parameters, it is difficult to draw a solid conclusion on this issue.

%Figures: Temp vs beta from bb fits 

\subsection{Submillimetre excess}\label{submmexcess}
A submm excess has been observed in the past in several dwarf galaxies \citep{Galliano2003, Galliano2005, Galametz2009, Bot2010, Grossi2010}. It has been called ``excess'' because the current available models are unable to fully explain the submm emission of these galaxies. In most models, $\beta_{theo}$ =  2 is often assumed in order to mimic the optical properties of the dust grain mixture of the Galaxy. In spiral galaxies, a modified blackbody with a fixed $\beta_{theo}$ to 2 reproduces well the FIR emission \citep{Bendo2003, Bendo2010}. In the colour-colour diagrams we hinted that a low $\beta_{obs}$ may be the sign of a possible presence of an excess emission adding its contribution to a $\beta_{theo}$ =  2 submm SED. \cite{Boselli2012} also showed that the {\it F$_{250}$/F$_{500}$} colour was more consistent with an effective emissivity index of 1.5 for the lowest metallicity galaxies in the HRS sample. Here we want to determine, systematically, which galaxies of the DGS and KINGFISH samples present an excess. 
A modified blackbody with a fixed emissivity index $\beta_{theo}$ of 2.0 is fit to the data for both DGS and KINGFISH samples (109 galaxies in total). Here again, we use the 70 \mic\ point only if the modelled flux is larger than the observed flux.

We take the relative residual at 500 \mic\ to be: 

\begin{equation}
{\it R(500)} = \frac{L_{\nu}^{observed}(500) - L_{\nu}^{model}(500)}{L_{\nu}^{model}(500)}
\end{equation}

In order to define a residual, {\it R}, at 500 \mic\ the galaxy must be detected out to 500 \mic. This, unfortunately, reduces our sample to 78 galaxies due to the high number of faint galaxies in the DGS sample. 

Following the same procedure as in Section \ref{bbfitserrors}, we randomly perturb the fluxes within the errors bars and perform fits of the perturbed SEDs (300 for each galaxy). A distribution of {\it R(500)} is generated and the 66.67\% confidence level of the distribution gives the error on the residual at 500 \mic: {\it $\Delta${\it R(500)}}. The values of {\it R(500)} and {\it $\Delta${\it R(500)}} are listed in Table \ref{excess}.

\begin{figure}[h!tbp]
\begin{center}
\includegraphics[width=8.8cm]{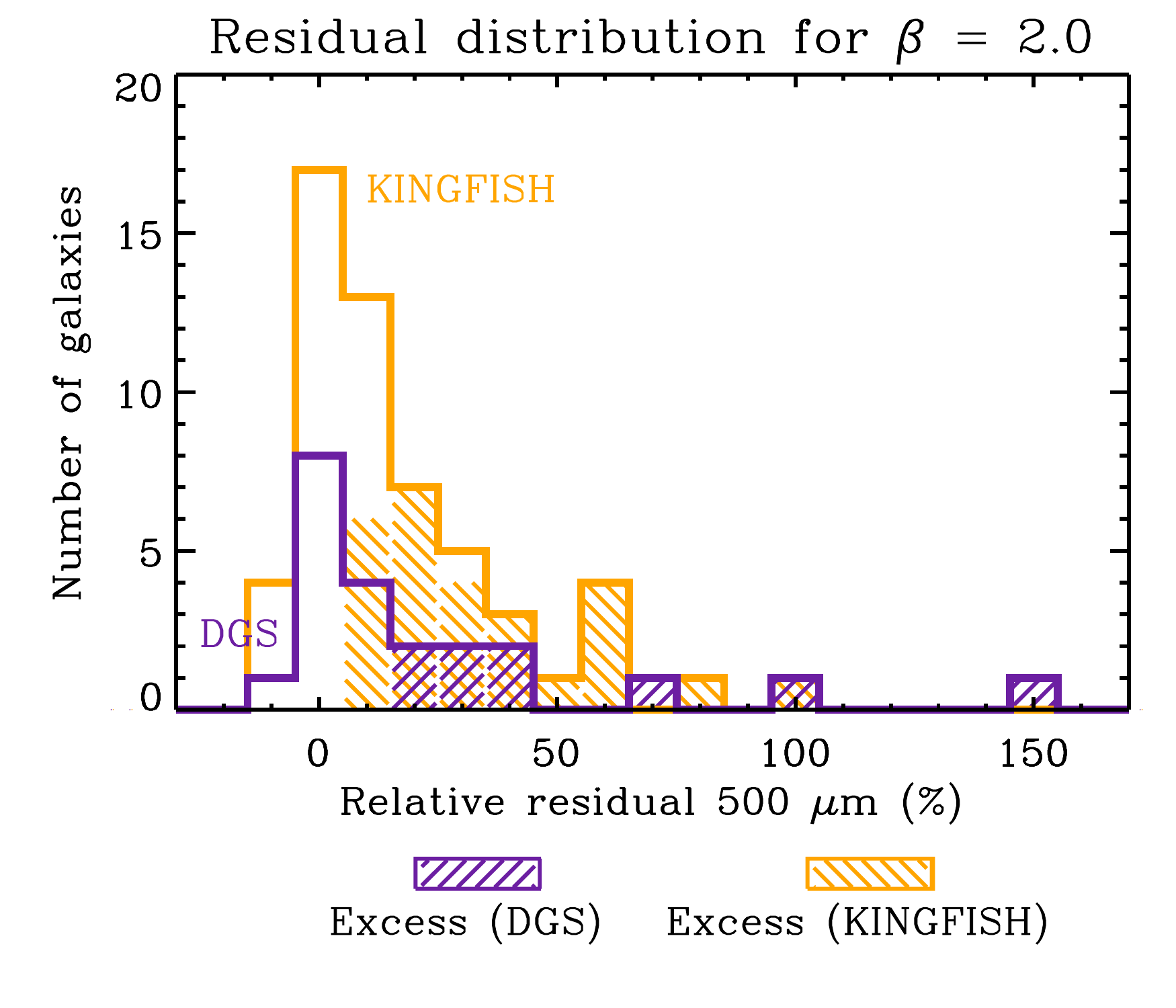}
\caption{Relative residual distribution at 500 \mic\ for modified blackbody fits with a fixed $\beta_{theo}$ of 2.0 for DGS (purple) and KINGFISH (orange) samples. Galaxies for which the residual at 500 \mic\ is greater than the corresponding error bar ({\it R(500)} $>$ {\it $\Delta${\it R(500)}}) have been marked by hashed cells.}
\label{ResidualHisto}
\end{center}
\end{figure}

A galaxy is then flagged with ``excess'' if the relative residual at 500 \mic\ is greater than the corresponding error: {\it R(500)}~$>${\it $\Delta${\it R(500)}} (see Table \ref{excess}). As the 500 \mic\ point is included in the fit, the procedure will also try to achieve a good fit of the 500 \mic\ point, and this will give lower {\it R(500)} than if the 500 \mic\ point was not included in the fit. That is why we fix our ``excess'' criterion to a 1$\sigma$ detection only.
For both samples, the {\it R(500)} distribution is shown in Figure \ref{ResidualHisto}, and excess galaxies are indicated by hashed cells.

%----------------------- Residuals table-----------------------------------------------
\begin{table}[h!tbp]
\caption{Table of relative residuals at 500 \mic\ for a modified blackbody fit with $\beta_{theo}$ fixed to 2.0 for the DGS sample. A column with the $\beta_{obs}$ values from Table \ref{bbfitsparam} have been added.}
\begin{center}
\label{excess}
\begin{tabular}{l || c || ccc}
\hline
\hline
&{\bf $\beta_{obs}$} &  &{\bf $\beta_{theo}$ = 2.0} & \\
\hline
Source & & {\it R(500)} (\%) & {\it $\Delta${\it R(500)}}  (\%) & Excess ? \\
\hline
Haro11       &     1.96&            27.6    &        10.8    &       yes   \\
Haro2        &     2.38&            -6.4    &        13.7    &             \\
Haro3        &     2.15&            -2.3    &        11.8    &             \\
He2-10       &     2.24&            -3.3    &         9.8    &             \\
HS0052+2536  &     1.20&           154.2    &        19.6    &       yes   \\
IC10         &     2.25&             0.1    &        26.9    &             \\
IIZw40       &     1.71&             2.2    &         8.0    &             \\
Mrk1089      &     2.34&            -3.6    &        14.2    &             \\
Mrk930       &     2.22&             8.3    &        28.9    &             \\
NGC1140      &     2.17&            -3.6    &        17.3    &             \\
NGC1569      &     2.20&             1.4    &         9.2    &             \\
NGC1705      &     1.16&            42.5    &        17.6    &       yes   \\
NGC2366      &     0.96&            39.6    &        12.4    &       yes   \\
NGC4214      &     1.39&            17.9    &         8.2    &       yes   \\
NGC4449      &     2.18&             1.2    &         8.9    &             \\
NGC4861      &     1.36&            32.5    &        15.7    &       yes   \\
NGC5253      &     1.86&             8.5    &         8.5    &             \\
NGC625       &     1.33&            21.6    &         9.4    &       yes   \\
NGC6822      &     0.99&            96.8    &        19.7    &       yes   \\
UM311        &     1.58&             9.4    &         9.5    &             \\
UM448        &     1.99&             6.9    &        16.2    &             \\
VIIZw403     &     1.57&            71.7    &        23.0    &       yes   \\
\hline
\end{tabular}
\end{center}
\end{table}

Out of 78 galaxies, 45\% present an excess at 500 \mic\ {\revised with respect to a  $\beta_{theo}$ =  2 modified blackbody}: nine are from DGS and 26 from KINGFISH. It is interesting to note that eight out of the nine KINGFISH galaxies of Irregular type (Im, I0 or Sm) detected at 500 \mic, are among the 26 ``excess'' KINGFISH galaxies. The one missing is HoII which has a very large error bar on the 500 \mic\ flux and thus a very wide {\it R(500)} distribution. 
\cite{Dale2012} looked at the residual at 500 \mic\ for a Draine \& Li (2007) model fit \citep[see][for details]{Dale2012} and also found that most of these Irregular galaxies presented an excess at 500 \mic. They mention a dozen KINGFISH galaxies with a {\it R(500)} above 60\%. However as their study is based on a different model than ours, we will not go deeper into any further comparison.

\begin{figure}[h!tbp]
\begin{center}
\includegraphics[width=8.8cm]{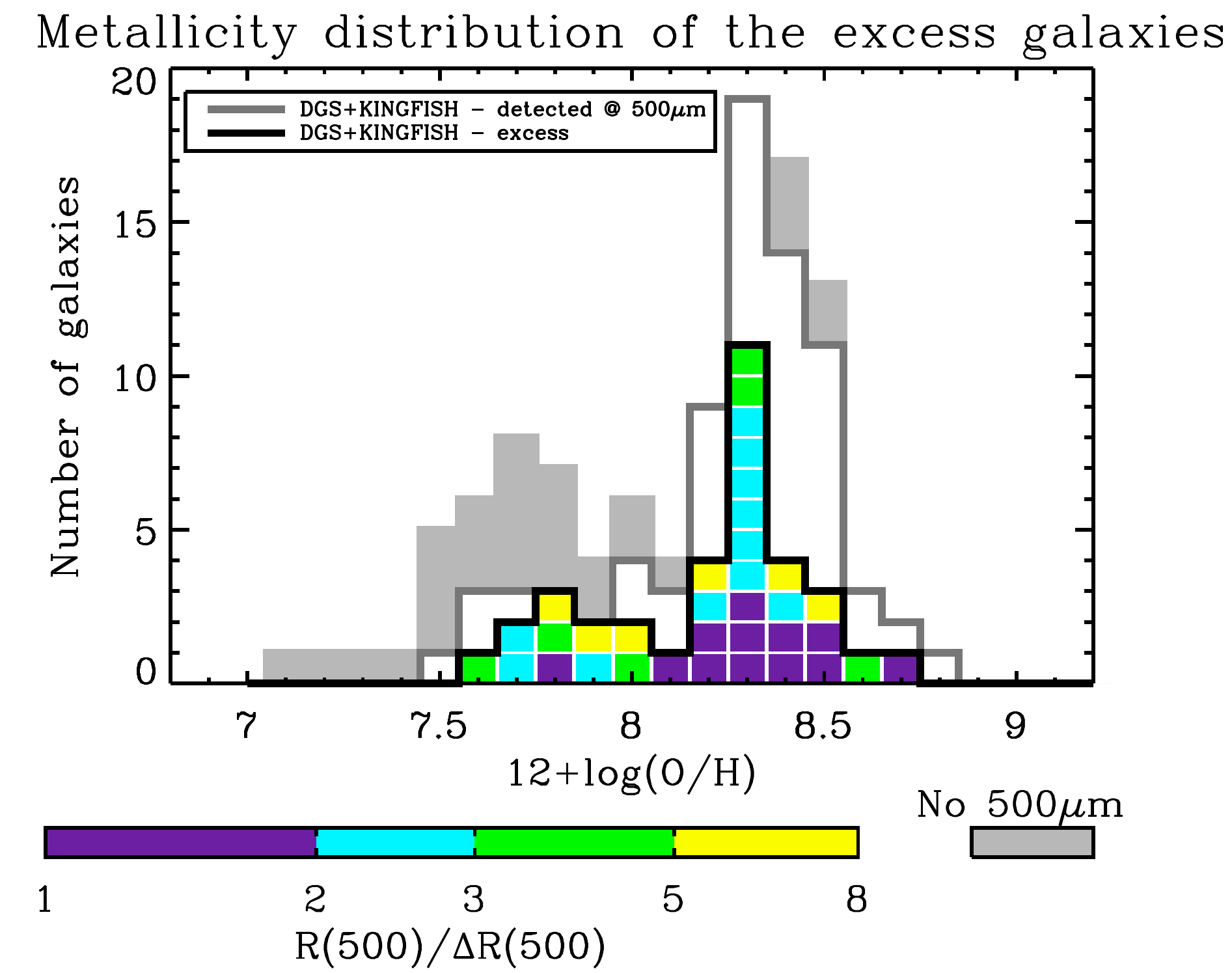}
\caption{Metallicity distribution for the excess galaxies in black. The colours mark the presence of an excess at 500 \mic\ and code the intensity of this excess: {\it R(500)}/{\it $\Delta${\it R(500)}}. The metallicity distribution for the DGS \& KINGFISH {\revised galaxies detected at 500 \mic}\ is outlined in grey. The grey cells mark all of the galaxies for which no detection is available at 500 \mic.}
\label{ResidualMetallicity}
\end{center}
\end{figure}

Figure \ref{ResidualMetallicity} shows the metallicity distribution of the 35 excess galaxies (black line) together with the joint metallicity distribution of DGS and KINGFISH samples (grey line). Note how the absence of detections at 500 \mic\ reduces the low-metallicity tail of the joint metallicity distribution. The metallicity distribution for the excess galaxies peaks around 12+log(O/H) $\sim$ 8.3 and is skewed towards the low-metallicity end of the distribution {\revised: 63 \% of the excess galaxies are galaxies with {\it Z} $<$ 0.4 \zsun, whereas, in the total distribution of galaxies detected at 500 \mic, (grey line on Fig. \ref{ResidualMetallicity}), only 49\% of the total number of galaxies are galaxies with {\it Z} $<$ 0.4 \zsun. Moreover, the proportion of excess galaxies in the [7.5 - 8.3] metallicity range is $\sim$ 53\% versus $\sim$ 28\% in the ]8.3 - 8.8] range.} This shows that the submm excess seems to occur mainly in metal-poor galaxies, {\revised at least when a $\beta_{theo}$ =  2 modified blackbody model is used.}
The colours on Fig. \ref{ResidualMetallicity} code the signal-to-noise ratio of the residual at 500 \mic\ for the excess galaxies. There seems to be a dichotomy in the distribution around 12+log(O/H) $\sim$ 8.3, with the strongest excesses being detected in the lowest metallicity galaxies.

\begin{figure}[h!tbp]
\begin{center}
\includegraphics[width=8.8cm]{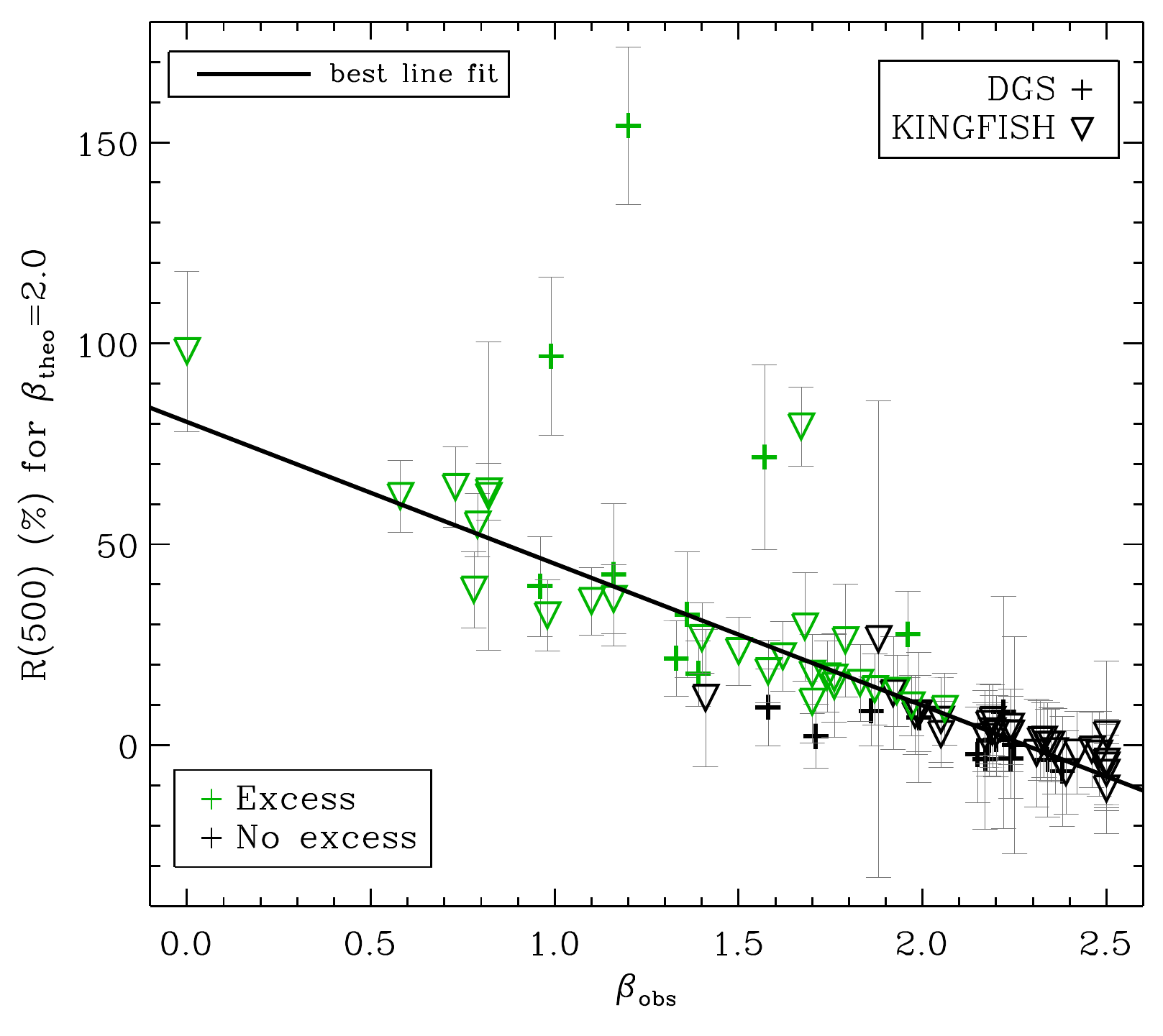}
\caption{Relative residual at 500 \mic\ versus $\beta_{obs}$ (from Section \ref{bbfitsprop}) for DGS (crosses) and KINGFISH (downward triangles) galaxies. The green symbols mark the galaxies presenting an excess at 500 \mic. The best fit line is indicated as a solid black line, and corresponds to: R(500)= (80.4 $\pm$ 4.1) + (-35.3 $\pm$ 2.2) $\times$ $\beta_{obs}$.}
\label{ResidualBeta}
\end{center}
\end{figure}

On Fig. \ref{ResidualBeta}, we have a clear anti-correlation between {\it R(500)} and $\beta_{obs}$ from the modified blackbody fits from Section \ref{bbfitsprop}: $\rho$=-0.78. For galaxies with a ``naturally'' flatter slope (i.e. a low $\beta_{obs}$), forcing a steeper slope (i.e. fixing $\beta_{theo}=2.0$) will naturally increase the residuals at the longest wavelengths, thus generating the correlation between {\it R(500)} and $\beta_{obs}$.

 %Figures: Residual histogramms for beta = 2.0 for KF+DGS 

% Figures: Intesity of excess in Z histo

%Figures: Residual vs beta obs

All of the galaxies showing an excess (i.e. {\it R(500)} $>$ {\it $\Delta${\it R(500)}}) have indeed a low $\beta_{obs}$ ($\beta_{obs} \leq$ 2.0) (Fig. \ref{ResidualBeta}). It is also interesting to note that this corresponds to 80\% of the 44 galaxies with $\beta_{obs} \leq$ 2.0. On the colour-colour diagram of Fig. \ref{diagramPACS_SPIRE}, all of the excess galaxies fall on the left side of the $\beta_{theo}$~=~2.0 line. Moreover, all of the galaxies falling on the left side of the $\beta_{theo}$~=~1.5 line, except one, present an excess. This is coherent with what is observed on Fig. \ref{ResidualBeta},
%This means that one can detect with a very high probability an excess at 500 \mic\ just with a simple modified blackbody model, and a FIR/submm colour-colour diagram. 
and can be useful to select potential targets for FIR/submm follow-up observations.

\subsection{A word of caution: submm excess appearing beyond 500 \mic}\label{submmexcesslong}
The previous analysis offers some tools to detect a submm excess in a galaxy. However as we are considering only \hersc\ wavelengths, any galaxy for which a submm excess is appearing beyond \hersc\ wavelengths would not be detected here. This is illustrated with two galaxies of the DGS sample {\revised with observations beyond 500 \mic}, Haro 11 and II Zw 40, both modelled with modified blackbodies, with the same procedure as in Section \ref{submmexcess}. Haro 11 falls to the left side of the $\beta$ = 1.5 line and has been identified as an ``excess'' galaxy in Table \ref{excess} whereas II Zw 40 falls to the right side of the $\beta$ = 1.5 line (Figure \ref{diagramPACS_SPIRE}) and does not present any excess at 500 \mic\ when using only \hersc\ bands (Table \ref{excess}). 

\begin{figure}[h!tbp]
\begin{center}
\includegraphics[width=8.8cm]{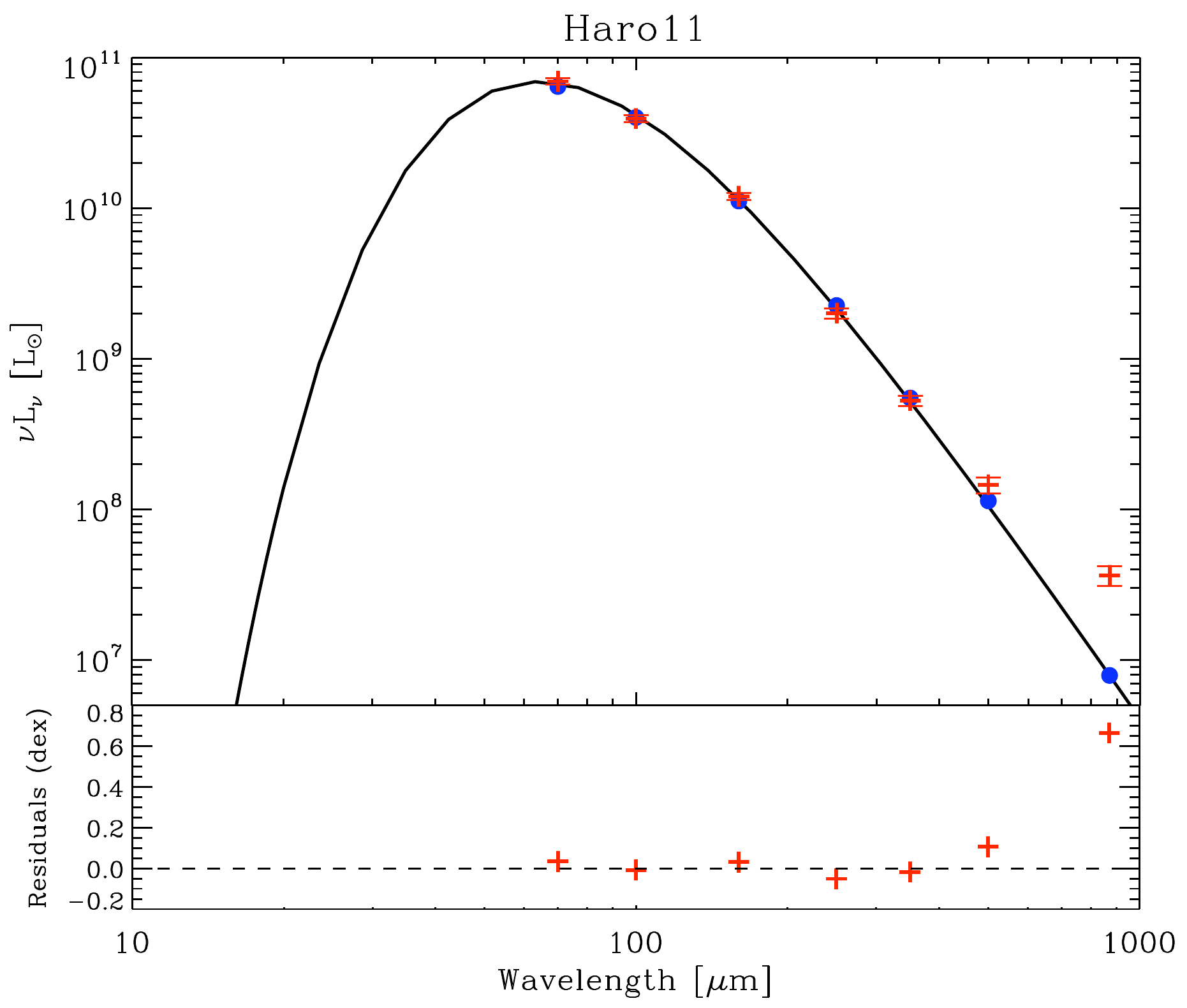}
\includegraphics[width=8.8cm]{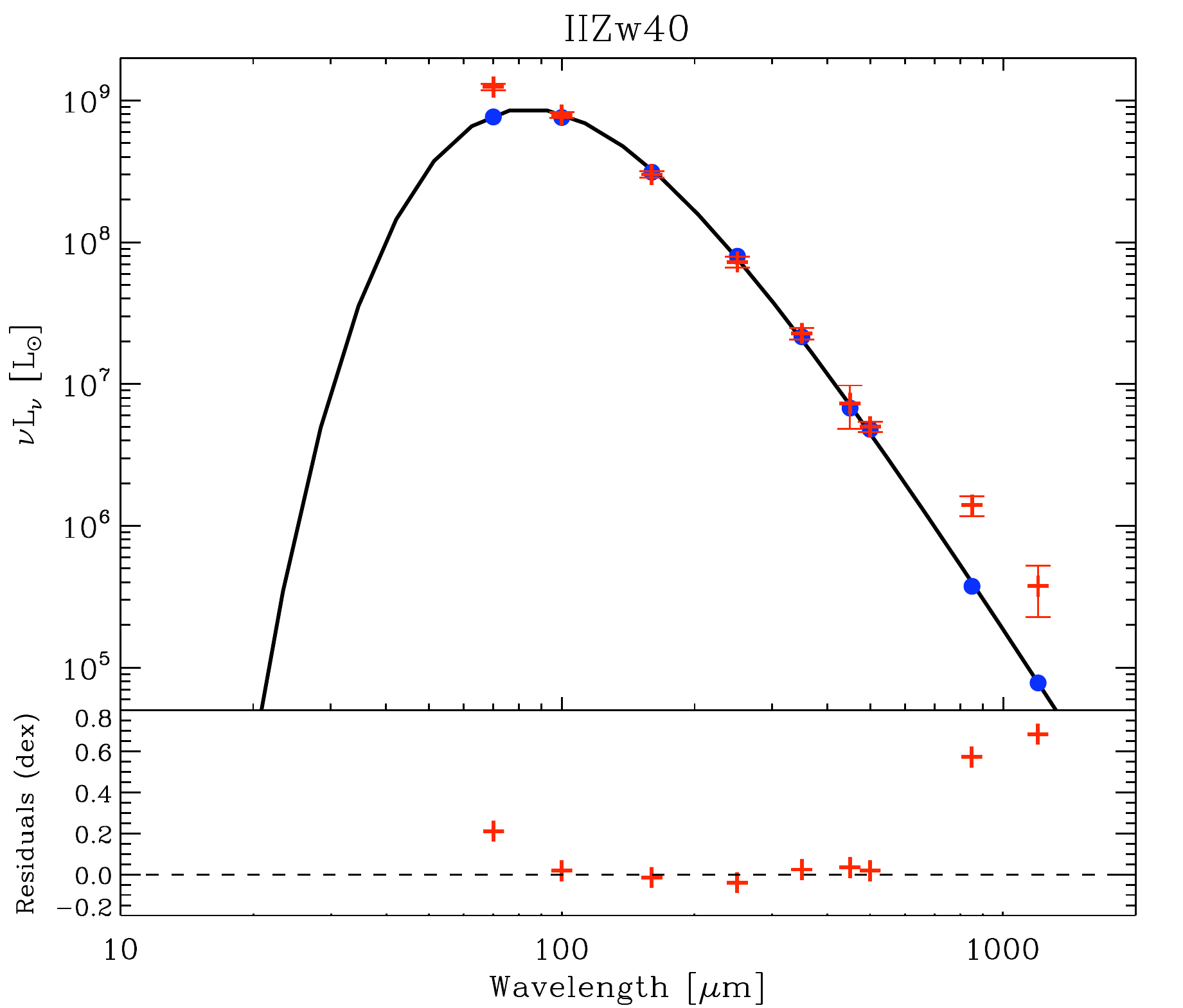}
\caption{SEDs of Haro 11 {\it (top)} and II Zw 40 {\it (bottom)}. They have been obtained with a modified blackbody model with a fixed $\beta_{theo}$~=~2.0. The 870 \mic\ point for Haro 11 is from \lab\ \citep{Galametz2009}. For II Zw 40, the 450 and 850 \mic\ points are from SCUBA, and the 1.2 mm point is from MAMBO \citep{Galliano2005}. All submm points have been corrected for non-dust contamination (free-free and synchrotron radiations and CO line contamination). The filled blue circles are the modelled fluxes in each band. The red crosses are the observations. The total SEDs are displayed in black. The bottom panel of each plot indicates the residuals from the fit.}
\label{submmSEDs}
\end{center}
\end{figure}

As shown in Figure \ref{submmSEDs}, Haro 11 presents an excess at 500 \mic\ ({\it R(500)} $\sim$ 28\% $\pm$ 13\%), confirmed at 870 \mic\ ({\it R(870)} $\sim$ 360\% $\pm$ 15\%). However the submm excess is clearly appearing at longer wavelengths ($\geq$ 500 \mic) for II Zw 40 when including observations beyond 500 \mic. At 500 \mic\, {\it R(500)} $\sim$ 2 $\pm$ 8 \%, and {\it R(450)} $\sim$ 6 $\pm$ 34 \% but at 850 and 1200 \mic\ we have {\it R(850)} $\sim$ 265 $\pm$ 16 \% and {\it R(1200)} $\sim$ 370 $\pm$ 40 \%. This illustrates the need for submm data, to complement the existing \hersc\ data. For several galaxies of our sample, new submm observations at 870 \mic\ with \lab\ will soon be available. 

% Figures: BB Haro11 IIZw40 with submm

This submm excess has been one of the main sources of uncertainty in dust modelling in dwarf galaxies for the past few years, especially on the dust mass parameter. Here we show that this excess does not seem that uncommon even in moderately metal-poor environments and thus understanding its origin is crucial to get accurate dust parameters. Several explanations have already been proposed to investigate the origin of this excess, although not completely satisfactory.

\cite{Galliano2003, Galliano2005}, \cite{Galametz2009, Galametz2010, Galametz2011} modelled the submm excess they detected in their metal-poor galaxies with a very cold dust (VCD) component. They added to their SED models an extra modified blackbody with a submm emissivity index of 1 and a low dust temperature ($\sim$ 10K). Their additional component could explain the break observed in the submm domain in some of their SEDs but it led to very low gas-to-dust mass ratios, considering the observed gas mass, compared to that expected from chemical evolution and from the amount of available metals in the ISM.

Several studies have shown that fast rotating very small dust grains from ionized gas regions in many galaxies were producing centimetre (cm) radio emission \citep{Ferrara1994, Draine2012}. As shown by recent studies \citep{YsardVerstraete2010, Ysard2012}, the peak of the ``spinning'' dust emission depends on many parameters such as the radiation field intensity, the dust size distribution, dipole moment distribution, physical parameters of the gas phase, etc. This hypothesis was tested to explain the submm/cm excess by \cite{Bot2010}, \cite{Israel2010} and \cite{PlanckCollaboration2011XX} in the Large and Small Magellanic Clouds. The spinning dust model seemed sufficient for the mm/cm excess but another effect is required to explain the submm/mm excess. Indeed, the usual spinning dust models do not normally produce much emission in the submm domain but rather at longer wavelengths (as illustrated in \cite{Murphy2010} for NGC 6946). Moreover, PAHs have been assumed to be the carriers of this spinning dust emission \citep[as shown in][]{DraineLazarian1998} and this may seem contradictory with the weakness of the PAH features observed in low-metallicity galaxies \citep{Engelbracht2005}.

\cite{Paradis2009} showed that the FIR-submm excess in dense molecular clouds of the Galaxy could be explained by fractal aggregation of amorphous individual grains which induces changes in the dust optical properties in the submm. This grain coagulation effect had already been suggested in the past by \cite{Bazell1990} and \cite{Stepnik2001}.

\cite{Meny2007} proposed a new model for FIR/submm dust emission based on the physical properties of disordered matter. They consider the interaction of electromagnetic waves with the acoustic oscillations in a disordered charge distribution (DCD) and a distribution of low energy two level tunnelling states (TLS). This interaction results in an emission spectrum strongly dependent on the temperature and in an enhanced submm/mm emission compared to more classical models. The emissivity index is no longer constant over this wavelength range. This model has been successfully applied by \cite{Bot2010} to explain the mm excess in the Large Magellanic Cloud (LMC). However, when applied to the Small Magellanic Cloud (SMC), DCD/TLS effects alone do not reproduce the excess well \citep{Bot2010}.

A recent work by \cite{Draine2012} on the SMC focuses on magnetic grains as a possible source of submm excess. They consider nanoparticles of metallic iron, magnetite and maghemite that could be free fliers in the ISM or inclusions on larger dust grains. Magnetic grains indeed have an enhanced absorption cross section at submm wavelengths and part of the submm excess could be due to thermal emission from magnetic grain material. They show that a combination of a normal dust mixture (amorphous silicates and carbonaceous grains), spinning dust and magnetic dust could account for the observed SED in the submm/mm range of the SMC. 

The various explanations for the submm excess presented here will be explored for the DGS galaxies for which we have submm data, in further studies.

 %===============================================================
% Conclusion and summary
%===============================================================

\section{Conclusions}
We present here the new \hersc\ photometry data for the Dwarf Galaxy Survey, dedicated to the study of the gas and dust properties in the ISMs of 48 low-metallicity galaxies. 
We perform the data reduction by adapting the map making procedures for extended vs compact sources.  
We derive flux densities from aperture photometry or from a timeline fitting procedure (for SPIRE point sources) and present here a catalogue of flux densities for the whole sample.
We compare our PACS flux densities with corresponding MIPS flux densities to assess their reliability and we found that the MIPS and PACS flux densities are compatible.

We present a first analysis of the FIR/submm behaviour of the DGS galaxies, with modified blackbody fits and several \hersc\ colour-colour diagrams, especially focusing on the comparison with more metal-rich galaxies from the KINGFISH sample, making a total of 109 galaxies, and on the appearance of the submm excess among the samples. We note that: 
 \begin{itemize}
 	 \item{Dwarf galaxies present different dust properties than more metal-rich galaxies as in the KINGFISH sample and this can be quantified with modified blackbody fits, to get at the dust temperatures, emissivity indices, dust masses and FIR luminosities. The differences in the spread of KINGFISH and DGS galaxies on \hersc\ colour-colour diagrams qualitatively reflect these differences in the FIR dust properties and SED shape.}
 	\item{For modified blackbody fits, the range in dust temperature is 22-98 K and the median dust temperature of the DGS sample is $\sim$ 32 K. This is warmer than what is observed in more metal-rich galaxies ({\it T}$_{KINGFISH} \sim$ 23 K), and we see a trend of increasing dust temperature with decreasing metallicity. SEDs of lower metallicity galaxies peak at very short wavelengths, often between 50 and 100 \mic, and present overall warmer dust.}
	\item{Dwarf galaxies show a lower proportion of dust mass relative to stellar mass compared to more metal-rich galaxies (the median for the ratio {\it M$_{BB}$/M$_{star}$} is 0.02\% for DGS versus 0.1\% for KINGFISH). Despite their relatively lower dust masses, dwarf galaxies emit more in the FIR/submm than more metal-rich galaxies, per unit dust mass (about six times more for the DGS), reflecting the impact of the very energetic radiation environment on dust grains in dwarfs.}
	 \item{The range in $\beta_{obs}$ is 0.0-2.5 for the DGS and KINGFISH samples, with a median of $\sim$1.7 for the DGS and $\sim$1.9 for KINGFISH, lower than what is usually taken in SED models ($\beta_{theo}$~=~2.0). No clear trend between $\beta_{obs}$ and metallicity has been noted here. However galaxies with 1.0 $ \leq \beta_{obs} \leq$ 2.0 seem to be primarily metal-poor and moderately metal-poor galaxies (0.1~\zsun~$<$~{\it Z}~$<$~0.4 \zsun).}
	 \item{45\% of the DGS and KINGFISH galaxies harbour an excess at 500 \mic\ from a modified blackbody model with $\beta_{theo}$ = 2.0, when considering galaxies with detections at 500 \mic. This excess seems to appear mainly in lower metallicity galaxies ({\it Z} $<$ 0.4 \zsun), and the strongest excesses are detected in the most metal-poor galaxies. However the submm excess can sometimes only become apparent at wavelengths beyond \hersc\ bands, highlighting the need for submm data beyond 500 \mic.}
\end{itemize}

A following paper (R\'emy-Ruyer et al. 2013, in prep.) will present the systematic full dust modelling of all of the DGS galaxies with a semi-empirical model which allows for a more realistic description of the dust, and explore in detail the various dust properties of dwarf galaxies. The large ancillary data set, covering the whole IR-to-submm range, will be crucial when performing this multi-wavelength study of the DGS galaxies.

%===============================================================
% Acknowledgements
%===============================================================

\begin{acknowledgements}
{\revised The authors would like to thank the referee, Simone Bianchi, for his valuable comments that helped improve the quality of this paper. ARR is supported by a CFR grant from the AIM laboratory (Saclay, France). This research was, in part, made possible through the financial support of the Agence Nationale de la Recherche (ANR) through the programme SYMPATICO (Program Blanc Projet ANR-11-BS56-0023). IDL is a postdoctoral researcher of the FWO-Vlaanderen (Belgium).}

PACS has been developed by MPE (Germany); UVIE (Austria); KU Leuven, CSL, IMEC (Belgium); CEA, LAM (France); MPIA (Germany); INAF-IFSI/OAA/OAP/OAT, LENS, SISSA (Italy); IAC (Spain). This development has been supported by BMVIT (Austria), ESA-PRODEX (Belgium), CEA/CNES (France), DLR (Germany), ASI/INAF (Italy), and CICYT/MCYT (Spain). SPIRE has been developed by Cardiff University (UK); Univ. Lethbridge (Canada); NAOC (China); CEA, LAM (France); IFSI, Univ. Padua (Italy); IAC (Spain); SNSB (Sweden); Imperial College London, RAL, UCL-MSSL, UKATC, Univ. Sussex (UK) and Caltech, JPL, NHSC, Univ. Colorado (USA). This development has been supported by CSA (Canada); NAOC (China); CEA, CNES, CNRS (France); ASI (Italy); MCINN (Spain); Stockholm Observatory (Sweden); STFC (UK); and NASA (USA).

SPIRE has been developed by a consortium of institutes led by Cardiff Univ. (UK) and including: Univ. Lethbridge (Canada); NAOC (China); CEA, LAM (France); IFSI, Univ. Padua (Italy); IAC (Spain); Stockholm Observatory (Sweden); Imperial College London, RAL, UCL-MSSL, UKATC, Univ. Sussex (UK); and Caltech, JPL, NHSC, Univ. Colorado (USA). This development has been supported by national funding agencies: CSA (Canada); NAOC (China); CEA, CNES, CNRS (France); ASI (Italy); MCINN (Spain); SNSB (Sweden); STFC, UKSA (UK); and NASA (USA).
\end{acknowledgements}

%===============================================================
% Bibliography
%===============================================================

\bibliographystyle{aa}
\bibliography{DGS_photometry_submission_revised}

\begin{thebibliography}{139}
\expandafter\ifx\csname natexlab\endcsname\relax\def\natexlab#1{#1}\fi

\bibitem[{{Aloisi} {et~al.}(2007){Aloisi}, {Clementini}, {Tosi}, {Annibali},
  {Contreras}, {Fiorentino}, {Mack}, {Marconi}, {Musella}, {Saha}, {Sirianni},
  \& {van der Marel}}]{Aloisi2007}
{Aloisi}, A., {Clementini}, G., {Tosi}, M., {et~al.} 2007, \apjl, 667, L151

\bibitem[{{Aloisi} {et~al.}(2005){Aloisi}, {van der Marel}, {Mack},
  {Leitherer}, {Sirianni}, \& {Tosi}}]{Aloisi2005}
{Aloisi}, A., {van der Marel}, R.~P., {Mack}, J., {et~al.} 2005, \apjl, 631,
  L45

\bibitem[{{Amblard} {et~al.}(2010){Amblard}, {Cooray}, {Serra}, {Temi},
  {Barton}, {Negrello}, {Auld}, {Baes}, {Baldry}, {Bamford}, {Blain}, {Bock},
  {Bonfield}, {Burgarella}, {Buttiglione}, {Cameron}, {Cava}, {Clements},
  {Croom}, {Dariush}, {de Zotti}, {Driver}, {Dunlop}, {Dunne}, {Dye}, {Eales},
  {Frayer}, {Fritz}, {Gardner}, {Gonzalez-Nuevo}, {Herranz}, {Hill}, {Hopkins},
  {Hughes}, {Ibar}, {Ivison}, {Jarvis}, {Jones}, {Kelvin}, {Lagache}, {Leeuw},
  {Liske}, {Lopez-Caniego}, {Loveday}, {Maddox}, {Micha{\l}owski}, {Norberg},
  {Parkinson}, {Peacock}, {Pearson}, {Pascale}, {Pohlen}, {Popescu},
  {Prescott}, {Robotham}, {Rigby}, {Rodighiero}, {Samui}, {Sansom}, {Scott},
  {Serjeant}, {Sharp}, {Sibthorpe}, {Smith}, {Thompson}, {Tuffs}, {Valtchanov},
  {van Kampen}, {van der Werf}, {Verma}, {Vieira}, \& {Vlahakis}}]{Amblard2010}
{Amblard}, A., {Cooray}, A., {Serra}, P., {et~al.} 2010, \aap, 518, L9

\bibitem[{{Anderson} {et~al.}(2010){Anderson}, {Zavagno}, {Rod{\'o}n},
  {Russeil}, {Abergel}, {Ade}, {Andr{\'e}}, {Arab}, {Baluteau}, {Bernard},
  {Blagrave}, {Bontemps}, {Boulanger}, {Cohen}, {Compi{\`e}gne}, {Cox},
  {Dartois}, {Davis}, {Emery}, {Fulton}, {Gry}, {Habart}, {Huang}, {Joblin},
  {Jones}, {Kirk}, {Lagache}, {Lim}, {Madden}, {Makiwa}, {Martin},
  {Miville-Desch{\^e}nes}, {Molinari}, {Moseley}, {Motte}, {Naylor}, {Okumura},
  {Pinheiro Gon{\c c}alves}, {Polehampton}, {Saraceno}, {Sauvage}, {Sidher},
  {Spencer}, {Swinyard}, {Ward-Thompson}, \& {White}}]{Anderson2010}
{Anderson}, L.~D., {Zavagno}, A., {Rod{\'o}n}, J.~A., {et~al.} 2010, \aap, 518,
  L99

\bibitem[{{Aniano} {et~al.}(2012){Aniano}, {Draine}, {Calzetti}, {Dale},
  {Engelbracht}, {Gordon}, {Hunt}, {Kennicutt}, {Krause}, {Leroy}, {Rix},
  {Roussel}, {Sandstrom}, {Sauvage}, {Walter}, {Armus}, {Bolatto}, {Crocker},
  {Donovan Meyer}, {Galametz}, {Helou}, {Hinz}, {Johnson}, {Koda}, {Montiel},
  {Murphy}, {Skibba}, {Smith}, \& {Wolfire}}]{Aniano2012}
{Aniano}, G., {Draine}, B.~T., {Calzetti}, D., {et~al.} 2012, \apj, 756, 138

\bibitem[{{Asplund} {et~al.}(2009){Asplund}, {Grevesse}, {Sauval}, \&
  {Scott}}]{Asplund2009}
{Asplund}, M., {Grevesse}, N., {Sauval}, A.~J., \& {Scott}, P. 2009, \araa, 47,
  481

\bibitem[{{Auld} {et~al.}(2013){Auld}, {Bianchi}, {Smith}, {Davies}, {Bendo},
  {di Serego}, {Cortese}, {Baes}, {Bomans}, {Boquien}, {Boselli}, {Ciesla},
  {Clemens}, {Corbelli}, {De Looze}, {Fritz}, {Gavazzi}, {Pappalardo},
  {Grossi}, {Hunt}, {Madden}, {Magrini}, {Pohlen}, {Verstappen}, {Vlahakis},
  {Xilouris}, \& {Zibetti}}]{Auld2013}
{Auld}, R., {Bianchi}, S., {Smith}, M.~W.~L., {et~al.} 2013, \mnras, 428, 1880

\bibitem[{{Bazell} \& {Dwek}(1990)}]{Bazell1990}
{Bazell}, D. \& {Dwek}, E. 1990, \apj, 360, 142

\bibitem[{{Bendo} {et~al.}(2006){Bendo}, {Dale}, {Draine}, {Engelbracht},
  {Kennicutt}, {Calzetti}, {Gordon}, {Helou}, {Hollenbach}, {Li}, {Murphy},
  {Prescott}, \& {Smith}}]{Bendo2006}
{Bendo}, G.~J., {Dale}, D.~A., {Draine}, B.~T., {et~al.} 2006, \apj, 652, 283

\bibitem[{{Bendo} {et~al.}(2012){Bendo}, {Galliano}, \& {Madden}}]{Bendo2012}
{Bendo}, G.~J., {Galliano}, F., \& {Madden}, S.~C. 2012, \mnras, 423, 197

\bibitem[{{Bendo} {et~al.}(2003){Bendo}, {Joseph}, {Wells}, {Gallais}, {Haas},
  {Heras}, {Klaas}, {Laureijs}, {Leech}, {Lemke}, {Metcalfe}, {Rowan-Robinson},
  {Schulz}, \& {Telesco}}]{Bendo2003}
{Bendo}, G.~J., {Joseph}, R.~D., {Wells}, M., {et~al.} 2003, \aj, 125, 2361

\bibitem[{{Bendo} {et~al.}(2010){Bendo}, {Wilson}, {Pohlen}, {Sauvage}, {Auld},
  {Baes}, {Barlow}, {Bock}, {Boselli}, {Bradford}, {Buat}, {Castro-Rodriguez},
  {Chanial}, {Charlot}, {Ciesla}, {Clements}, {Cooray}, {Cormier}, {Cortese},
  {Davies}, {Dwek}, {Eales}, {Elbaz}, {Galametz}, {Galliano}, {Gear}, {Glenn},
  {Gomez}, {Griffin}, {Hony}, {Isaak}, {Levenson}, {Lu}, {Madden},
  {O'Halloran}, {Okumura}, {Oliver}, {Page}, {Panuzzo}, {Papageorgiou},
  {Parkin}, {Perez-Fournon}, {Rangwala}, {Rigby}, {Roussel}, {Rykala},
  {Sacchi}, {Schulz}, {Schirm}, {Smith}, {Spinoglio}, {Stevens}, {Sundar},
  {Symeonidis}, {Trichas}, {Vaccari}, {Vigroux}, {Wozniak}, {Wright}, \&
  {Zeilinger}}]{Bendo2010}
{Bendo}, G.~J., {Wilson}, C.~D., {Pohlen}, M., {et~al.} 2010, \aap, 518, L65

\bibitem[{{Bergvall} {et~al.}(2006){Bergvall}, {Zackrisson}, {Andersson},
  {Arnberg}, {Masegosa}, \& {{\"O}stlin}}]{Bergvall2006}
{Bergvall}, N., {Zackrisson}, E., {Andersson}, B.-G., {et~al.} 2006, \aap, 448,
  513

\bibitem[{{Bianchi}(2013)}]{Bianchi2013}
{Bianchi}, S. 2013, \aap, 552, A89

\bibitem[{{Boquien} {et~al.}(2011){Boquien}, {Calzetti}, {Combes}, {Henkel},
  {Israel}, {Kramer}, {Rela{\~n}o}, {Verley}, {van der Werf}, {Xilouris}, \&
  {HERM33ES Team}}]{Boquien2011}
{Boquien}, M., {Calzetti}, D., {Combes}, F., {et~al.} 2011, \aj, 142, 111

\bibitem[{{Bordalo} {et~al.}(2009){Bordalo}, {Plana}, \&
  {Telles}}]{Bordalo2009}
{Bordalo}, V., {Plana}, H., \& {Telles}, E. 2009, \apj, 696, 1668

\bibitem[{{Boselli} {et~al.}(2012){Boselli}, {Ciesla}, {Cortese}, {Buat},
  {Boquien}, {Bendo}, {Boissier}, {Eales}, {Gavazzi}, {Hughes}, {Pohlen},
  {Smith}, {Baes}, {Bianchi}, {Clements}, {Cooray}, {Davies}, {Gear}, {Madden},
  {Magrini}, {Panuzzo}, {Remy}, {Spinoglio}, \& {Zibetti}}]{Boselli2012}
{Boselli}, A., {Ciesla}, L., {Cortese}, L., {et~al.} 2012, \aap, 540, A54

\bibitem[{{Boselli} {et~al.}(2002){Boselli}, {Lequeux}, \&
  {Gavazzi}}]{Boselli2002}
{Boselli}, A., {Lequeux}, J., \& {Gavazzi}, G. 2002, \aap, 384, 33

\bibitem[{{Boselli} {et~al.}(2004){Boselli}, {Lequeux}, \&
  {Gavazzi}}]{Boselli2004}
{Boselli}, A., {Lequeux}, J., \& {Gavazzi}, G. 2004, \aap, 428, 409

\bibitem[{{Bot} {et~al.}(2010){Bot}, {Ysard}, {Paradis}, {Bernard}, {Lagache},
  {Israel}, \& {Wall}}]{Bot2010}
{Bot}, C., {Ysard}, N., {Paradis}, D., {et~al.} 2010, \aap, 523, A20+

\bibitem[{{Cannon} {et~al.}(2003){Cannon}, {Dohm-Palmer}, {Skillman}, {Bomans},
  {C{\^o}t{\'e}}, \& {Miller}}]{Cannon2003}
{Cannon}, J.~M., {Dohm-Palmer}, R.~C., {Skillman}, E.~D., {et~al.} 2003, \aj,
  126, 2806

\bibitem[{{Cantalupo} {et~al.}(2010){Cantalupo}, {Borrill}, {Jaffe}, {Kisner},
  \& {Stompor}}]{Cantalupo2010}
{Cantalupo}, C.~M., {Borrill}, J.~D., {Jaffe}, A.~H., {Kisner}, T.~S., \&
  {Stompor}, R. 2010, \apjs, 187, 212

\bibitem[{{Ciesla} {et~al.}(2012){Ciesla}, {Boselli}, {Smith}, {Bendo},
  {Cortese}, {Eales}, {Bianchi}, {Boquien}, {Buat}, {Davies}, {Pohlen},
  {Zibetti}, {Baes}, {Cooray}, {de Looze}, {di Serego Alighieri}, {Galametz},
  {Gomez}, {Lebouteiller}, {Madden}, {Pappalardo}, {Remy}, {Spinoglio},
  {Vaccari}, {Auld}, \& {Clements}}]{Ciesla2012}
{Ciesla}, L., {Boselli}, A., {Smith}, M.~W.~L., {et~al.} 2012, \aap, 543, A161

\bibitem[{{Cormier} {et~al.}(2012){Cormier}, {Lebouteiller}, {Madden}, {Abel},
  {Hony}, {Galliano}, {Baes}, {Barlow}, {Cooray}, {De Looze}, {Galametz},
  {Karczewski}, {Parkin}, {R{\'e}my}, {Sauvage}, {Spinoglio}, {Wilson}, \&
  {Wu}}]{Cormier2012}
{Cormier}, D., {Lebouteiller}, V., {Madden}, S.~C., {et~al.} 2012, \aap, 548,
  A20

\bibitem[{{Cormier} {et~al.}(2011){Cormier}, {Madden}, {Lebouteiller},
  {Galliano}, \& {Hony}}]{Cormier2011}
{Cormier}, D., {Madden}, S., {Lebouteiller}, V., {Galliano}, F., \& {Hony}, S.
  2011, in IAU Symposium, Vol. 280, IAU Symposium, 137P

\bibitem[{{Dale} {et~al.}(2012){Dale}, {Aniano}, {Engelbracht}, {Hinz},
  {Krause}, {Montiel}, {Roussel}, {Appleton}, {Armus}, {Beir{\~a}o}, {Bolatto},
  {Brandl}, {Calzetti}, {Crocker}, {Croxall}, {Draine}, {Galametz}, {Gordon},
  {Groves}, {Hao}, {Helou}, {Hunt}, {Johnson}, {Kennicutt}, {Koda}, {Leroy},
  {Li}, {Meidt}, {Miller}, {Murphy}, {Rahman}, {Rix}, {Sandstrom}, {Sauvage},
  {Schinnerer}, {Skibba}, {Smith}, {Tabatabaei}, {Walter}, {Wilson}, {Wolfire},
  \& {Zibetti}}]{Dale2012}
{Dale}, D.~A., {Aniano}, G., {Engelbracht}, C.~W., {et~al.} 2012, \apj, 745, 95

\bibitem[{{Dale} {et~al.}(2007){Dale}, {Gil de Paz}, {Gordon}, {Hanson},
  {Armus}, {Bendo}, {Bianchi}, {Block}, {Boissier}, {Boselli}, {Buckalew},
  {Buat}, {Burgarella}, {Calzetti}, {Cannon}, {Engelbracht}, {Helou},
  {Hollenbach}, {Jarrett}, {Kennicutt}, {Leitherer}, {Li}, {Madore}, {Martin},
  {Meyer}, {Murphy}, {Regan}, {Roussel}, {Smith}, {Sosey}, {Thilker}, \&
  {Walter}}]{Dale2007}
{Dale}, D.~A., {Gil de Paz}, A., {Gordon}, K.~D., {et~al.} 2007, \apj, 655, 863

\bibitem[{{de Vaucouleurs} {et~al.}(1991){de Vaucouleurs}, {de Vaucouleurs},
  {Corwin}, {Buta}, {Paturel}, \& {Fouque}}]{DeVaucouleurs1991}
{de Vaucouleurs}, G., {de Vaucouleurs}, A., {Corwin}, Jr., H.~G., {et~al.}
  1991, {Third Reference Catalogue of Bright Galaxies}, ed. {de Vaucouleurs,
  G., de Vaucouleurs, A., Corwin, H.~G., Jr., Buta, R.~J., Paturel, G., \&amp;
  Fouque, P.} (de Vaucouleurs, G.)

\bibitem[{{Draine} \& {Hensley}(2012)}]{Draine2012}
{Draine}, B.~T. \& {Hensley}, B. 2012, \apj, 757, 103

\bibitem[{{Draine} \& {Lazarian}(1998)}]{DraineLazarian1998}
{Draine}, B.~T. \& {Lazarian}, A. 1998, \apj, 508, 157

\bibitem[{{Dumke} {et~al.}(2004){Dumke}, {Krause}, \&
  {Wielebinski}}]{Dumke2004}
{Dumke}, M., {Krause}, M., \& {Wielebinski}, R. 2004, \aap, 414, 475

\bibitem[{{Dunne} {et~al.}(2011){Dunne}, {Gomez}, {da Cunha}, {Charlot}, {Dye},
  {Eales}, {Maddox}, {Rowlands}, {Smith}, {Auld}, {Baes}, {Bonfield}, {Bourne},
  {Buttiglione}, {Cava}, {Clements}, {Coppin}, {Cooray}, {Dariush}, {de Zotti},
  {Driver}, {Fritz}, {Geach}, {Hopwood}, {Ibar}, {Ivison}, {Jarvis}, {Kelvin},
  {Pascale}, {Pohlen}, {Popescu}, {Rigby}, {Robotham}, {Rodighiero}, {Sansom},
  {Serjeant}, {Temi}, {Thompson}, {Tuffs}, {van der Werf}, \&
  {Vlahakis}}]{Dunne2011}
{Dunne}, L., {Gomez}, H.~L., {da Cunha}, E., {et~al.} 2011, \mnras, 417, 1510

\bibitem[{{Dupac} {et~al.}(2003){Dupac}, {Bernard}, {Boudet}, {Giard},
  {Lamarre}, {M{\'e}ny}, {Pajot}, {Ristorcelli}, {Serra}, {Stepnik}, \&
  {Torre}}]{Dupac2003}
{Dupac}, X., {Bernard}, J.-P., {Boudet}, N., {et~al.} 2003, \aap, 404, L11

\bibitem[{{Engelbracht} {et~al.}(2005){Engelbracht}, {Gordon}, {Rieke},
  {Werner}, {Dale}, \& {Latter}}]{Engelbracht2005}
{Engelbracht}, C.~W., {Gordon}, K.~D., {Rieke}, G.~H., {et~al.} 2005, \apjl,
  628, L29

\bibitem[{{Engelbracht} {et~al.}(2008){Engelbracht}, {Rieke}, {Gordon},
  {Smith}, {Werner}, {Moustakas}, {Willmer}, \& {Vanzi}}]{Engelbracht2008}
{Engelbracht}, C.~W., {Rieke}, G.~H., {Gordon}, K.~D., {et~al.} 2008, \apj,
  678, 804

\bibitem[{{Eskew} {et~al.}(2012){Eskew}, {Zaritsky}, \& {Meidt}}]{Eskew2012}
{Eskew}, M., {Zaritsky}, D., \& {Meidt}, S. 2012, \aj, 143, 139

\bibitem[{{Ferrara} \& {Dettmar}(1994)}]{Ferrara1994}
{Ferrara}, A. \& {Dettmar}, R.-J. 1994, \apj, 427, 155

\bibitem[{{Galametz} {et~al.}(2012){Galametz}, {Kennicutt}, {Albrecht},
  {Aniano}, {Armus}, {Bertoldi}, {Calzetti}, {Crocker}, {Croxall}, {Dale},
  {Donovan Meyer}, {Draine}, {Engelbracht}, {Hinz}, {Roussel}, {Skibba},
  {Tabatabaei}, {Walter}, {Weiss}, {Wilson}, \& {Wolfire}}]{Galametz2012}
{Galametz}, M., {Kennicutt}, R.~C., {Albrecht}, M., {et~al.} 2012, \mnras, 425,
  763

\bibitem[{{Galametz} {et~al.}(2009){Galametz}, {Madden}, {Galliano}, {Hony},
  {Schuller}, {Beelen}, {Bendo}, {Sauvage}, {Lundgren}, \&
  {Billot}}]{Galametz2009}
{Galametz}, M., {Madden}, S., {Galliano}, F., {et~al.} 2009, \aap, 508, 645

\bibitem[{{Galametz} {et~al.}(2011){Galametz}, {Madden}, {Galliano}, {Hony},
  {Bendo}, \& {Sauvage}}]{Galametz2011}
{Galametz}, M., {Madden}, S.~C., {Galliano}, F., {et~al.} 2011, \aap, 532, A56

\bibitem[{{Galametz} {et~al.}(2010){Galametz}, {Madden}, {Galliano}, {Hony},
  {Sauvage}, {Pohlen}, {Bendo}, {Auld}, {Baes}, {Barlow}, {Bock}, {Boselli},
  {Bradford}, {Buat}, {Castro-Rodr{\'{\i}}guez}, {Chanial}, {Charlot},
  {Ciesla}, {Clements}, {Cooray}, {Cormier}, {Cortese}, {Davies}, {Dwek},
  {Eales}, {Elbaz}, {Gear}, {Glenn}, {Gomez}, {Griffin}, {Isaak}, {Levenson},
  {Lu}, {O'Halloran}, {Okumura}, {Oliver}, {Page}, {Panuzzo}, {Papageorgiou},
  {Parkin}, {P{\'e}rez-Fournon}, {Rangwala}, {Rigby}, {Roussel}, {Rykala},
  {Sacchi}, {Schulz}, {Schirm}, {Smith}, {Spinoglio}, {Stevens}, {Sundar},
  {Symeonidis}, {Trichas}, {Vaccari}, {Vigroux}, {Wilson}, {Wozniak}, {Wright},
  \& {Zeilinger}}]{Galametz2010}
{Galametz}, M., {Madden}, S.~C., {Galliano}, F., {et~al.} 2010, \aap, 518, L55

\bibitem[{{Galliano} {et~al.}(2008){Galliano}, {Dwek}, \&
  {Chanial}}]{Galliano2008}
{Galliano}, F., {Dwek}, E., \& {Chanial}, P. 2008, \apj, 672, 214

\bibitem[{{Galliano} {et~al.}(2011){Galliano}, {Hony}, {Bernard}, {Bot},
  {Madden}, {Roman-Duval}, {Galametz}, {Li}, {Meixner}, {Engelbracht},
  {Lebouteiller}, {Misselt}, {Montiel}, {Panuzzo}, {Reach}, \&
  {Skibba}}]{Galliano2011}
{Galliano}, F., {Hony}, S., {Bernard}, J.-P., {et~al.} 2011, \aap, 536, A88

\bibitem[{{Galliano} {et~al.}(2005){Galliano}, {Madden}, {Jones}, {Wilson}, \&
  {Bernard}}]{Galliano2005}
{Galliano}, F., {Madden}, S.~C., {Jones}, A.~P., {Wilson}, C.~D., \& {Bernard},
  J.-P. 2005, \aap, 434, 867

\bibitem[{{Galliano} {et~al.}(2003){Galliano}, {Madden}, {Jones}, {Wilson},
  {Bernard}, \& {Le Peintre}}]{Galliano2003}
{Galliano}, F., {Madden}, S.~C., {Jones}, A.~P., {et~al.} 2003, \aap, 407, 159

\bibitem[{{Gieren} {et~al.}(2006){Gieren}, {Pietrzy{\'n}ski}, {Nalewajko},
  {Soszy{\'n}ski}, {Bresolin}, {Kudritzki}, {Minniti}, \&
  {Romanowsky}}]{Gieren2006}
{Gieren}, W., {Pietrzy{\'n}ski}, G., {Nalewajko}, K., {et~al.} 2006, \apj, 647,
  1056

\bibitem[{{Griffin} {et~al.}(2010){Griffin}, {Abergel}, {Abreu}, {Ade},
  {Andr{\'e}}, {Augueres}, {Babbedge}, {Bae}, {Baillie}, {Baluteau}, {Barlow},
  {Bendo}, {Benielli}, {Bock}, {Bonhomme}, {Brisbin}, {Brockley-Blatt},
  {Caldwell}, {Cara}, {Castro-Rodriguez}, {Cerulli}, {Chanial}, {Chen},
  {Clark}, {Clements}, {Clerc}, {Coker}, {Communal}, {Conversi}, {Cox},
  {Crumb}, {Cunningham}, {Daly}, {Davis}, {de Antoni}, {Delderfield}, {Devin},
  {di Giorgio}, {Didschuns}, {Dohlen}, {Donati}, {Dowell}, {Dowell}, {Duband},
  {Dumaye}, {Emery}, {Ferlet}, {Ferrand}, {Fontignie}, {Fox}, {Franceschini},
  {Frerking}, {Fulton}, {Garcia}, {Gastaud}, {Gear}, {Glenn}, {Goizel},
  {Griffin}, {Grundy}, {Guest}, {Guillemet}, {Hargrave}, {Harwit}, {Hastings},
  {Hatziminaoglou}, {Herman}, {Hinde}, {Hristov}, {Huang}, {Imhof}, {Isaak},
  {Israelsson}, {Ivison}, {Jennings}, {Kiernan}, {King}, {Lange}, {Latter},
  {Laurent}, {Laurent}, {Leeks}, {Lellouch}, {Levenson}, {Li}, {Li},
  {Lilienthal}, {Lim}, {Liu}, {Lu}, {Madden}, {Mainetti}, {Marliani}, {McKay},
  {Mercier}, {Molinari}, {Morris}, {Moseley}, {Mulder}, {Mur}, {Naylor},
  {Nguyen}, {O'Halloran}, {Oliver}, {Olofsson}, {Olofsson}, {Orfei}, {Page},
  {Pain}, {Panuzzo}, {Papageorgiou}, {Parks}, {Parr-Burman}, {Pearce},
  {Pearson}, {P{\'e}rez-Fournon}, {Pinsard}, {Pisano}, {Podosek}, {Pohlen},
  {Polehampton}, {Pouliquen}, {Rigopoulou}, {Rizzo}, {Roseboom}, {Roussel},
  {Rowan-Robinson}, {Rownd}, {Saraceno}, {Sauvage}, {Savage}, {Savini},
  {Sawyer}, {Scharmberg}, {Schmitt}, {Schneider}, {Schulz}, {Schwartz},
  {Shafer}, {Shupe}, {Sibthorpe}, {Sidher}, {Smith}, {Smith}, {Smith},
  {Spencer}, {Stobie}, {Sudiwala}, {Sukhatme}, {Surace}, {Stevens}, {Swinyard},
  {Trichas}, {Tourette}, {Triou}, {Tseng}, {Tucker}, {Turner}, {Vaccari},
  {Valtchanov}, {Vigroux}, {Virique}, {Voellmer}, {Walker}, {Ward}, {Waskett},
  {Weilert}, {Wesson}, {White}, {Whitehouse}, {Wilson}, {Winter}, {Woodcraft},
  {Wright}, {Xu}, {Zavagno}, {Zemcov}, {Zhang}, \& {Zonca}}]{Griffin2010}
{Griffin}, M.~J., {Abergel}, A., {Abreu}, A., {et~al.} 2010, \aap, 518, L3

\bibitem[{{Grocholski} {et~al.}(2012){Grocholski}, {van der Marel}, {Aloisi},
  {Annibali}, {Greggio}, \& {Tosi}}]{Grocholski2012}
{Grocholski}, A.~J., {van der Marel}, R.~P., {Aloisi}, A., {et~al.} 2012, \aj,
  143, 117

\bibitem[{{Grossi} {et~al.}(2010){Grossi}, {Hunt}, {Madden}, {Vlahakis},
  {Bomans}, {Baes}, {Bendo}, {Bianchi}, {Boselli}, {Clemens}, {Corbelli},
  {Cortese}, {Dariush}, {Davies}, {de Looze}, {di Serego Alighieri}, {Fadda},
  {Fritz}, {Garcia-Appadoo}, {Gavazzi}, {Giovanardi}, {Hughes}, {Jones},
  {Pierini}, {Pohlen}, {Sabatini}, {Smith}, {Verstappen}, {Xilouris}, \&
  {Zibetti}}]{Grossi2010}
{Grossi}, M., {Hunt}, L.~K., {Madden}, S., {et~al.} 2010, \aap, 518, L52

\bibitem[{{Guseva} {et~al.}(2012){Guseva}, {Izotov}, {Fricke}, \&
  {Henkel}}]{Guseva2012}
{Guseva}, N.~G., {Izotov}, Y.~I., {Fricke}, K.~J., \& {Henkel}, C. 2012, \aap,
  541, A115

\bibitem[{{Guseva} {et~al.}(2007){Guseva}, {Izotov}, {Papaderos}, \&
  {Fricke}}]{Guseva2007}
{Guseva}, N.~G., {Izotov}, Y.~I., {Papaderos}, P., \& {Fricke}, K.~J. 2007,
  \aap, 464, 885

\bibitem[{{Guseva} {et~al.}(2000){Guseva}, {Izotov}, \& {Thuan}}]{Guseva2000}
{Guseva}, N.~G., {Izotov}, Y.~I., \& {Thuan}, T.~X. 2000, \apj, 531, 776

\bibitem[{{Guseva} {et~al.}(2003{\natexlab{a}}){Guseva}, {Papaderos}, {Izotov},
  {Green}, {Fricke}, {Thuan}, \& {Noeske}}]{Guseva2003a}
{Guseva}, N.~G., {Papaderos}, P., {Izotov}, Y.~I., {et~al.} 2003{\natexlab{a}},
  \aap, 407, 91

\bibitem[{{Guseva} {et~al.}(2003{\natexlab{b}}){Guseva}, {Papaderos}, {Izotov},
  {Green}, {Fricke}, {Thuan}, \& {Noeske}}]{Guseva2003b}
{Guseva}, N.~G., {Papaderos}, P., {Izotov}, Y.~I., {et~al.} 2003{\natexlab{b}},
  \aap, 407, 105

\bibitem[{{Israel} {et~al.}(1996){Israel}, {Maloney}, {Geis}, {Herrmann},
  {Madden}, {Poglitsch}, \& {Stacey}}]{Israel1996}
{Israel}, F.~P., {Maloney}, P.~R., {Geis}, N., {et~al.} 1996, \apj, 465, 738

\bibitem[{{Israel} {et~al.}(2010){Israel}, {Wall}, {Raban}, {Reach}, {Bot},
  {Oonk}, {Ysard}, \& {Bernard}}]{Israel2010}
{Israel}, F.~P., {Wall}, W.~F., {Raban}, D., {et~al.} 2010, \aap, 519, A67

\bibitem[{{Izotov} {et~al.}(1999){Izotov}, {Chaffee}, {Foltz}, {Green},
  {Guseva}, \& {Thuan}}]{Izotov1999}
{Izotov}, Y.~I., {Chaffee}, F.~H., {Foltz}, C.~B., {et~al.} 1999, \apj, 527,
  757

\bibitem[{{Izotov} {et~al.}(2004){Izotov}, {Papaderos}, {Guseva}, {Fricke}, \&
  {Thuan}}]{Izotov2004}
{Izotov}, Y.~I., {Papaderos}, P., {Guseva}, N.~G., {Fricke}, K.~J., \& {Thuan},
  T.~X. 2004, \aap, 421, 539

\bibitem[{{Izotov} {et~al.}(2006){Izotov}, {Stasi{\'n}ska}, {Meynet}, {Guseva},
  \& {Thuan}}]{Izotov2006}
{Izotov}, Y.~I., {Stasi{\'n}ska}, G., {Meynet}, G., {Guseva}, N.~G., \&
  {Thuan}, T.~X. 2006, \aap, 448, 955

\bibitem[{{Izotov} \& {Thuan}(1998)}]{Izotov1998}
{Izotov}, Y.~I. \& {Thuan}, T.~X. 1998, \apj, 500, 188

\bibitem[{{Izotov} {et~al.}(1994){Izotov}, {Thuan}, \&
  {Lipovetsky}}]{Izotov1994}
{Izotov}, Y.~I., {Thuan}, T.~X., \& {Lipovetsky}, V.~A. 1994, \apj, 435, 647

\bibitem[{{Izotov} {et~al.}(1997){Izotov}, {Thuan}, \&
  {Lipovetsky}}]{Izotov1997}
{Izotov}, Y.~I., {Thuan}, T.~X., \& {Lipovetsky}, V.~A. 1997, \apjs, 108, 1

\bibitem[{{Izotov} {et~al.}(2007){Izotov}, {Thuan}, \&
  {Stasi{\'n}ska}}]{Izotov2007}
{Izotov}, Y.~I., {Thuan}, T.~X., \& {Stasi{\'n}ska}, G. 2007, \apj, 662, 15

\bibitem[{{Juvela} \& {Ysard}(2012{\natexlab{a}})}]{JuvelaYsard2012b}
{Juvela}, M. \& {Ysard}, N. 2012{\natexlab{a}}, \aap, 541, A33

\bibitem[{{Juvela} \& {Ysard}(2012{\natexlab{b}})}]{JuvelaYsard2012a}
{Juvela}, M. \& {Ysard}, N. 2012{\natexlab{b}}, \aap, 539, A71

\bibitem[{{Karachentsev} {et~al.}(2002){Karachentsev}, {Dolphin}, {Geisler},
  {Grebel}, {Guhathakurta}, {Hodge}, {Karachentseva}, {Sarajedini}, {Seitzer},
  \& {Sharina}}]{Karachentsev2002}
{Karachentsev}, I.~D., {Dolphin}, A.~E., {Geisler}, D., {et~al.} 2002, \aap,
  383, 125

\bibitem[{{Karachentsev} {et~al.}(2004){Karachentsev}, {Karachentseva},
  {Huchtmeier}, \& {Makarov}}]{Karachentsev2004}
{Karachentsev}, I.~D., {Karachentseva}, V.~E., {Huchtmeier}, W.~K., \&
  {Makarov}, D.~I. 2004, \aj, 127, 2031

\bibitem[{{Karachentsev} {et~al.}(2003){Karachentsev}, {Sharina}, {Dolphin},
  {Grebel}, {Geisler}, {Guhathakurta}, {Hodge}, {Karachentseva}, {Sarajedini},
  \& {Seitzer}}]{Karachentsev2003}
{Karachentsev}, I.~D., {Sharina}, M.~E., {Dolphin}, A.~E., {et~al.} 2003, \aap,
  398, 467

\bibitem[{{Kelly} {et~al.}(2012){Kelly}, {Shetty}, {Stutz}, {Kauffmann},
  {Goodman}, \& {Launhardt}}]{Kelly2012}
{Kelly}, B.~C., {Shetty}, R., {Stutz}, A.~M., {et~al.} 2012, \apj, 752, 55

\bibitem[{{Kennicutt} {et~al.}(2011){Kennicutt}, {Calzetti}, {Aniano},
  {Appleton}, {Armus}, {Beir{\~a}o}, {Bolatto}, {Brandl}, {Crocker}, {Croxall},
  {Dale}, {Meyer}, {Draine}, {Engelbracht}, {Galametz}, {Gordon}, {Groves},
  {Hao}, {Helou}, {Hinz}, {Hunt}, {Johnson}, {Koda}, {Krause}, {Leroy}, {Li},
  {Meidt}, {Montiel}, {Murphy}, {Rahman}, {Rix}, {Roussel}, {Sandstrom},
  {Sauvage}, {Schinnerer}, {Skibba}, {Smith}, {Srinivasan}, {Vigroux},
  {Walter}, {Wilson}, {Wolfire}, \& {Zibetti}}]{Kennicutt2011}
{Kennicutt}, R.~C., {Calzetti}, D., {Aniano}, G., {et~al.} 2011, \pasp, 123,
  1347

\bibitem[{{Kennicutt}(1998)}]{Kennicutt1998}
{Kennicutt}, Jr., R.~C. 1998, \apj, 498, 541

\bibitem[{{Kennicutt} {et~al.}(2003){Kennicutt}, {Armus}, {Bendo}, {Calzetti},
  {Dale}, {Draine}, {Engelbracht}, {Gordon}, {Grauer}, {Helou}, {Hollenbach},
  {Jarrett}, {Kewley}, {Leitherer}, {Li}, {Malhotra}, {Regan}, {Rieke},
  {Rieke}, {Roussel}, {Smith}, {Thornley}, \& {Walter}}]{Kennicutt2003}
{Kennicutt}, Jr., R.~C., {Armus}, L., {Bendo}, G., {et~al.} 2003, \pasp, 115,
  928

\bibitem[{{Kim} {et~al.}(2009){Kim}, {Kim}, {Hwang}, {Lee}, {Im}, {Karoji},
  {Noumaru}, \& {Tanaka}}]{Kim2009}
{Kim}, M., {Kim}, E., {Hwang}, N., {et~al.} 2009, \apj, 703, 816

\bibitem[{{Kobulnicky} {et~al.}(1999){Kobulnicky}, {Kennicutt}, \&
  {Pizagno}}]{Kobulnicky1999}
{Kobulnicky}, H.~A., {Kennicutt}, Jr., R.~C., \& {Pizagno}, J.~L. 1999, \apj,
  514, 544

\bibitem[{{Kobulnicky} \& {Skillman}(1997)}]{Kobulnicky1997}
{Kobulnicky}, H.~A. \& {Skillman}, E.~D. 1997, \apj, 489, 636

\bibitem[{{Kong} {et~al.}(2002){Kong}, {Cheng}, {Weiss}, \&
  {Charlot}}]{Kong2002}
{Kong}, X., {Cheng}, F.~Z., {Weiss}, A., \& {Charlot}, S. 2002, \aap, 396, 503

\bibitem[{{Lebouteiller} {et~al.}(2012){Lebouteiller}, {Cormier}, {Madden},
  {Galliano}, {Indebetouw}, {Abel}, {Sauvage}, {Hony}, {Contursi}, {Poglitsch},
  {R{\'e}my}, {Sturm}, \& {Wu}}]{Lebouteiller2012}
{Lebouteiller}, V., {Cormier}, D., {Madden}, S.~C., {et~al.} 2012, \aap, 548,
  A91

\bibitem[{{Lee} \& {Skillman}(2004)}]{LeeSkillman2004}
{Lee}, H. \& {Skillman}, E.~D. 2004, \apj, 614, 698

\bibitem[{{Lee} {et~al.}(2006){Lee}, {Skillman}, \& {Venn}}]{Lee2006}
{Lee}, H., {Skillman}, E.~D., \& {Venn}, K.~A. 2006, \apj, 642, 813

\bibitem[{{Lequeux} {et~al.}(1979){Lequeux}, {Peimbert}, {Rayo}, {Serrano}, \&
  {Torres-Peimbert}}]{Lequeux1979}
{Lequeux}, J., {Peimbert}, M., {Rayo}, J.~F., {Serrano}, A., \&
  {Torres-Peimbert}, S. 1979, \aap, 80, 155

\bibitem[{{Leroy} {et~al.}(2011){Leroy}, {Bolatto}, {Gordon}, {Sandstrom},
  {Gratier}, {Rosolowsky}, {Engelbracht}, {Mizuno}, {Corbelli}, {Fukui}, \&
  {Kawamura}}]{Leroy2011}
{Leroy}, A.~K., {Bolatto}, A., {Gordon}, K., {et~al.} 2011, \apj, 737, 12

\bibitem[{{Leroy} {et~al.}(2009){Leroy}, {Walter}, {Bigiel}, {Usero}, {Weiss},
  {Brinks}, {de Blok}, {Kennicutt}, {Schuster}, {Kramer}, {Wiesemeyer}, \&
  {Roussel}}]{Leroy2009}
{Leroy}, A.~K., {Walter}, F., {Bigiel}, F., {et~al.} 2009, \aj, 137, 4670

\bibitem[{{L{\'o}pez-S{\'a}nchez} {et~al.}(2004){L{\'o}pez-S{\'a}nchez},
  {Esteban}, \& {Rodr{\'{\i}}guez}}]{LopezSanchez2004}
{L{\'o}pez-S{\'a}nchez}, {\'A}.~R., {Esteban}, C., \& {Rodr{\'{\i}}guez}, M.
  2004, \apjs, 153, 243

\bibitem[{{Lynds} {et~al.}(1998){Lynds}, {Tolstoy}, {O'Neil}, \&
  {Hunter}}]{Lynds1998}
{Lynds}, R., {Tolstoy}, E., {O'Neil}, Jr., E.~J., \& {Hunter}, D.~A. 1998, \aj,
  116, 146

\bibitem[{{Madden}(2000)}]{Madden2000}
{Madden}, S.~C. 2000, \nar, 44, 249

\bibitem[{{Madden} {et~al.}(2006){Madden}, {Galliano}, {Jones}, \&
  {Sauvage}}]{Madden2006}
{Madden}, S.~C., {Galliano}, F., {Jones}, A.~P., \& {Sauvage}, M. 2006, \aap,
  446, 877

\bibitem[{{Madden} {et~al.}(1997){Madden}, {Poglitsch}, {Geis}, {Stacey}, \&
  {Townes}}]{Madden1997}
{Madden}, S.~C., {Poglitsch}, A., {Geis}, N., {Stacey}, G.~J., \& {Townes},
  C.~H. 1997, \apj, 483, 200

\bibitem[{{Madden} {et~al.}(2012){Madden}, {R{\'e}my}, {Galliano}, {Galametz},
  {Bendo}, {Cormier}, {Lebouteiller}, {Hony}, \& {Hony}}]{Madden2012}
{Madden}, S.~C., {R{\'e}my}, A., {Galliano}, F., {et~al.} 2012, in IAU
  Symposium, Vol. 284, IAU Symposium, ed. R.~J. {Tuffs} \& C.~C. {Popescu},
  141--148

\bibitem[{{Madden} {et~al.}(2013){Madden}, {R{\'e}my-Ruyer}, {Galametz},
  {Cormier}, {Lebouteiller}, {Galliano}, {Hony}, {Bendo}, {Smith}, {Pohlen},
  {Roussel}, {Sauvage}, {Wu}, {Sturm}, {Poglitsch}, {Contursi}, {Doublier},
  {Baes}, {Barlow}, {Boselli}, {Boquien}, {Carlson}, {Ciesla}, {Cooray},
  {Cortese}, {De Looze}, {Irwin}, {Isaak}, {Kamenetzky}, {Karczewski}, {Lu},
  {MacHattie}, {Halloran}, {Parkin}, {Rangwala}, {Schirm}, {Schulz},
  {Spinoglio}, {Vaccari}, {Wilson}, \& {Wozniak}}]{Madden2013}
{Madden}, S.~C., {R{\'e}my-Ruyer}, A., {Galametz}, M., {et~al.} 2013, PASP,
  125, 600

\bibitem[{{Magrini} \& {Gon{\c c}alves}(2009)}]{Magrini2009}
{Magrini}, L. \& {Gon{\c c}alves}, D.~R. 2009, \mnras, 398, 280

\bibitem[{{Masegosa} {et~al.}(1994){Masegosa}, {Moles}, \&
  {Campos-Aguilar}}]{Masegosa1994}
{Masegosa}, J., {Moles}, M., \& {Campos-Aguilar}, A. 1994, \apj, 420, 576

\bibitem[{{McCall} {et~al.}(1985){McCall}, {Rybski}, \& {Shields}}]{McCall1985}
{McCall}, M.~L., {Rybski}, P.~M., \& {Shields}, G.~A. 1985, \apjs, 57, 1

\bibitem[{{Meny} {et~al.}(2007){Meny}, {Gromov}, {Boudet}, {Bernard},
  {Paradis}, \& {Nayral}}]{Meny2007}
{Meny}, C., {Gromov}, V., {Boudet}, N., {et~al.} 2007, \aap, 468, 171

\bibitem[{{Moles} {et~al.}(1994){Moles}, {Marquez}, {Masegosa}, {del Olmo},
  {Perea}, \& {Arp}}]{Moles1994}
{Moles}, M., {Marquez}, I., {Masegosa}, J., {et~al.} 1994, \apj, 432, 135

\bibitem[{{Moll} {et~al.}(2007){Moll}, {Mengel}, {de Grijs}, {Smith}, \&
  {Crowther}}]{Moll2007}
{Moll}, S.~L., {Mengel}, S., {de Grijs}, R., {Smith}, L.~J., \& {Crowther},
  P.~A. 2007, \mnras, 382, 1877

\bibitem[{{Mould} {et~al.}(2000){Mould}, {Huchra}, {Freedman}, {Kennicutt},
  {Ferrarese}, {Ford}, {Gibson}, {Graham}, {Hughes}, {Illingworth}, {Kelson},
  {Macri}, {Madore}, {Sakai}, {Sebo}, {Silbermann}, \& {Stetson}}]{Mould2000}
{Mould}, J.~R., {Huchra}, J.~P., {Freedman}, W.~L., {et~al.} 2000, \apj, 529,
  786

\bibitem[{{Murphy} {et~al.}(2010){Murphy}, {Helou}, {Condon}, {Schinnerer},
  {Turner}, {Beck}, {Mason}, {Chary}, \& {Armus}}]{Murphy2010}
{Murphy}, E.~J., {Helou}, G., {Condon}, J.~J., {et~al.} 2010, \apjl, 709, L108

\bibitem[{{O'Halloran} {et~al.}(2006){O'Halloran}, {Satyapal}, \&
  {Dudik}}]{OHalloran2006}
{O'Halloran}, B., {Satyapal}, S., \& {Dudik}, R.~P. 2006, \apj, 641, 795

\bibitem[{{Ott}(2010)}]{Ott2010}
{Ott}, S. 2010, in Astronomical Society of the Pacific Conference Series, Vol.
  434, Astronomical Data Analysis Software and Systems XIX, ed. {Y.~Mizumoto,
  K.-I.~Morita, \&amp; M.~Ohishi}, 139

\bibitem[{{Papadopoulos} {et~al.}(2004){Papadopoulos}, {Thi}, \&
  {Viti}}]{Papadopoulos2004}
{Papadopoulos}, P.~P., {Thi}, W.-F., \& {Viti}, S. 2004, \mnras, 351, 147

\bibitem[{{Paradis} {et~al.}(2009){Paradis}, {Bernard}, \&
  {M{\'e}ny}}]{Paradis2009}
{Paradis}, D., {Bernard}, J.-P., \& {M{\'e}ny}, C. 2009, \aap, 506, 745

\bibitem[{{Paradis} {et~al.}(2010){Paradis}, {Veneziani}, {Noriega-Crespo},
  {Paladini}, {Piacentini}, {Bernard}, {de Bernardis}, {Calzoletti},
  {Faustini}, {Martin}, {Masi}, {Montier}, {Natoli}, {Ristorcelli}, {Thompson},
  {Traficante}, \& {Molinari}}]{Paradis2010}
{Paradis}, D., {Veneziani}, M., {Noriega-Crespo}, A., {et~al.} 2010, \aap, 520,
  L8

\bibitem[{{Pilbratt} {et~al.}(2010){Pilbratt}, {Riedinger}, {Passvogel},
  {Crone}, {Doyle}, {Gageur}, {Heras}, {Jewell}, {Metcalfe}, {Ott}, \&
  {Schmidt}}]{Pilbratt2010}
{Pilbratt}, G.~L., {Riedinger}, J.~R., {Passvogel}, T., {et~al.} 2010, \aap,
  518, L1

\bibitem[{{Pilyugin} \& {Thuan}(2005)}]{PilyuginThuan2005}
{Pilyugin}, L.~S. \& {Thuan}, T.~X. 2005, \apj, 631, 231

\bibitem[{{Planck Collaboration} {et~al.}(2011{\natexlab{a}}){Planck
  Collaboration}, {Ade}, {Aghanim}, {Arnaud}, {Ashdown}, {Aumont},
  {Baccigalupi}, {Balbi}, {Banday}, {Barreiro}, \&
  et~al.}]{PlanckCollaboration2011XIX}
{Planck Collaboration}, {Ade}, P.~A.~R., {Aghanim}, N., {et~al.}
  2011{\natexlab{a}}, \aap, 536, A19

\bibitem[{{Planck Collaboration} {et~al.}(2011{\natexlab{b}}){Planck
  Collaboration}, {Ade}, {Aghanim}, {Arnaud}, {Ashdown}, {Aumont},
  {Baccigalupi}, {Balbi}, {Banday}, {Barreiro}, \&
  et~al.}]{PlanckCollaboration2011XX}
{Planck Collaboration}, {Ade}, P.~A.~R., {Aghanim}, N., {et~al.}
  2011{\natexlab{b}}, \aap, 536, A20

\bibitem[{{Poglitsch} {et~al.}(1995){Poglitsch}, {Krabbe}, {Madden}, {Nikola},
  {Geis}, {Johansson}, {Stacey}, \& {Sternberg}}]{Poglitsch1995}
{Poglitsch}, A., {Krabbe}, A., {Madden}, S.~C., {et~al.} 1995, \apj, 454, 293

\bibitem[{{Poglitsch} {et~al.}(2010){Poglitsch}, {Waelkens}, {Geis},
  {Feuchtgruber}, {Vandenbussche}, {Rodriguez}, {Krause}, {Renotte}, {van
  Hoof}, {Saraceno}, {Cepa}, {Kerschbaum}, {Agn{\`e}se}, {Ali}, {Altieri},
  {Andreani}, {Augueres}, {Balog}, {Barl}, {Bauer}, {Belbachir}, {Benedettini},
  {Billot}, {Boulade}, {Bischof}, {Blommaert}, {Callut}, {Cara}, {Cerulli},
  {Cesarsky}, {Contursi}, {Creten}, {De Meester}, {Doublier}, {Doumayrou},
  {Duband}, {Exter}, {Genzel}, {Gillis}, {Gr{\"o}zinger}, {Henning},
  {Herreros}, {Huygen}, {Inguscio}, {Jakob}, {Jamar}, {Jean}, {de Jong},
  {Katterloher}, {Kiss}, {Klaas}, {Lemke}, {Lutz}, {Madden}, {Marquet},
  {Martignac}, {Mazy}, {Merken}, {Montfort}, {Morbidelli}, {M{\"u}ller},
  {Nielbock}, {Okumura}, {Orfei}, {Ottensamer}, {Pezzuto}, {Popesso},
  {Putzeys}, {Regibo}, {Reveret}, {Royer}, {Sauvage}, {Schreiber}, {Stegmaier},
  {Schmitt}, {Schubert}, {Sturm}, {Thiel}, {Tofani}, {Vavrek}, {Wetzstein},
  {Wieprecht}, \& {Wiezorrek}}]{Poglitsch2010}
{Poglitsch}, A., {Waelkens}, C., {Geis}, N., {et~al.} 2010, \aap, 518, L2

\bibitem[{{Popescu} \& {Hopp}(2000)}]{Popescu2000}
{Popescu}, C.~C. \& {Hopp}, U. 2000, \aaps, 142, 247

\bibitem[{{Pustilnik} {et~al.}(2003){Pustilnik}, {Kniazev}, {Pramskij},
  {Ugryumov}, \& {Masegosa}}]{Pustilnik2003}
{Pustilnik}, S.~A., {Kniazev}, A.~Y., {Pramskij}, A.~G., {Ugryumov}, A.~V., \&
  {Masegosa}, J. 2003, \aap, 409, 917

\bibitem[{{Reach} {et~al.}(1995){Reach}, {Dwek}, {Fixsen}, {Hewagama},
  {Mather}, {Shafer}, {Banday}, {Bennett}, {Cheng}, {Eplee}, {Leisawitz},
  {Lubin}, {Read}, {Rosen}, {Shuman}, {Smoot}, {Sodroski}, \&
  {Wright}}]{Reach1995}
{Reach}, W.~T., {Dwek}, E., {Fixsen}, D.~J., {et~al.} 1995, \apj, 451, 188

\bibitem[{Roussel(2012)}]{Roussel2012}
Roussel, H. 2012, ArXiv:1205.2576

\bibitem[{{Sauvage} {et~al.}(1990){Sauvage}, {Vigroux}, \&
  {Thuan}}]{Sauvage1990}
{Sauvage}, M., {Vigroux}, L., \& {Thuan}, T.~X. 1990, \aap, 237, 296

\bibitem[{{Saviane} {et~al.}(2008){Saviane}, {Ivanov}, {Held}, {Alloin},
  {Rich}, {Bresolin}, \& {Rizzi}}]{Saviane2008}
{Saviane}, I., {Ivanov}, V.~D., {Held}, E.~V., {et~al.} 2008, \aap, 487, 901

\bibitem[{{Schruba} {et~al.}(2012){Schruba}, {Leroy}, {Walter}, {Bigiel},
  {Brinks}, {de Blok}, {Kramer}, {Rosolowsky}, {Sandstrom}, {Schuster},
  {Usero}, {Weiss}, \& {Wiesemeyer}}]{Schruba2012}
{Schruba}, A., {Leroy}, A.~K., {Walter}, F., {et~al.} 2012, \aj, 143, 138

\bibitem[{{Schulte-Ladbeck} {et~al.}(2001){Schulte-Ladbeck}, {Hopp}, {Greggio},
  {Crone}, \& {Drozdovsky}}]{Schulte-Ladbeck2001}
{Schulte-Ladbeck}, R.~E., {Hopp}, U., {Greggio}, L., {Crone}, M.~M., \&
  {Drozdovsky}, I.~O. 2001, \aj, 121, 3007

\bibitem[{{Sharina} {et~al.}(1996){Sharina}, {Karachentsev}, \&
  {Tikhonov}}]{Sharina1996}
{Sharina}, M.~E., {Karachentsev}, I.~D., \& {Tikhonov}, N.~A. 1996, \aaps, 119,
  499

\bibitem[{{Shetty} {et~al.}(2009{\natexlab{a}}){Shetty}, {Kauffmann}, {Schnee},
  \& {Goodman}}]{Shetty2009a}
{Shetty}, R., {Kauffmann}, J., {Schnee}, S., \& {Goodman}, A.~A.
  2009{\natexlab{a}}, \apj, 696, 676

\bibitem[{{Shetty} {et~al.}(2009{\natexlab{b}}){Shetty}, {Kauffmann}, {Schnee},
  {Goodman}, \& {Ercolano}}]{Shetty2009b}
{Shetty}, R., {Kauffmann}, J., {Schnee}, S., {Goodman}, A.~A., \& {Ercolano},
  B. 2009{\natexlab{b}}, \apj, 696, 2234

\bibitem[{{Skibba} {et~al.}(2011){Skibba}, {Engelbracht}, {Dale}, {Hinz},
  {Zibetti}, {Crocker}, {Groves}, {Hunt}, {Johnson}, {Meidt}, {Murphy},
  {Appleton}, {Armus}, {Bolatto}, {Brandl}, {Calzetti}, {Croxall}, {Galametz},
  {Gordon}, {Kennicutt}, {Koda}, {Krause}, {Montiel}, {Rix}, {Roussel},
  {Sandstrom}, {Sauvage}, {Schinnerer}, {Smith}, {Walter}, {Wilson}, \&
  {Wolfire}}]{Skibba2011}
{Skibba}, R.~A., {Engelbracht}, C.~W., {Dale}, D., {et~al.} 2011, \apj, 738, 89

\bibitem[{{Skillman} {et~al.}(2003){Skillman}, {C{\^o}t{\'e}}, \&
  {Miller}}]{Skillman2003}
{Skillman}, E.~D., {C{\^o}t{\'e}}, S., \& {Miller}, B.~W. 2003, \aj, 125, 610

\bibitem[{{Smith} {et~al.}(2012{\natexlab{a}}){Smith}, {Eales}, {Gomez},
  {Roman-Duval}, {Fritz}, {Braun}, {Baes}, {Bendo}, {Blommaert}, {Boquien},
  {Boselli}, {Clements}, {Cooray}, {Cortese}, {de Looze}, {Ford}, {Gear},
  {Gentile}, {Gordon}, {Kirk}, {Lebouteiller}, {Madden}, {Mentuch},
  {O'Halloran}, {Page}, {Schulz}, {Spinoglio}, {Verstappen}, {Wilson}, \&
  {Thilker}}]{Smith2012b}
{Smith}, M.~W.~L., {Eales}, S.~A., {Gomez}, H.~L., {et~al.} 2012{\natexlab{a}},
  \apj, 756, 40

\bibitem[{{Smith} {et~al.}(2012{\natexlab{b}}){Smith}, {Gomez}, {Eales},
  {Ciesla}, {Boselli}, {Cortese}, {Bendo}, {Baes}, {Bianchi}, {Clemens},
  {Clements}, {Cooray}, {Davies}, {de Looze}, {di Serego Alighieri}, {Fritz},
  {Gavazzi}, {Gear}, {Madden}, {Mentuch}, {Panuzzo}, {Pohlen}, {Spinoglio},
  {Verstappen}, {Vlahakis}, {Wilson}, \& {Xilouris}}]{Smith2012a}
{Smith}, M.~W.~L., {Gomez}, H.~L., {Eales}, S.~A., {et~al.} 2012{\natexlab{b}},
  \apj, 748, 123

\bibitem[{{Stepnik} {et~al.}(2001){Stepnik}, {Abergel}, {Bernard}, {Boulanger},
  {Jones}, {Lagache}, {Lamarre}, {Pajot}, {Le Peintre}, {Giard}, {Meny},
  {Ristorcelli}, {Serra}, {Cambresy}, \& {Torre}}]{Stepnik2001}
{Stepnik}, B., {Abergel}, A., {Bernard}, J.-P., {et~al.} 2001, in Astronomical
  Society of the Pacific Conference Series, Vol. 243, From Darkness to Light:
  Origin and Evolution of Young Stellar Clusters, ed. {T.~Montmerle \&amp;
  P.~Andr{\'e}}, 47

\bibitem[{{Thuan} {et~al.}(1995){Thuan}, {Izotov}, \& {Lipovetsky}}]{Thuan1995}
{Thuan}, T.~X., {Izotov}, Y.~I., \& {Lipovetsky}, V.~A. 1995, \apj, 445, 108

\bibitem[{{Tosi} {et~al.}(2001){Tosi}, {Sabbi}, {Bellazzini}, {Aloisi},
  {Greggio}, {Leitherer}, \& {Montegriffo}}]{Tosi2001}
{Tosi}, M., {Sabbi}, E., {Bellazzini}, M., {et~al.} 2001, \aj, 122, 1271

\bibitem[{{Tully}(1988)}]{Tully1988}
{Tully}, R.~B. 1988, {Nearby galaxies catalog} (Cambridge University Press)

\bibitem[{{Ugryumov} {et~al.}(2003){Ugryumov}, {Engels}, {Pustilnik},
  {Kniazev}, {Pramskij}, \& {Hagen}}]{Ugryumov2003}
{Ugryumov}, A.~V., {Engels}, D., {Pustilnik}, S.~A., {et~al.} 2003, \aap, 397,
  463

\bibitem[{{van Zee} \& {Haynes}(2006)}]{VanZeeHaynes2006}
{van Zee}, L. \& {Haynes}, M.~P. 2006, \apj, 636, 214

\bibitem[{{van Zee} {et~al.}(1996){van Zee}, {Haynes}, {Salzer}, \&
  {Broeils}}]{VanZee1996}
{van Zee}, L., {Haynes}, M.~P., {Salzer}, J.~J., \& {Broeils}, A.~H. 1996, \aj,
  112, 129

\bibitem[{{Walter} {et~al.}(2007){Walter}, {Cannon}, {Roussel}, {Bendo},
  {Calzetti}, {Dale}, {Draine}, {Helou}, {Kennicutt}, {Moustakas}, {Rieke},
  {Armus}, {Engelbracht}, {Gordon}, {Hollenbach}, {Lee}, {Li}, {Meyer},
  {Murphy}, {Regan}, {Smith}, {Brinks}, {de Blok}, {Bigiel}, \&
  {Thornley}}]{Walter2007}
{Walter}, F., {Cannon}, J.~M., {Roussel}, H., {et~al.} 2007, \apj, 661, 102

\bibitem[{{Wilson}(1995)}]{Wilson1995}
{Wilson}, C.~D. 1995, \apjl, 448, L97

\bibitem[{{Wilson}(2005)}]{Wilson2005}
{Wilson}, C.~D. 2005, in IAU Symposium, Vol. 231, Astrochemistry: Recent
  Successes and Current Challenges, ed. D.~C. {Lis}, G.~A. {Blake}, \&
  E.~{Herbst}, 271--280

\bibitem[{{Yang} \& {Phillips}(2007)}]{YangPhillips2007}
{Yang}, M. \& {Phillips}, T. 2007, \apj, 662, 284

\bibitem[{{Ysard} {et~al.}(2012){Ysard}, {Juvela}, {Demyk}, {Guillet},
  {Abergel}, {Bernard}, {Malinen}, {M{\'e}ny}, {Montier}, {Paradis},
  {Ristorcelli}, \& {Verstraete}}]{Ysard2012}
{Ysard}, N., {Juvela}, M., {Demyk}, K., {et~al.} 2012, \aap, 542, A21

\bibitem[{{Ysard} \& {Verstraete}(2010)}]{YsardVerstraete2010}
{Ysard}, N. \& {Verstraete}, L. 2010, \aap, 509, A12

\bibitem[{{Zhu} {et~al.}(2009){Zhu}, {Papadopoulos}, {Xilouris}, {Kuno}, \&
  {Lisenfeld}}]{Zhu2009}
{Zhu}, M., {Papadopoulos}, P.~P., {Xilouris}, E.~M., {Kuno}, N., \&
  {Lisenfeld}, U. 2009, \apj, 706, 941

\bibitem[{{Zibetti} {et~al.}(2009){Zibetti}, {Charlot}, \& {Rix}}]{Zibetti2009}
{Zibetti}, S., {Charlot}, S., \& {Rix}, H.-W. 2009, \mnras, 400, 1181

\bibitem[{{Zubko} {et~al.}(2004){Zubko}, {Dwek}, \& {Arendt}}]{Zubko2004}
{Zubko}, V., {Dwek}, E., \& {Arendt}, R.~G. 2004, \apjs, 152, 211

\end{thebibliography}

%===============================================================
% Appendix
%===============================================================

\appendix
\section{Modified blackbody fits for the Dwarf Galaxy Survey}

% Figures: Modified blackbody fits for all DGS
\begin{figure*}[h!tbp]
\begin{center}
\includegraphics[height=23cm, width=16cm]{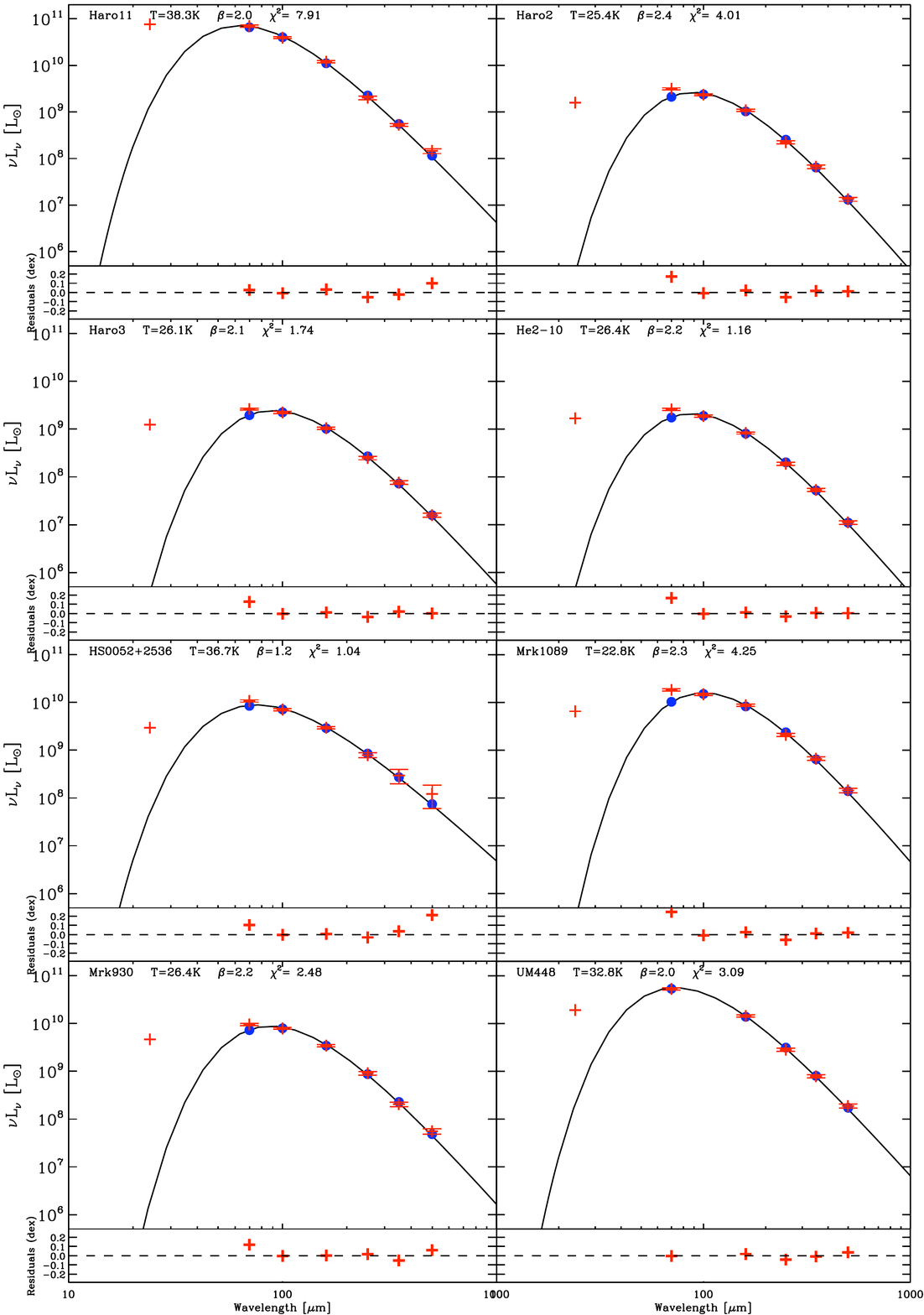}
\caption{Modified blackbody fits of the 70 to 500 \mic\ range for the DGS. The solid black line is the modelled modified blackbody, the blue circles are the modelled points. The red symbols are the observations: crosses are for detections and diamonds are for upper limits. We overlay the MIPS 24 \mic\ point from \cite{Bendo2012}. The  {\it T} and $\beta$ parameters are indicated on the top of each plot along with the $\chi^2$ value of the fit. The bottom panel of each plot indicates the residuals from the fit.}
\end{center}
\end{figure*}

\addtocounter{figure}{-1}
\begin{figure*}[h!tbp]
\begin{center}
\includegraphics[height=23cm, width=16cm]{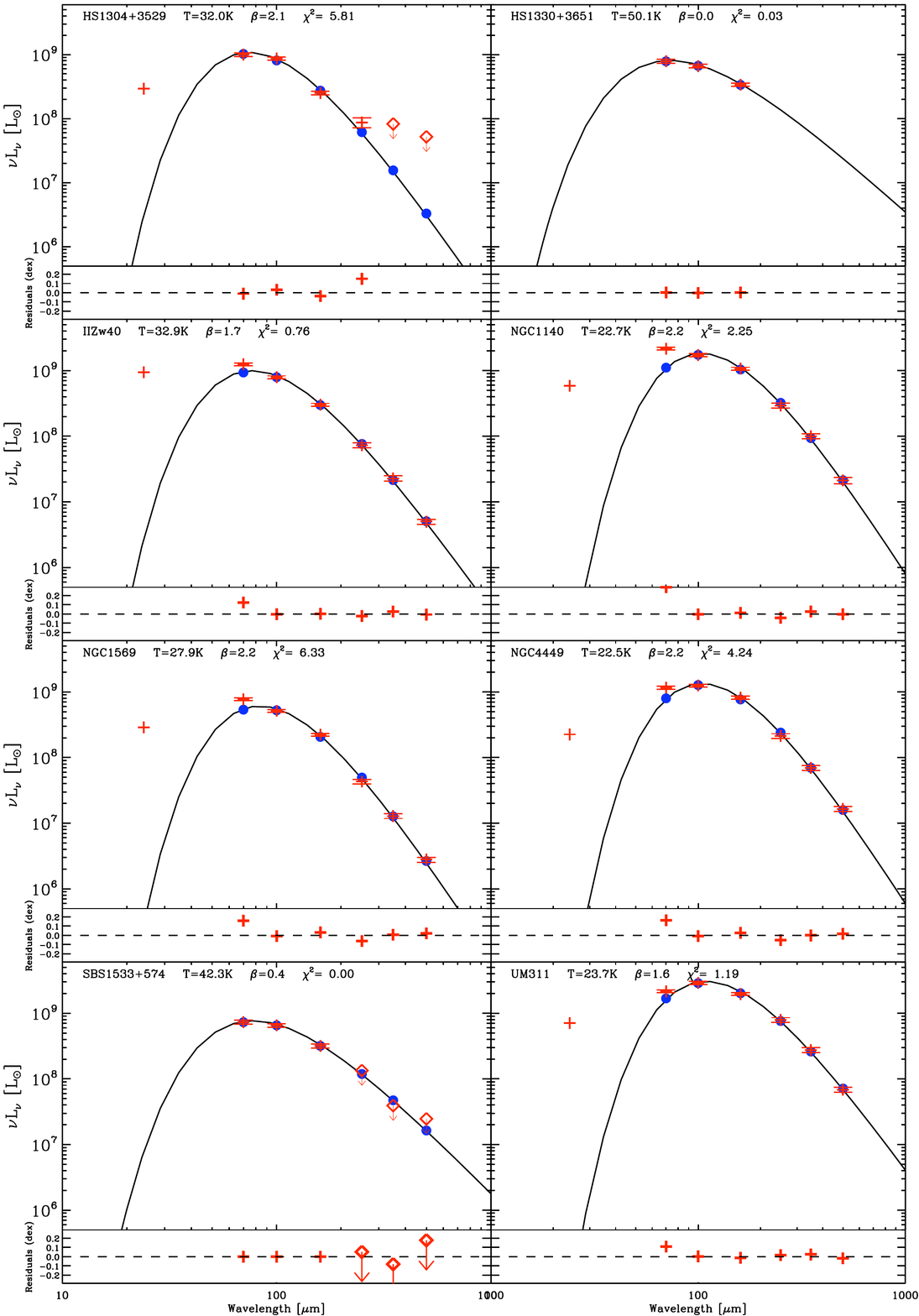}
\caption{ {\it (continued)} Modified blackbody fits of the 70 to 500 \mic\ range for the DGS.
}
\end{center}
\end{figure*}

\addtocounter{figure}{-1}
\begin{figure*}[h!tbp]
\begin{center}
\includegraphics[height=23cm, width=16cm]{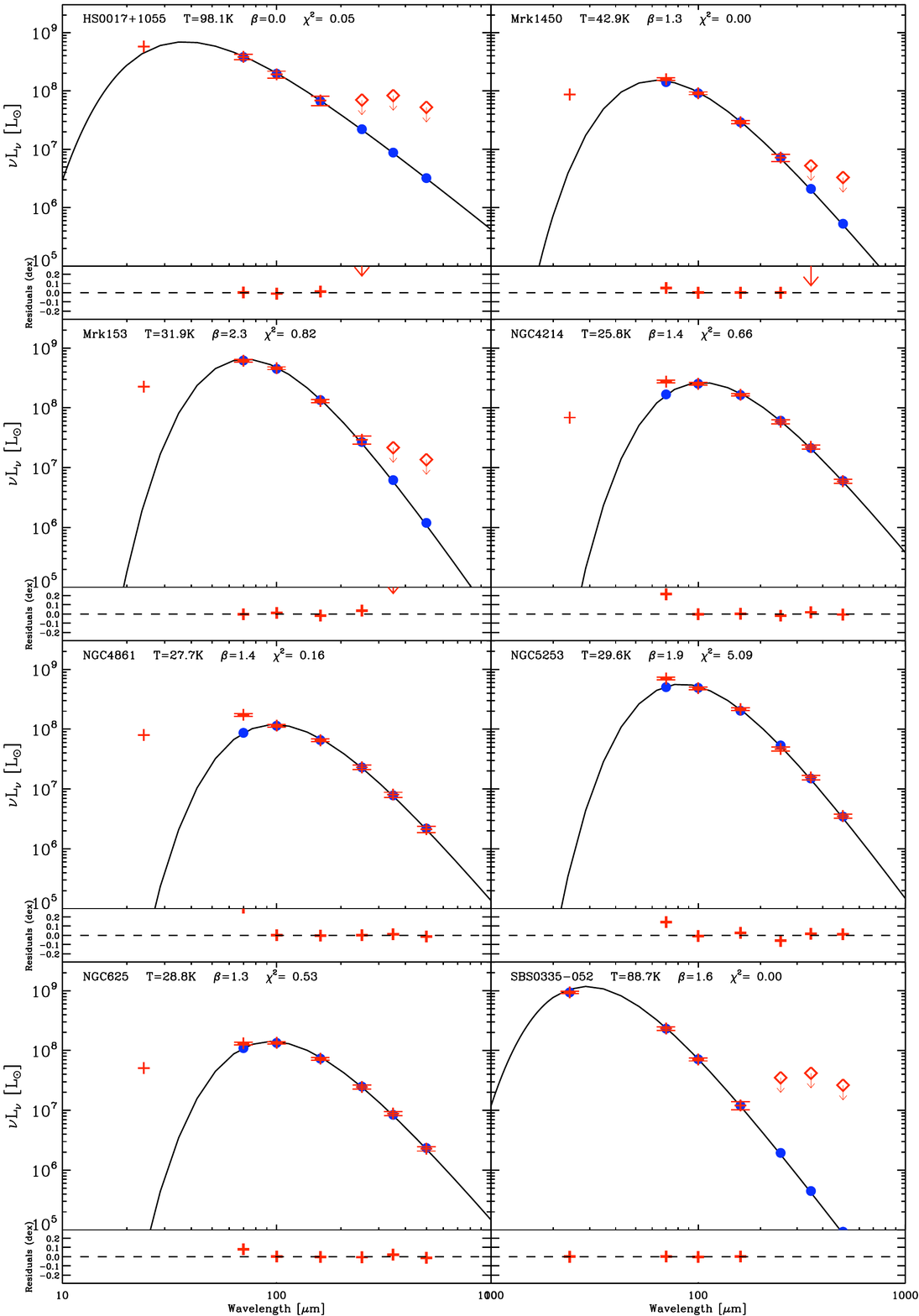}
\caption{{\it (continued)} Modified blackbody fits of the 70 to 500 \mic\ range for the DGS.
Note: for SBS0335-052, we included the 24 \mic\ point in the fit as the 24 \mic\ point fell below the modelled modified blackbody when we just overlaid it on the plot.}
\end{center}
\end{figure*}

\addtocounter{figure}{-1}
\begin{figure*}[h!tbp]
\begin{center}
\includegraphics[height=23cm, width=16cm]{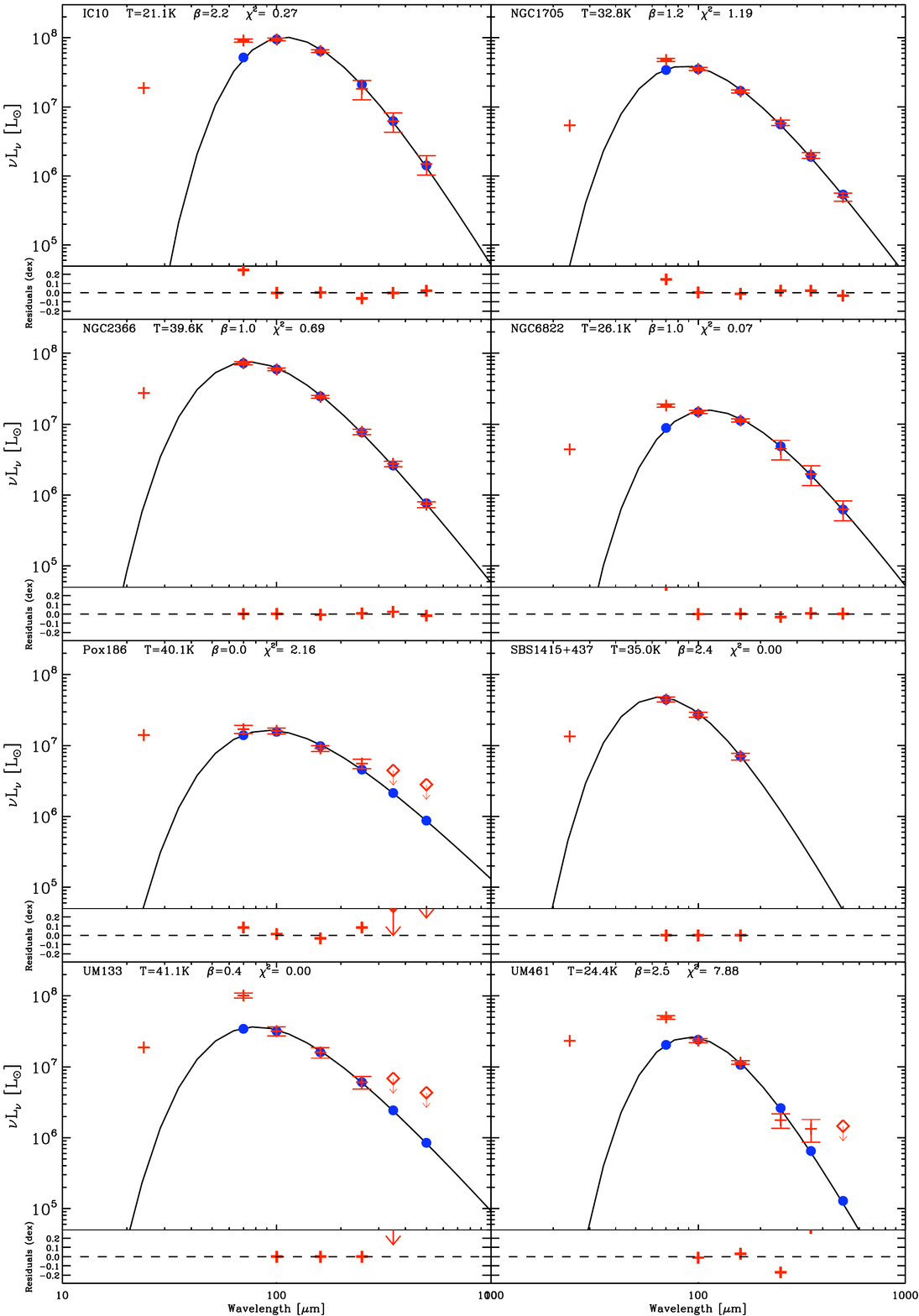}
\caption{{\it (continued)} Modified blackbody fits of the 70 to 500 \mic\ range for the DGS.}
\end{center}
\end{figure*}

\addtocounter{figure}{-1}
\begin{figure*}[h!tbp]
\begin{center}
\includegraphics[width=16cm]{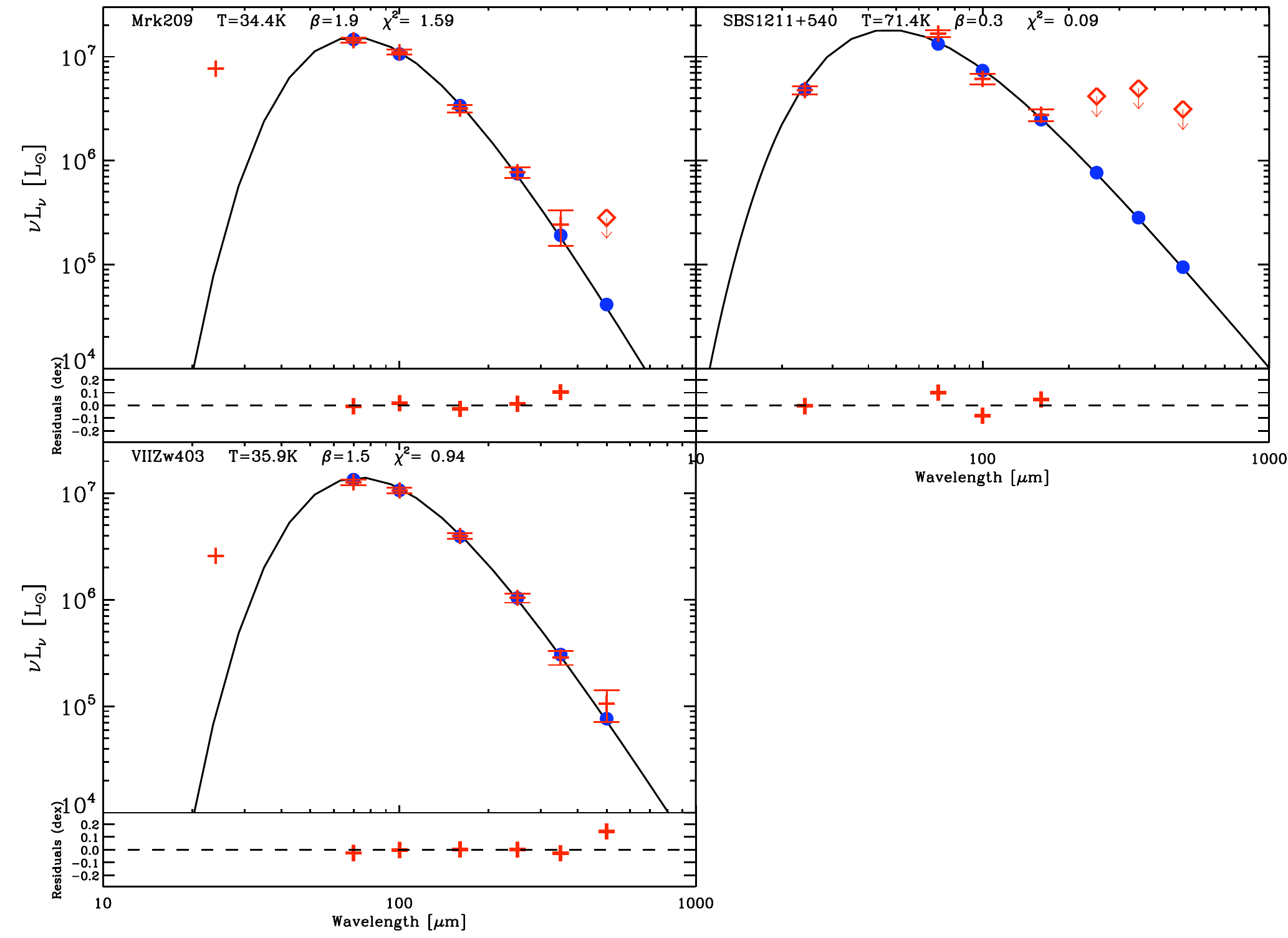}
\caption{{\it (continued)} Modified blackbody fits of the 70 to 500 \mic\ range for the DGS.
Note: for SBS1211+540, we included the 24 \mic\ point in the fit as the 24 \mic\ point fell below the modelled modified blackbody when we just overlaid it on the plot.}
\end{center}
\end{figure*}

\end{document}